%% file: rot01.tex
\input zmacxxx

\input epsf
\def\psib{\psi}

\def\Abbar{\mkern0.5mu\raise0.5pt\hbox{$\not$}\mkern-0.8mu{\bf A}\mkern.1mu}
\def\Ds{\mkern-0.5mu\raise0.5pt\hbox{$\not$}\mkern-.6mu {\bf D}\mkern.1mu}

\preprint{T02/012}

\title{Chiral Anomalies and Topology*}

\centerline{J.~ZINN-JUSTIN}
\centerline{\it
CEA-Saclay, Service de Physique Th\'eorique**,}\centerline{\it  F-91191 Gif-sur-Yvette
 Cedex, FRANCE} 
\medskip
\centerline{\it and}
\bigskip
\centerline{\it Institut de Math\'ematiques de Jussieu--Chevaleret,}
\centerline{\it Universit\'e de Paris VII}
\medskip
\centerline{\it email: zinn@spht.saclay.cea.fr}
\footnote{}{${}^*$Contribution to the  School ``Topology and Geometry in Physics", Rot an der Rot 24--28 September 2001.
}

\footnote{}{${}^{**}$Laboratoire de la Direction des
Sciences de la Mati\`ere du 
Commissariat \`a l'Energie Atomique}

\abstract
When a quantum field theory has a symmetry, global or local like in gauge theories,
in the tree or classical approximation  formal manipulations  lead to believe that the symmetry can also be implemented in the full quantum theory, provided one uses the proper quantization rules. While this is often true, it is not a general property and therefore requires a proof because simple formal manipulations ignore the unavoidable divergences of perturbation theory. \par
The existence of invariant regularizations allows  solving the problem in  most cases but the combination of gauge symmetry and chiral fermions 
leads to subtle issues.  Depending on the specific group and  field content,  {\it anomalies}\/  are found: obstructions to the quantization of chiral gauge symmetries.
Because  anomalies take the form of local polynomials in the fields, are  linked to local group transformations, but vanish for global (rigid) transformations one discovers  that they have a {\it topological}\/ character.
\par
In these notes we review various perturbative and non-perturbative regularization techniques, and show that they leave room for possible anomalies when both gauge fields and chiral fermions are present. We determine the form of anomalies in  simple  examples. We relate anomalies to the index of the Dirac operator in a gauge background. We exhibit gauge instantons that contribute to the anomaly in the example of the $CP(N-1)$ models and $SU(2)$ gauge theories. We briefly mentioned a few physical consequences. \par
For many years the problem of anomalies had been discussed only within
the framework of perturbation theory. New non-perturbative solutions based on lattice regularization have recently been proposed. We describe the so-called overlap and domain wall fermion formulations.
\par
 
\endabstract
\listcontent
\bigskip
\section Symmetries, regularization, anomalies

{\it Divergences.} Symmetries of the classical lagrangian or tree approximation do not always translate into symmetries of the corresponding complete quantum theory. Indeed quantum field theories are  affected by UV divergences that invalidate simple algebraic proofs.
\par 
The origin of UV divergences in field theory is double. First a field contains an infinite number of degrees of freedom. The corresponding divergences  are directly related to renormalization group and reflect the property that even in renormalizable quantum field
theories degrees of freedom remain coupled on all scales.\par 
However another of type of divergences can appear, which is related to the order between quantum operators  and the transition between  classical
and quantum hamiltonians. Such divergences are already present  in perturbation theory in ordinary quantum mechanics,
for instance in the quantization of the geodesic motion on a manifold (like a sphere).
Even in the case of forces linear in the velocities (like a coupling to a magnetic field)  finite ambiguities are found. In local quantum field theory the problem is even more severe because
for a scalar field for example the commutation between field operator $\hat\phi$ and conjugate
momentum $\hat\pi$ takes the form
$$[\hat\phi(x),\hat\pi(y)]=i\hbar\, \delta ^{d-1}(x-y).$$
In a local theory all operators are taken at the same point and thus the commutator is divergent except
in quantum mechanics ($d=1$ with our conventions).\par
Divergences of this nature thus are present as soon as derivative couplings are involved, or when fermions are present. They reflect the property that the knowledge of the classical theory is not sufficient, in general, to completely determine the 
quantized theory.
\smallskip
{\it Regularization.} Regularization is a useful  intermediate step in the renormalization program that consists in modifying the initial theory at short distance, large momentum or otherwise to render perturbation theory finite.  
Note  that from the point of view of Particle Physics  all these modifications alter in some essential way the physical properties of the theory, and thus can only be considered as intermediate steps in the removal of  divergences.\par
When a regularization can be found which preserves the symmetry of the initial classical action,  a symmetric quantum field theory can be constructed.\par 
Momentum cut-off regularization schemes,  based on modifying propagators
at large momenta, are specifically designed
to cut the infinite number of degrees of freedom. With some care these methods will preserve formal symmetries of the un-renormalized theory that correspond to global (space-independent) linear group transformations. 
Problems may, however, arise when the symmetries correspond to non-linear or
local transformations, like in the examples of non-linear $\sigma$ models or
gauge theories, due to the unavoidable presence of derivative couplings.
It is easy to verify that in this case regularizations that only cut momenta do not in general   provide a complete regularization. \par
The addition of regulator fields has in general the same effect as modifying propagators, but offers a few new possibilities, in particular when regulator fields have the wrong spin--statistics connection. Fermion loops in a gauge background can be regularized by such a method.\par
Other methods have to be explored. In many examples dimensional~regularization solves the problem because then the commutator between field and conjugated momentum taken at the same point vanishes. 
However in the case of chiral fermions dimensional regularization fails because chiral properties are specific to even dimensions. \par
Of particular interest is the method of lattice regularization,
because it can be used, beyond perturbation theory, either to discuss the
existence of a quantum field theory, or to determine physical properties of field theories by non-perturbative  numerical techniques. One verifies that
such a regularization indeed specifies an order between quantum operators. It 
therefore  solves the ordering problem in non-linear
$\sigma$-models or non-abelian gauge theories. However, again it fails in presence of chiral
fermions:  the manifestation of this difficulty takes the form of a doubling of the fermion degrees of freedom. Until recently this had prevented a straightforward numerical study of chiral theories.
\smallskip
{\it Anomalies.} 
That no conventional regularization scheme can be found in the case of gauge
theories with chiral fermions is not surprising since we know examples of
theories with anomalies, i.~e.~theories in which a local symmetry present at the
in the tree or classical approximation cannot be implemented in the full quantum theory. Note that  this creates obstructions to the construction of chiral gauge theories because exact gauge symmetry, and thus  the absence of anomaly, is essential for the physical consistency of a gauge theory. \par
Note that we study in these lectures only local anomalies, which can be determined by perturbative calculations;  peculiar global non-perturbative anomalies have also been exhibited.\par
Anomalies are local quantities because they are consequences of short distance
singularities. They are responses to local (space-dependent) group
transformations but vanish for a class of space-independent transformations.
This gives them a topological character that is further confirmed
by their relations with the index of the Dirac operator in a gauge background.
\par
The recently discovered solutions of the Ginsparg--Wilson relation and the method of overlap fermions seem to provide an unconventional solution to the regularization problem in gauge theories with chiral fermions on the lattice. They evade the doubling problem
of fermion because chiral transformations are  no longer strictly local on the lattice, and
relate the problem of anomalies with the invariance of the fermion measure.
The absence of anomalies can then be verified directly on the lattice, and this seems to confirm that the theories that had been discovered anomaly-free in perturbation theory
are also anomaly-free in the non-perturbative lattice construction. 
Therefore the specific problem of lattice fermions was in essence technical rather than reflecting an inconsistency of chiral gauge theories beyond perturbation theory, as one may have feared.\par 
Finally since these new regularization schemes have a natural implementation in five dimensions under the form of domain wall fermions, it again
opens the door to speculations about higher space dimensions. 
\par 
We first discuss the advantages and shortcomings, from the point of view of symmetries,  of three regularization schemes, momentum cut-off, dimensional and lattice regularization. We show that they leave room for possible anomalies when both gauge fields and chiral fermions are present. \par
We then recall the origin and the form
of anomalies, beginning with the simplest example of the so-called abelian anomaly,
i.e.~the anomaly in the conservation of the abelian axial current in gauge theories.
We relate anomalies to the index of a covariant Dirac operator in the background of a gauge field.\par
In the two-dimensional $CP(N-1)$ models and in four-dimensional non-abelian gauge theories we exhibit gauge instantons. We show that they can be classified in terms of a topological charge, space integral of the chiral anomaly. 
The existence of gauge field configurations that contribute to the anomaly
has direct physical implications, like possible strong CP violation and the solution to the $U(1)$ problem.
\par
We  examine the form of the anomaly for a general axial current, and infer conditions for gauge theories that couple differently to fermion chiral components to be anomaly-free. 
 A few  physical applications are also briefly mentioned.\par
Finally the formalism of overlap fermions on the lattice and the role of the Ginsparg--Wilson relation are explained. The alternative construction of domain wall fermions is explained, starting from the basic mechanism of zero-modes in supersymmetric quantum mechanics.\par
{\it Conventions.} Throughout these notes we work in {\it euclidean space}\/ (with imaginary or euclidean time), and this also implies a formalism of {\it euclidean fermions}. 
\section Momentum cut-off regularization

We first discuss methods that work in the continuum (compared to lattice methods) and at fixed dimension (unlike dimensional regularization). The idea then is to modify  field propagators beyond a large  momentum cut-off  to render all Feynman diagrams  convergent. 
However the regularization has to satisfy one important condition:
the inverse of the regularized propagator must remain a {\it smooth}\/ function of the
momentum ${\bf p}$. Indeed  singularities in   momentum
variables generate, after Fourier transformation, contributions to the 
large distance  behaviour of the propagator, and regularization should modify
the theory only at short distance. 
\subsection Matter fields: propagator modification

{\it Scalar fields.} A simple modification of the propagator improves the convergence of Feynman diagrams at large momentum. For example in the case of the action of the  self-coupled scalar field,
$$ {\cal S}(\phi)= \int \d^{d}x \left[\ud\phi (x) (-\nabla_x^2 +m^2 )\phi (x)+V_{\rm I}\bigl(\phi(x)\bigr) \right] , 
\eqnd\eactscdiv $$
the  propagator in Fourier space $1/(m^2+p^2)$ can be replaced by
$$\Delta  _{\rm B}(p)= \left({1 \over p^2+m^2} \right)_{ {\rm reg.}},$$
with
$$\Delta ^{-1}_{\rm B}(p)= (p^2+m^2)\prod_{i=1}^n (1+p^2/M_i^2).\eqnd\epropaul $$
The masses $M_i$ are proportional to the momentum cut-off $ \Lambda $,
$$M_i=\alpha _i\Lambda\,,\quad   \alpha _i>0\,.$$
If the degree $ n $ is chosen large enough  all diagrams become convergent.
In the formal large cut-off limit, at parameters $\alpha $ fixed, the initial propagator is recovered. This is the spirit of momentum or Pauli--Villars's regularization.\par
Note that such a propagator cannot be derived from a hermitian hamiltonian.
Hermiticity of the hamiltonian  implies that if the propagator is, as above, a
rational function, it must be a sum of poles with positive residues and thus
cannot decrease faster than $1/p^{2}$.\par
While this modification can be implemented also in Minkowski space because the regularized propagators decreases in all complex $p^2$ directions (except real negative), 
in euclidean time more general modifications are possible. Schwinger's proper time representation suggests: 
$$ \Delta _{\rm B}(p)= \int
^{\infty}_{0}\d t\,\rho  (t\Lambda^2 )\e^{-t  (p^2+m^2 )}, \eqnd \epropreg  $$
in which the function $ \rho  (t  ) $ is positive (to ensure that $\Delta _{\rm B}(p)$ does not vanish and thus is invertible) and satisfies the condition 
$$ |1-\rho(t)|<C\e^{-\sigma t}\ (\sigma >0) \ {\rm for}\ t \to +\infty\,.$$
By choosing a function $\rho(t)$ that decreases fast enough for $t\to0$
the behaviour of the propagator can be arbitrarily improved. If  $\rho (t  )=O(t^n)$ the behaviour \epropaul\ is recovered. Another example is:
$$ \rho (t)=\theta(t- 1), \eqnn $$
$ \theta  (t  ) $ being the step function, which leads to exponential decrease:
$$  \Delta _{\rm B}(p)={  \e^{- (p
^{2}+m^{2}) / \Lambda^ 2 } \over p^2+m^2}. \eqnd \erhot  $$
As the example shows, it is thus possible to find in this
more general class  propagators without unphysical
singularities, but  they do not follow from a hamiltonian
formalism because continuation to real time becomes impossible.
\smallskip
{\it Spin $1/2$ fermions.}
For spin $1/2$ fermions  similar methods are applicable. 
To the free Dirac action 
$${\cal S}_{{\rm F}0}=\int\d^d x\,\bar\psi(x) (\sla{\partial }+m)   \psi(x)  \  , \eqnn  $$
corresponds in Fourier representation the propagator $1/( m+i\sla{p})$. We  can replace it  by 
the regularized propagator $ \Delta_{\rm F}(p)$,
$$ \Delta ^{-1}_{\rm F}(p)= (m+i\sla{p}) \prod_{i=1}^n (1+p^2/M_i^2)  .\eqnd\eferegul $$
Note that we use the standard notation $ \sla{p} \equiv p_\mu \gamma _\mu $, with euclidean fermion conventions, analytic continuation to imaginary or euclidean time of the usual Minkowski fermions, and  hermitian matrices $\gamma _\mu$. 
\smallskip
{\it Remarks.}
Momentum cut-off regularizations have several advantages: one can work at fixed dimension and in the continuum. However, two potential weaknesses have to be stressed:\par
(i) The generating functional of correlation function ${\cal Z}(J)$ obtained by adding to the action \eactscdiv\ a source term for fields 
$${\cal S}(\phi)\mapsto {\cal S}(\phi)-\int \d^d x\,J(x)\phi(x),$$
can be written, 
$$ {\cal Z} (J) =\det^{1/2}(\Delta _{\rm B})  \exp\left[-{\cal V}_{\rm I}
\left({\delta / \delta J } \right) \right] \exp\left(\ud\int\d^d x\,\d^d y\, J(x) \Delta_{\rm B}(x-y)
J(y)\right), \eqnd\eregpert $$  
where the determinant is generated by the gaussian integration, and
$${\cal V}_{\rm I}(\phi ) \equiv \int\d^d x\,V_{\rm I}\bigl(\phi(x)\bigr). $$
None of the momentum cut-off  regularizations  described so far can deal with the determinant. As long as the determinant  is a divergent constant that cancels in normalized correlation functions this is not a problem, but in the case of a determinant  in the background of an external field (which generates a set of  one-loop diagrams)  this may become a serious issue.\par
(ii) This problem is related to another one: Some models have even in simple
quantum mechanics divergences or ambiguities due to problem of order between quantum operators in products of position and momentum variables. A class of Feynman diagrams then cannot be regularized by this method. 
Quantum field theories where this problem occurs include models with non-linear  or gauge symmetries.  
\medskip
{\it Global linear symmetries.} To implement symmetries of the classical
action in the quantum theory we need a regularization scheme that preserves
the symmetry. This requires some care but can always be achieved for linear
global symmetries, i.e.~symmetries that correspond to transformations of the
fields of the form
$$\phi_R(x)=R\, \phi(x)\,, $$
where $R$ is a constant matrix. The main reason is that in the quantum hamiltonian
field operators and conjugate momenta are not mixed by the transformation, and therefore the order of operators is to some extent irrelevant. To take an example directly relevant here,
a theory with massless fermions may, in four dimensions, have a chiral symmetry
$$\psi_\theta (x) =e^{i\theta \gamma_5}\psi (x) ,\qquad \bar \psi_\theta(x)
=\bar \psi (x)\e^{i\theta \gamma_5}\,. $$  
The substitution \eferegul~(for $m=0$) preserves chiral symmetry.
Note the importance here of being able to work in fixed dimension four because
chiral symmetry is defined only in even dimensions. In particular the invariance of the integration measure $[\d\bar\psi (x)\d\psi (x)]$ relies on the property that $\tr \gamma_5=0$.
\subsection Regulator fields

Regularization in the form \epropaul\ or \eferegul\ has 
another equivalent formulation based on the introduction of regulator fields.
\smallskip
{\it The scalar case.}
 In the case of scalar fields  to regularize the action   \eactscdiv\ for the scalar field $\phi$
one introduces additional dynamical fields $ \phi_{r}$, $ r=1,\ldots ,r_{\rm max}$,
and considers the modified action $ {\cal S}_{ \rm reg. } (\phi ,\phi_{k}
 )$:  
$$ \eqalignno{ {\cal S}_{  \rm reg. }  (\phi ,\phi_{r} ) &=
\int \d^{d}x \left[{ 1 \over 2}\phi \left(-\nabla^2 +m^2\right)\phi + \sum{1 \over 2 z _r}\phi_r \left(-\nabla^2 +M^2_r \right)\phi_r \right.& \cr & \quad \left.+V_{\rm I} (\phi +\textstyle{ \sum_r}\phi_r ) \right] .&
\eqnd \eactregu  \cr}$$ 
With the action \eactregu~any internal $ \phi $ propagator is replaced
by the sum of the $ \phi $ propagator and all the $\phi_r $ propagators $
 z _r / (p^2+M^2_r  )$. For an appropriate choice of the constants
$z _r$,  after integration over the regulator fields, the form \epropaul\ is recovered. Note that the condition of cancellation of the $1/p^2$ contribution at large momentum implies
$$1+\sum_r z_r=0\,. $$
Therefore not all $z_r$ can be positive and  some of the fields $\phi_r$ necessarily are unphysical.
\smallskip
{\it Fermions.} The fermion inverse propagator \eferegul\ can be written
$$\Delta ^{-1}_{\rm F}(p)= (m+i\sla{p}) \prod_{r=1}^{r_{\rm max}}(1+i\sla{p}/M_r)(1-i\sla{p}/M_r).$$
This indicates that again the same form can be obtained by a set of regulator fields $\{\bar\psi_{r\pm}\psi_{r\pm}\}$.
One replaces the kinetic part of the action by
$$\eqalignno{\int\d^d x\,\bar\psi(x) (\sla{\partial }+m)   \psi(x) &\mapsto
\int\d^d x\,\bar\psi(x) (\sla{\partial }+m) \psi(x) \cr&\quad+\sum_{r,\epsilon =\pm} {1\over z_{r\epsilon }}
\int\d^d x\,\bar\psi_{r\epsilon }(x) (\sla{\partial }+\epsilon M_r) \psi_{r\epsilon }(x).&\eqnn \cr}$$
In the same way in the interaction the fields $\psi$ and $\bar\psi$ are replaced
by the sums
$$\psi \mapsto \psi +\sum_{r,\epsilon }\psi_{r\epsilon }\,,\quad \bar \psi \mapsto \bar\psi +\sum_{r,\epsilon }\bar\psi_{r\epsilon }\,.$$
Again for a proper choice of the constants $z_r$, after integration over the regulator fields
the form \eferegul\ is recovered. Note in particular that for $m=0$ one finds
$z_{r+}=z_{r-}$. This indicates how chiral symmetry is preserved by the regularization, although the regulators are massive: by fermion doubling. The fermions $\psi _+$ and $\psi _-$ are chiral partners.
For a pair $\psi\equiv(\psi _+,\psi_-)$, $\bar\psi\equiv(\bar\psi _+,\bar\psi_-)$ the action can be written
$$\int\d^d x\,\bar\psi(x) \left(\sla{\partial }\otimes{\bf 1}+M{\bf 1}\otimes \sigma _3  \right)\psi(x),$$
where the first matrix $\bf 1$ and the Pauli matrix $\sigma _3$ act in $\pm$ space.
The spinors then transform like
$$\psi_\theta (x) =e^{i\theta \gamma_5\otimes \sigma _1}\psi (x) ,\qquad \bar \psi_\theta(x)
=\bar \psi (x)\e^{i\theta \gamma_5\otimes \sigma _1}\,, $$  
because $\sigma _1$ anticommutes with $\sigma _3$.\par
\subsection Abelian gauge theory

The problem of matter in presence of a gauge field can be decomposed into two steps, first matter
in an external gauge field, and then the integration over the gauge field.
For  gauge fields we choose  a covariant gauge, in such a way that power counting is the same as for scalar fields.
\medskip
{\it Charged fermions in a gauge background.} The new problem that arises in presence of a gauge
field is that only covariant derivatives are allowed, because gauge invariance is essential for the physical consistency of the theory. The regularized action in a gauge background now reads 
$$ {\cal S}  (\bar \psi ,\psi ,A   ) = \int \d^{d}x\,
\bar \psi (x) \left(m + \Dbar\right)\prod_r \left(1- {\Dbar^2
/  M_r^2}\right)  \psi (x), \eqnn $$ 
where ${\rm D}_\mu$ is the covariant derivative
$${\rm D}_\mu=\partial _\mu+ie A_\mu\,.$$
Note that up to this point the regularization, unlike dimensional or lattice regularizations,
preserves a possible chiral symmetry for $m=0$. \par
The higher order covariant derivatives  however generate new, more singular, gauge interactions and it is no longer clear whether the theory can be rendered finite. 
\par
Fermion correlation functions in the gauge background   are generated by:
$${\cal Z} (\bar\eta , \eta;A  )  = \int \left[
\d  \psi (x) \d \bar \psi (x) \right]
\exp\left[- {\cal S}( \bar\psi ,\psi,A)+\int \d^{d}x \, \bar \eta (x)\psi (x)+\bar \psi (x)\eta (x)
\right] ,  \eqnd\eFIZfer $$ 
where $\bar\eta ,\eta $ are Grassmann sources. Integrating over fermions explicitly we obtain
$$\eqalignno{  {\cal Z} (\bar\eta , \eta;A  ) & ={\cal Z}_0 ( A  )  \exp\left[-\int\d^d x\,\d^d y\,  \bar \eta(y)\Delta_{\rm F}(A;y,x)\eta(x)\right],& \eqnn \cr
{\cal Z}_0 ( A  ) & ={\cal N}\det \left[\left(m + \Dbar\right)\prod_r \left(1- {\Dbar^2
/  M_r^2}\right) \right],\cr }$$
where $\cal N$ is a gauge field-independent normalization and $\Delta_{\rm F}(A;y,x)$
the fermion propagator in an external gauge field.\par
Diagrams constructed from $\Delta_{\rm F}(A;y,x)$ belong to loops with gauge field propagators,
and therefore can be rendered finite if the gauge field propagator can be improved, a condition that we check below.
The other problem involves the determinant that generates closed fermion loops in a gauge background.
Using $\ln\det=\tr\ln$ we find
$$\ln {\cal Z}_0 ( A  ) =\tr\ln\left(m + \Dbar\right)+\sum_r \tr\ln \left(1- {\Dbar^2
/  M_r^2}\right)-(A=0),$$
or using the anticommutation of $\gamma _5$ with $\Dbar$
$$\det(\Dbar+m)= \det \gamma _5(\Dbar+m) \gamma _5=\det(m-\Dbar), $$
$$\ln {\cal Z}_0 ( A  ) =\ud\tr\ln\left(m^2 -\Dbar^2\right)+\sum_r \tr\ln \left(1- {\Dbar^2
/  M_r^2}\right)-(A=0),$$
We see that the regularization has no effect from the point of view of power counting
on the determinant, and therefore on one-loop diagrams of the form of fermion closed loops with external gauge fields, a problem that requires an additional regularization. 
\smallskip
{\it The fermion determinant.}
The fermion determinant can finally be regularized by adding to the action a
boson  regulator field  with fermion spin, and therefore a propagator similar to $\Delta_{\rm F}$
but with different masses
$${\cal S}_{\rm B}(\bar\phi ,\phi ;A)=  \int \d^{d}x\,
\bar \phi   (x) \left(M_0^{\rm B} + \Dbar\right)\prod_{r=1} \left(1- {\Dbar^2
/  (M^{\rm B}_r)^2}\right)  \phi  (x). \eqnn $$ 
The integration over the boson ghost fields $\bar\phi ,\phi $ adds to $\ln{\cal Z}_0$
the quantity
$$\delta \ln {\cal Z}_0 (A) =-\ud\tr\ln\left((M_0^{\rm B})^2 - \Dbar^2\right)-\sum_{r=1} \tr\ln \left(1- {\Dbar^2 /  (M^{\rm B}_r)^2}\right)-(A=0).$$
Expanding in inverse powers of $\Dbar$ one adjusts the masses to cancel as many powers as possible.
However, the unpaired  initial fermion mass $m$ is the source of a problem. The corresponding determinant can only be regularized with an unpaired  boson $M_0^{\rm B}$. In the chiral limit
$m=0$ we have two options: either we give a chiral charge to the boson field and the mass $M_0^{\rm B}$ breaks chiral symmetry, or we leave it invariant in a chiral transformation.
Then we find the determinant of the transformed operator
$$\e^{i\theta \gamma _5(x)}\Dbar  \e^{i\theta \gamma _5(x)} (\Dbar+M_0^{\rm B})^{-1}.$$
For $\theta (x)$ constant $\e^{i\theta \gamma _5} $ anticommutes with $\Dbar$ and cancels. Otherwise
a non-trivial contribution remains. The method thus suggests possible difficulties  with space-dependent chiral transformations.\par
Since actually the problem reduces to the study of a determinant in an external background
one can study it directly, as we will starting with section \label{\ssAbAnom}.  
One examines whether it is possible to define some regularized form in a way consistent with chiral symmetry. When this is possible one then inserts the one-loop renormalized diagrams in the general diagrams regularized by the preceding cut-off  methods. 
\smallskip
{\it Boson determinant in  a gauge background.} The boson determinant can be 
regularized by introducing a massive scalar charged fermion. It can also be
expressed in terms of the statistical operator using Schwinger's representation
($\tr\ln=\ln\det$)
$$\ln \det H-\ln \det H_0=\tr\int_0^\infty {\d t\over t}\left[\e^{-t H_0}-\e^{-t H}\right],$$
where the operator $H$ is analogous to a non-relativistic hamiltonian in a magnetic field, 
$$H=-{\rm D}_\mu {\rm D}_\mu +m^2,\quad H_0=-\nabla ^2+m^2.$$
The integral over time is regularized by cutting it for $t$ small, integrating from
$t=1/\Lambda ^2$.
\smallskip
{\it The gauge propagator.} For the free gauge action in a covariant gauge usual derivatives can be used because in an abelian theory the gauge field is neutral.
The tensor $ F_{\mu \nu} $ is gauge invariant and the action for the scalar $\partial_\mu A_\mu$ is arbitrary.  Therefore the  large momentum behaviour of the gauge field propagator can be arbitrarily improved.
\subsection Non-abelian gauge theories

Compared with the abelian case, the new features of the non-abelian gauge  action are the presence of gauge field self-interactions and ghost terms. For future purpose we define our notation. We introduce the covariant derivative, as acting on matter field,
$${\bf D}_\mu =\partial _\mu+ {\bf A}_\mu(x)\,, \eqnd\enonabcov   $$
where ${\bf A}_\mu$ is an anti-hermitian matrix, and  the curvature tensor
${\bf F}_{\mu \nu}$ 
$${\bf F}_{\mu \nu}=[{\bf D}_\mu,{\bf D}_\nu]= \partial_\mu {\bf A}_\nu -\partial _\nu {\bf A}_\mu +[{\bf A}_\mu ,{\bf A}_\nu ].\eqnd\enonabcur $$
The pure gauge action then is
$${\cal S}({\bf A}_\mu)= -{1 \over 4g^2}\tr \int \d^d x\,\tr{\bf F}_{\mu \nu}(x) {\bf  F}_{\mu \nu}(x) ,\eqnn
$$ 
In the covariant gauge
$${\cal S}_{\rm gauge}({\bf A}_\mu)=-{1\over 2\xi}\int\d^d x\,\tr(\partial _\mu {\bf A}_\mu)^2, $$
the ghost field action takes the form:
$${\cal S}_{\rm ghost}({\bf A}_\mu,\bar{\bf C},{\bf C})=- \int \d^d x\,\tr\bar{\bf C}\, \partial_{\mu} \left( \partial _\mu{\bf C} +[{\bf A}_\mu , {\bf C}] \right).   $$
The ghost fields thus have a simple $\delta_{ab}/p^2$ propagator and
canonical dimension one in four dimensions.  \par
The problem of regularization in non-abelian gauge
theories has several features in common with the abelian case, as well as with
the non-linear $\sigma$-model. The regularized gauge action takes the form:
$${\cal S}_{\rm reg.} ({\bf A}_{\mu} ) =- \int \d^d x\,\tr{\bf F}_{\mu\nu}
P\left(\left.{\bf D}^2\right/ \Lambda^2\right){\bf F}_{\mu\nu}\, , \eqnn $$
in which $P$ is a polynomial of arbitrary degree. In the same way the gauge
function $\partial_{\mu}{\bf A}_{\mu}$ is changed into:
$$\partial_{\mu}{\bf A}_{\mu} \longmapsto \left.Q\left(\partial^2 \right/
\Lambda^2 \right) \partial_{\mu}{\bf A}_{\mu}\, , \eqnn $$
in which $Q$ is a polynomial of same degree as $P$. As a consequence both the
gauge field propagator and the ghost propagator can be arbitrarily  improved. However, as in the abelian case, the covariant derivatives
generate new interactions that are more singular. It is easy to verify that
the power counting of one-loop diagrams is unchanged while higher order
diagrams can be made convergent by taking the degrees of $P$ and $Q$ large
enough:  Regularization by higher derivatives takes care of all diagrams except, as  in all geometric models, some one-loop diagrams (and thus subdiagrams). \par 
As with charged matter the one-loop diagrams have to be examined separately.
For fermion matter  it is however still possible as, in the abelian
case, to add a set of regulator fields, massive fermions and bosons with
spin.  In the chiral situation
 the problem of the compatibility between the gauge symmetry and the quantization is reduced to an explicit verification of the WT identities for the one-loop
diagrams. Note that the preservation of gauge symmetry is necessary 
for the cancellation of unphysical states in physical amplitudes, and
thus essential to the physical relevance of the quantum field theory.
\section Other regularization schemes

The other regularization schemes we now discuss have the common property
that they modify in some essential way the structure of space--time,
dimension regularization because it relies on defining Feynman diagrams for non-integer dimensions, lattice regularization because continuum space
is replaced by a discrete lattice.
\subsection Dimensional regularization

Dimensional regularization involves continuation of Feynman diagrams in the parameter $ d $ ($ d $ is
the space dimension) to arbitrary complex values, and therefore seems to have no meaning outside perturbation theory. However this regularization
very often leads to the simplest perturbative calculations.\par
In addition it solves with the problem of commutation of quantum operators in local field theories. Indeed commutators for example in the case of a scalar field  take the form
$$[\hat\phi(x),\hat \pi (y) ]=i\hbar\, \delta^{d-1}(x-y)= i \hbar (2\pi)^{1-d}\int \d^{d-1}p\,\e^{ip(x-y)}\,, $$
where $\hat\pi(x)$ is the momentum conjugated to the field $\hat \phi(x)$. As we have already stressed, in a local theory all fields are taken at same point, and therefore a commutation in the product $\hat\phi(x) \hat \pi ( x)$
generates a divergent contribution (for $d>1$) proportional to 
$$\delta^{d-1}(0)=(2\pi)^{1-d}\int \d^{d-1}p\,.$$
The rule of dimensional regularization imply the consistency of the change of variables $p\mapsto \lambda p$ and thus $ \int\d^d p=0$, in contrast to momentum regularization where it is proportional to a power of the cut-off. 
Therefore the order between operators becomes irrelevant because the commutator vanishes. Dimensional regularization thus is applicable to geometric models where these problems of quantization occur, like  non-linear $\sigma $ models or gauge theories.\par 
Its use, however, requires   some care in massless theories. For instance in a massless theory in two dimensions integrals of the form
$  \int  \d^d k/ k^2 $    are met. They again vanish in dimensional regularization  for the same reason. However here they correspond to an unwanted cancellation between UV and IR logarithmic divergences. 
\par
More important here, it is not applicable when some essential property of the field
theory is specific to the initial dimension. An example is provided by theories containing fermions in which parity symmetry is violated.
\medskip
{\it Fermions.} 
For fermions belonging to the
fundamental representation of the spin group ${\rm Spin}(d) $ the strategy is the same.
The spin problem can be reduced to the calculation of traces of $ \gamma $
matrices. Therefore only one additional prescription for the trace of the
unit matrix is needed. There is no natural 
continuation since odd and even dimensions behave differently. Since no algebraic manipulation depends on the explicit value of the trace, any smooth continuation in the neighbourhood of the relevant dimension is satisfactory. A convenient choice  is to
take the trace constant. In even dimension as long as only $ \gamma_{\mu} $
matrices are involved no other problem arises. However no dimensional continuation that preserves all properties of $\gamma_{d+1}$ can be found.  This leads to serious difficulties if  $ \gamma_{d+1} $  in the calculation of Feynman diagrams has to be expressed in terms of the product of all other $\gamma $ matrices. For example in four dimensions $\gamma_5$ is related to the other $\gamma$ matrices by
$$4!\, \gamma_5= -\epsilon_{\mu_1\ldots \mu_4}\gamma_{\mu_1}\ldots \gamma_{\mu_4}\,,\eqnd\egamv $$
where $\epsilon_{\mu_1\cdots \mu_4}$ is the complete antisymmetric tensor
with $\epsilon_ {1234}=1$. Problems therefore arise in the case of gauge theories with chiral fermions, because the special properties of $\gamma_5$ are involved as we recall below.
This difficulty is the source of chiral anomalies.\par 
Since perturbation theory involves the calculation of traces, one possibility is to define $\gamma_5$ near four dimensions by 
$$\gamma_5=E_{\mu_1\ldots \mu_4}\gamma_{\mu_1}\ldots \gamma_{\mu_4}\,,\eqnd\eEgams$$
where $ E_{\mu \nu \rho \sigma }$ is a completely antisymmetric tensor,
which reduces to $-\epsilon_ {\mu \nu \rho \sigma }/4!$ in four dimensions. It is easy to  then verify that, with this definition, $\gamma_5$ anticommutes with the other $\gamma_{\mu}$ matrices only in four dimensions. If for example we  evaluate the product
$\gamma_\nu \gamma_5 \gamma_\nu$ in $d$ dimensions, we find:
$$\gamma_{\nu} \gamma_5 \gamma_{\nu}=(d-8) \gamma_5\, .$$
Anticommuting properties of the $\gamma_5$ would have led to a factor $-d$
instead. Multiplied by a pole in $1/(d-4)$ consequence of  UV divergences in one-loop diagrams,  the additional contribution proportional to $d-4$ may lead to a finite difference with the formal result.\par
The other option would be to keep the anticommuting property of $\gamma _5$,
but the preceding example  shows that this is contradictory with a form \eEgams.
Actually one verifies that the only consistent prescription for generic dimensions then is that the traces of $\gamma _5$ with any product of $\gamma_\mu $ matrices vanishes, and therefore is useless.\par
Finally an alternative possibility consists in breaking $O(d)$ symmetry and keeping  the four $\gamma $ matrices of $d=4$.
\subsection Lattice regularization

We have explained that Pauli--Villars's regularization does
not work for field theories in which the geometric properties generate  interactions  like models on homogeneous spaces (for example the non-linear
$\sigma$-model) or gauge theories. In these theories some divergences are
related to the problem of quantization and order of operators, which already appears in simple quantum mechanics.  Other regularization methods then are needed. In many cases lattice regularization may be used. 
\smallskip
{\it Lattice field theory.} To each site $x$ of the lattice are attached field variables corresponding to fields in the continuum. To the action $ {\cal S} $ 
in the continuum corresponds a lattice action, the energy of lattice field configurations in the language of classical statistical physics.
The functional integral becomes a sum over configurations and the regularized partition function is the partition function of a lattice model. \par
All expressions in these notes will refer implicitly to a hypercubic lattice and we denote by $ a $  the lattice spacing. \par
The advantages of lattice regularization are:
\smallskip
(i ) Lattice regularization indeed corresponds to a specific choice of quantization.
\par
(ii) It is the only established regularization that for gauge theories and other geometric models has a meaning
outside perturbation theory. For instance the regularized
functional integral can be calculated by numerical methods, like
stochastic methods (Monte-Carlo type simulations) or strong coupling
expansions. 
\par 
(iii) It preserves most global or local symmetries with the exception
of the space $ O(d) $ symmetry that is replaced by a
hypercubic symmetry (but this turns out not to be a major difficulty) and fermion chirality, which turns out to be a more serious problem, as we will show. 
\smallskip
The main disadvantage is that it leads to rather complicated
perturbative calculations.  
\subsection Boson field theories

{\it Scalar fields.} To the action \eactscdiv\ for a  scalar field $\phi$ 
in the continuum corresponds a lattice action,  which is obtained in the following way: The euclidean lagrangian density becomes a function of lattice variables
$\phi(x)$, where $x$ now is a lattice site. Locality can be implemented by  considering   lattice lagrangians that depend only on a site and  its neighbours
(though this is a too strong requirement; lattice interactions decreasing exponentially with distance are also local).
Derivatives $ \partial_{\mu}\phi $  of the continuum are replaced by finite differences, for example:
$$ \partial_{\mu}\phi \mapsto \nabla^{\rm lat.}_{\mu}\phi =
\left[ \phi (x+an_{\mu} )-\phi (x)\right]/a\,, \eqnd\elatderiv $$
where $a$ is the lattice spacing and $ n_{\mu} $ the unit vector in the $ \mu $ direction.   
The lattice action then is the sum over lattice sites.\par 
With the choice \elatderiv\ the propagator $ \Delta_{a} (p) $ for the Fourier components of a massive scalar field   is given by
$$ \Delta^{-1}_{a} (p)=  m^2+{2 \over a^2}
\sum^{d}_{\mu =1} \bigl(1- \cos\left(ap_{\mu}\right) \bigr)
. \eqnn $$
It is a periodic function of the components $p_\mu$ of the momentum vector  with period $ 2\pi /a$. In the small lattice spacing limit  the continuum propagator is recovered:
$$ \Delta_{a} ^{-1}(p)=m^2+p^2-{\textstyle{1\over 12}}
\sum_{\mu}a^2p^{4}_{\mu} + O \left(p^{6}_{\mu} \right). \eqnn $$
In particular hypercubic symmetry implies  $O(d) $ symmetry at order $ p^2$.  
\medskip
{\it Gauge theories.} 
Lattice regularization defines unambiguously a quantum theory. Therefore, once one has realized that  gauge  fields should be replaced by link variables corresponding to parallel transport along links of the lattice, one can regularize a gauge theory. \par
The link variables ${\bf U}_{xy} $ are group elements associated with the links joining the sites
$x$ and $y$ on the lattice. The regularized form of $ \int \d x\, F^2_{\mu \nu} $ is
the product of link variables along a closed curve on the lattice, the
simplest being a square on a hypercubic lattice, leading to  the well-known plaquette action, each square forming a plaquette.
The typical gauge invariant lattice action corresponding to the continuum
action of a gauge field coupled to scalar bosons then has the form: 
$$  {\cal S}  ({\bf U}, \phi^* ,\phi  ) = \beta \sum_{\rm plaquettes} \tr
{\bf U}_{xy}{\bf U}_{yz}{\bf U}_{zt }{\bf U}_{tx} +
\kappa \sum_{ \rm links}\phi^*_x {\bf U}_{xy}\phi_y + \sum_{\rm sites} V(\phi^*_x\phi_x),
\eqnd\eaplaq $$
where $ x$, $y$,... denotes lattice sites,  and $ \beta $ and $ \kappa $ are coupling constants. The action \eaplaq~is invariant under independent group transformations 
on each lattice site, lattice equivalents of the gauge transformations of the continuum theory.
The measure of integration over the gauge variables is the group invariant
measure on each site. Note that on the lattice
and in a finite volume the gauge invariant action leads to a well-defined
partition function because the gauge group is compact. However in the continuum or infinite volume limits
the compact character of the group is lost.  Even on the lattice regularized perturbation theory is defined only  after gauge  fixing.   \par
We finally note that on the lattice the difficulties with the regularization  do not come  from the gauge field directly, but involve the gauge field only through the integration over chiral fermions.
\subsection Fermion and the doubling problem
 
We now review a few problems specific to relativistic fermions  on the lattice. We consider the free action for a Dirac fermion:\sslbl\sslattfer 
$$ {\cal S}  (\bar \psi ,\psi  )= \int \d^{d}x\,\bar \psi (x)
\left(\sla\partial +m \right)\psi (x). $$
To regularize this action by a lattice and preserve chiral properties
in the massless limit one can replace $\partial_{\mu}\psi (x) $ by 
$$ \nabla^{\rm lat.}_\mu\psi(x)= \left[\psi \left(x+an_{\mu} \right)-\psi \left(x-an
_{\mu} \right) \right]/2a. $$
Then  the inverse of the fermion propagator $\Delta$ for the Fourier
components $\tilde \psi(p)$ of the field is: 
$$ \Delta^{-1} (p)=  m+i \sum_{\mu}\gamma_{\mu}{
\sin ap_{\mu} \over a}\, \eqnd\eferprop $$
a periodic function of the components $p_\mu$ of the momentum vector. 
A problem then arises: the equations relevant to the small lattice spacing
limit, 
$$ \sin(a\,p_{\mu})=0 $$
have each two solutions $ p_{\mu}=0 $ and $ p_{\mu}=\pi /a $
within one period, i.e.~within the Brillouin zone $2\pi/a$. Therefore the
propagator \eferprop~propagates $ 2^d $ fermions. To remove this degeneracy it
is possible to add to the regularized action an additional scalar term $
\delta {\cal S} $ involving second derivatives:
$$ \delta {\cal S}  (\bar \psi ,\psi )=\ud M  \sum_{x,\mu}
\left[ 2\bar \psi (x)\psi (x)-\bar \psi
\left(x+an_{\mu} \right)\psi (x)-\bar \psi (x)\psi
\left(x+an_{\mu} \right) \right] . \eqnd{\ewilsfer} $$
After Fourier transformation the modified Dirac operator $D_W$ reads
$$D_W(p)=m+M \sum_{\mu} \left(1-\cos ap_{\mu} \right)+{i \over a} \sum_{\mu}
\gamma_{\mu}\sin ap_\mu \,.\eqnd\eWferDop $$
The fermion propagator becomes:
$$ \Delta (p)= D^\dagger_W(p) \left(D_W(p)D^\dagger_W(p)\right)^{-1}, $$
with:
$$ D_W(p)D^\dagger_W(p)= \left[ m+M \sum_{\mu} \left(1- \cos ap_{\mu} \right)
\right]^2+{1 \over a^2}  \sum_{\mu} \sin^2 ap_{\mu} \,.
$$
Therefore the degeneracy between the different states has been lifted. For
each component $ p_{\mu} $ that takes the value $ \pi /a $ the mass is
increased by $ M $. If $ M $ is of order $ 1/a $ the spurious states are
eliminated in the continuum limit. This is the recipe of Wilson's 
fermions. \par
However a  problem arises if one wants to construct a theory
with massless fermions and chiral symmetry. Chiral symmetry
implies for the Dirac operator $D(p)$ anticommutation with $\gamma _5$
$$ \{D(p),\gamma _5\}=0\,,\eqnn $$
and therefore both the mass term and the term \ewilsfer~are excluded. 
It remains  possible to add various counter-terms and try to adjust them
to recover chiral symmetry in the continuum limit. But then there is no {\it a priori}\/ guarantee that this is indeed possible and moreover calculations are
plagued by fine tuning problems and cancellations of unnecessary UV
divergences.   
\par
One could also think about modifying the fermion propagator by adding
terms connecting fermions separated by more than one lattice spacing. But it has been proven that this does not solve the doubling problem.
In fact this doubling of the number of fermion degrees of freedom is directly
related to the problem of anomalies.  \par 
Since the most naive form of the propagator yields $ 2^{d} $
fermion states, one tries in practical calculations to reduce this number to
a smaller multiple of two, using for instance the idea of staggered fermions introduced
by Kogut and Susskind. \par
However  the general picture has recently changed with the
discovery of the properties of overlap fermions and solutions of the
Ginsparg--Wilson relation, a topic we postpone, and on which we will come back in section \label{\ssrotGW}.
\section The abelian anomaly

We have pointed out  that none of the standard regularization 
methods can deal in a straightforward way with one-loop diagrams in the case of  gauge
fields coupled to chiral fermions. We now show that indeed chiral symmetric gauge theories, involving gauge fields coupled to massless fermions, can be found  where the axial current is not conserved. The divergence of the axial current in a chiral theory, when it does not vanish, is called an {\it anomaly}. Anomalies in particular lead to obstructions to the construction of 
gauge theories when the gauge field couples differently to the two fermion chiral components. \par
Several examples are physically important
like  the theory
of weak electromagnetic interactions,  the electromagnetic decay of the $\pi_0$ meson, or the $U(1)$ problem. We first discuss the abelian axial current, in four dimensions (the generalization to all even dimensions then is straightforward), and then the general non-abelian situation.\sslbl\ssAbAnom\par
\subsection Abelian axial current and abelian vector gauge field

The only possible source of anomalies are one-loop fermion diagrams in gauge theories when chiral properties are involved. This reduces the problem to the discussion of fermions in background gauge fields, or equivalently to the properties of the determinant of the gauge covariant Dirac operator.\par
We thus consider the QED-like fermion action ${\cal S} (\bar\psi ,\psi;A )$ for massless Dirac fermions $\psi,\bar\psi$ in the background of an abelian  gauge field $A_\mu$
$$ {\cal S} (\bar \psi ,\psi ;A ) = - \int \d  ^{4}x\, \bar
\psi(x  )\Dbar  \psi  (x ),\quad \Dbar \equiv \sla{\partial }+ie \Abar(x)\,,  \eqnd\eQEDfer $$
and the corresponding  functional integral
$$ {\cal Z} (A_ \mu )=\int \left[\d \psi \d \bar\psi\right]
\exp\left[- {\cal S} (\psi ,\bar\psi;A)\right] =\det  \Dbar\,.  \eqnn $$
We can find regularizations that preserve gauge invariance, 
$$\psi(x)=\e^{i\Lambda (x)}\psi'(x),\quad \bar\psi(x)=\e^{-i\Lambda (x)}\bar\psi'(x),\quad 
A_\mu(x)=-{1\over e}\partial _\nu \Lambda (x)+A'_\mu(x), \eqnd\eAbgautr $$
and since the fermions are massless,  chiral symmetry. We would therefore naively expect the corresponding axial current to be conserved. However the proof of current conservation involves space-dependent  chiral transformations, and therefore steps that cannot be regularized without breaking the local symmetry.  \par 
Under a  space-dependent chiral transformation
$$  \psi _{\theta}  (x ) = \e^{i\theta(x) \gamma _{5}}\psi (x), \qquad \bar \psi _{\theta}  (x  ) =\bar \psi  (x )
\e^{i\theta(x) \gamma _{5}}  , \eqnd\espachir $$
the action becomes
$$ {\cal S}_\theta   (\bar \psi ,\psi ;A) = - \int \d  ^{4}x\, \bar
\psi_\theta (x  )\Dbar  \psi_\theta   (x )= {\cal S} (\bar \psi ,\psi;A  )+ \int\d^4 x \,\partial_{\mu}\theta(x)J_{\mu}^5(x) ,$$
where $J_{\mu}^5(x)$, the coefficient of $ \partial _\mu \theta $, 
is the axial current,  
$$ J_{\mu}^5(x)=i \bar\psi(x)\gamma_5 \gamma_{\mu}\psi(x). \eqnn $$
After the transformation \espachir\ ${\cal Z} (A_ \mu )$ becomes:
$${\cal Z} (A_ \mu ,\theta )=\det\left[\e^{i\gamma_5
\theta(x)}\Dbar\e^{i\gamma_5 \theta(x)}\right] . \eqnn $$
Note that $\ln[{\cal Z} (A_ \mu ,\theta )]$ is the generating functional of connected
$\partial _\mu J_{\mu}^5$ correlation functions in an external field $A_\mu$.\par
Since $\e^{i \gamma_5 \theta}$ has a determinant that is unity,  one would naively conclude that ${\cal Z} (A_ \mu  ,\theta )= {\cal Z} (A_ \mu   )$  and therefore that the current  $J_{\mu}^5(x)$ is conserved. This is a conclusion we now check by an explicit calculation of the expectation value of $\partial_{\mu} J_{\mu}^5(x)$ in the case
of the action \eQEDfer.  
\smallskip
{\it Remarks.}\par
(i) For any regularization that is consistent with the hermiticity of $\gamma_5$ 
$$\left| {\cal Z} (A_ \mu ,\theta  )\right|^2 =\det\left[\e^{i\gamma_5
\theta(x)}\Dbar\e^{i\gamma_5 \theta(x)}\right] \det\left[\e^{-i\gamma_5
\theta(x)}\Dbar^\dagger\e^{-i\gamma_5 \theta(x)}\right]  = \det
\hskip2pt(\Dbar\Dbar^{\dagger}) ,$$  
and thus $ | {\cal Z} (A_ \mu ,\theta  ) |$ is independent of $\theta $.  Therefore an anomaly can  appear only in the imaginary part of $\ln {\cal Z}$.\par 
(ii) We have shown that one can find a regularization with regulator fields such that gauge invariance is maintained, and the determinant is independent of $\theta $ for $\theta (x)$ constant.\par
(iii) If the regularization is gauge invariant ${\cal Z} (A_ \mu  ,\theta )$ is also gauge invariant.
Therefore a possible anomaly will also be gauge invariant.\par
(iv) $\ln{\cal Z} (A_ \mu  ,\theta )$ is connected and 1PI. Short distance singularities thus take the form of local polynomials in the fields and sources. Since a possible anomaly is a short distance
effect it must also take the form of a local polynomial of $A_\mu$ and $\partial _\mu \theta $ constrained by   parity and power counting, $A_\mu$ and $\partial _\mu \theta $  having dimension one and no mass being available, 
$$\ln{\cal Z} (A_ \mu  ,\theta )- \ln{\cal Z} (A_ \mu  , 0 )=i\int\d^4x\,{\cal L}(A,\partial \theta;x ),$$
where ${\cal L}$ is the sum of monomials of dimension four. At order $\theta $ only one is available:
$${\cal L}(A,\partial \theta;x )\propto e^2\epsilon _{\mu \nu \rho \sigma }\partial _\mu \theta (x)A_\nu(x)\partial _\rho A_\sigma (x), $$
where $\epsilon_ {\mu \nu \rho \sigma }$ is the complete antisymmetric tensor
with $\epsilon _{1234}=1$.
A simple integration by parts and anti-symmetrization shows that
$$\int\d^4x\,{\cal L}(A,\partial \theta;x )\propto e^2\epsilon _{\mu \nu \rho \sigma }
\int\d^4x\,F_{\mu\nu}(x)F_{\rho\sigma}(x),$$
an expression that is gauge invariant.
\par 
The coefficient of $\theta (x)$ is the expectation value  in an external gauge field of $\partial_{\mu}J_{\mu}^5(x)$, the divergence of the axial current. It is thus determined   up to a multiplicative constant,  
 $$\left<\partial_{\lambda}J_{\lambda}^5(x)\right> \propto e^2
\epsilon_{\mu\nu\rho\sigma}\partial_{\mu}A_{\nu} (x)\partial_{\rho}A_{\sigma}(x)\propto e^2 \epsilon_{\mu\nu\rho\sigma}F_{\mu\nu}(x)F_{\rho\sigma}(x)\,,
$$
where $F_{\mu\nu}=\partial _\mu A_\nu -\partial _\nu  A_\mu $ is the electromagnetic tensor, and  we denote by $\left<\bullet\right>$ expectation values with respect to the measure $\e^{-{\cal S}(\bar\psi,\psi;A)}$.\par
Since the  possible anomaly is independent up to a multiplicative factor of the regularization, it must indeed be a gauge invariant local function of $A_\mu$. \par 
To find the  multiplicative factor, which is the only regularization dependent feature, it is  sufficient to calculate the coefficient of the
term quadratic in $A$   in the expansion of
$\left<\partial_{\lambda}J_{\lambda}^5(x)\right>$ in powers of $A$. We
define the three-point function in momentum representation: 
$$\eqalignno{\Gamma^{(3)}_{\lambda\mu\nu} (k;p_1,p_2 ) & =\left.
{\delta \over \delta A_{\mu} (p_1 )}{\delta \over \delta
A_{\nu} (p_2 )}\left< J_{\lambda}^5(k)\right>\right\vert_{A=0},&
 \eqnd \ethreept  \cr &=\left. {\delta \over \delta A_{\mu} (p_1
 )}{\delta \over \delta A_{\nu} (p_2 )}i\tr\left[\gamma_5
\gamma_{\lambda}\Dbar^{-1}(k)\right] \right\vert_{A=0}. & \cr}$$  
$\Gamma^{(3)}$ is the sum of the two Feynman diagrams of figure
\label{\figSMweiv}. \par 
\midinsert
\moveleft.45mm\vbox{\elevenpoint\hfil
$k,\lambda$\kern18.65mm\lower6.4mm\hbox{$q$}
\kern15.5mm\raise10mm\hbox{$p_1,\mu$}\kern-6.65mm
\lower10.7mm\hbox{$p_2,\nu$}
\kern7.85mm
$k,\lambda$\kern18.65mm\lower6.4mm\hbox{$q$}
\kern15.5mm\raise10mm\hbox{$p_1,\mu$}\kern-6.65mm
\lower10.7mm\hbox{$p_2,\nu$}
\hfil
\vskip1.8mm
\hfil(a)\kern52.8mm (b)\hfil}
\vskip-27.5mm
\epsfxsize=92.5mm
\epsfysize=20.8mm
\centerline{\epsfbox{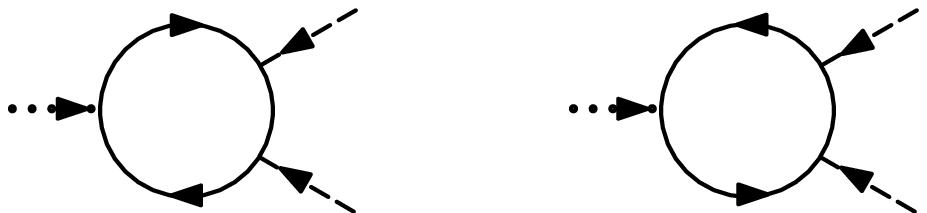}}
\figure{9.mm}{Anomalous diagrams.}
\figlbl\figSMweiv
\endinsert
The contribution of diagram (a)  is:  
$${\rm(a)}\mapsto {e^2 \over (2\pi)^4}\tr\left[\int \d^4
q\, \gamma_5  \gamma _\lambda \left(\sla{q}+ \sla{k}\right)^{-1} \gamma_{\mu} \left(\sla{q}-
\sla{p}_2\right)^{-1} \gamma_{\nu} \sla{q}^{-1}\right],\eqnn $$
and the contribution of diagram (b) is obtained by exchanging $p_1,\gamma _\nu
\leftrightarrow p_2,\gamma_\nu$. \par
Power counting tells us that the function  $\Gamma^{(3)}$ may have a linear divergence
 that, due to the presence of the $\gamma_5$ factor, must be proportional to $\epsilon_{\lambda\mu\nu\rho}$, symmetric in the exchange $p_1,\gamma _\nu
\leftrightarrow p_2,\gamma_\nu$, and thus proportional to
$$  \epsilon_{\lambda\mu\nu\rho}\left(p_1-p_2\right)_{\rho}\,.
\eqnd \eambigu $$ 
On the other hand by commuting $\gamma _5$ we notice that $\Gamma^{(3)}$ is formally a  symmetric function of the three sets of external arguments. A divergence breaks the symmetry between  external arguments. Therefore a symmetric regularization of the kind we will adopt in the first calculation leads to a finite result. 
The result is not ambiguous because a possible ambiguity again is proportional to \eambigu.\par
Similarly if the regularization is consistent with   gauge invariance the  vector current is conserved  
$$p_{1\mu}\Gamma^{(3)}_{\lambda\mu\nu}\left(k;p_1,p_2\right)=0\, .\eqnn $$
Applied to the divergent part the equation implies
$$-p_{1\mu}p_{2\rho} \epsilon_{\lambda\mu\nu\rho}=0\,,$$
which cannot be satisfied. Therefore the sum of the two diagrams is finite.  Finite ambiguities must also have the form \eambigu~and thus are also forbidden by gauge invariance. All  regularizations consistent with   gauge invariance must give the same answer.\par
Therefore there are two possibilities:\par
(i) The divergence $k_{\lambda}\Gamma^{(3)}_{\lambda\mu\nu} \left(k;p_1,p_2\right) $  in a  regularization respecting the symmetry between the three arguments vanishes. Then  both $\Gamma^{(3)}$  is gauge invariant  and the axial current is conserved. \par
(ii) The divergence of the symmetric regularization does not vanish.
Then it is possible to add to $\Gamma^{(3)}$ a term proportional to
\eambigu~to restore gauge invariance but  this term breaks  the symmetry between external momenta:  the axial current is not conserved, an anomaly is present.
\subsection Explicit calculation 

{\it Momentum regularization.} The calculation can be done using one of the various gauge invariant regularizations, for
example momentum cut-off regularization or dimensional regularization with
$\gamma_5$ being defined as in dimension four and thus no longer anticommuting
with other $\gamma$ matrices. 
Instead we choose a regularization that
preserves the symmetry between the three external arguments and global chiral symmetry, but breaks gauge invariance,
modifying the fermion propagator: 
$$(\sla{q})^{-1}\longmapsto (\sla{q})^{-1}\rho(\varepsilon q^2), $$
where $\varepsilon$ is the regularization parameter ($\varepsilon\to 0_+$), $\rho(z)$ is a positive differentiable function  such that $\rho(0)=1$, and decreasing fast enough for   $z\to+\infty $, at least like  $1/z$.\par  
Then  current conservation and gauge invariance  are compatible only if the divergence $k_{\lambda}\Gamma^{(3)}_{\lambda\mu\nu} (k;p_1,p_2 )$ vanishes. \par
It is convenient to consider directly the contribution
$C^{(2)}(k)$ of order $A^2$ to
$\left<k_{\lambda}J_{\lambda}^5(k)\right>$, which sums the two diagrams:\sslbl\sssAnomc 
$$\eqalignno{C^{(2)}(k) & = e^2 \int \d^4 p_1\,
\d^4 p_2\, A_{\mu} ( p_1 ) A_{\nu} ( p_2 ) \int{ \d^4 q
\over (2\pi)^4} \rho\bigl(\varepsilon (q+k)^2\bigr) \rho \bigl(\varepsilon
(q-p_2)^2\bigr)\rho\bigl(\varepsilon q^2 \bigr)& \cr & \qquad\times \tr\left[ \gamma_5
\sla{k}(\sla{q}+ \sla{k})^{-1} \gamma_{\mu} (\sla{q}-
\sla{p}_2)^{-1} \gamma_{\nu}\sla{q}^{-1}\right] ,& \eqnn\cr}$$  
because the calculation 
then suggests how the method generalizes to arbitrary even
dimensions.\par
We first transform the expression, using the identity:
$$\sla{k} (\sla{q}+ \sla{k} )^{-1}= 1 -\sla{q} (\sla{q}+
\sla{k} )^{-1}. \eqnd \ediagide  $$
Then
$$C^{(2)}(k) =C_1^{(2)}(k)+C_2^{(2)}(k),$$
with
$$\eqalignno{C^{(2)}_1(k) & = e^2 \int \d^4 p_1\,
\d^4 p_2\, A_{\mu} ( p_1 ) A_{\nu} ( p_2 ) \int{ \d^4 q
\over (2\pi)^4} \rho\bigl(\varepsilon (q+k)^2\bigr) \rho \bigl(\varepsilon
(q-p_2)^2\bigr)\rho\bigl(\varepsilon q^2 \bigr)& \cr & \qquad\times \tr\left[ \gamma_5 \gamma_{\mu} (\sla{q}-
\sla{p}_2)^{-1} \gamma_{\nu}\sla{q}^{-1}\right] ,& \eqnn\cr}$$  
and
$$\eqalignno{C^{(2)}_2(k) & = -e^2 \int \d^4 p_1\,
\d^4 p_2\, A_{\mu} ( p_1 ) A_{\nu} ( p_2 ) \int{ \d^4 q
\over (2\pi)^4} \rho\bigl(\varepsilon (q+k)^2\bigr) \rho \bigl(\varepsilon
(q-p_2)^2\bigr)\rho\bigl(\varepsilon q^2 \bigr)& \cr & \qquad\times \tr\left[ \gamma_5
\sla{q}(\sla{q}+ \sla{k})^{-1} \gamma_{\mu} (\sla{q}-
\sla{p}_2)^{-1} \gamma_{\nu}\sla{q}^{-1}\right].& \eqnn\cr}$$  
In  $C^{(2)}_2(k) $ we  use
the cyclic property of the trace and the {\it  commutation}\/ of  $\gamma_{\nu}\sla{q}^{-1}$ and $\gamma_5$ to cancel the propagator $\sla{q}^{-1}$ and obtain
$$\eqalignno{C^{(2)}_2(k) & = - e^2 \int \d^4 p_1\,
\d^4 p_2\, A_{\mu} ( p_1 ) A_{\nu} ( p_2 ) \int{ \d^4 q
\over (2\pi)^4} \rho\bigl(\varepsilon (q+k)^2\bigr) \rho \bigl(\varepsilon
(q-p_2)^2\bigr)\rho\bigl(\varepsilon q^2 \bigr)& \cr & \qquad\times \tr\left[ 
 \gamma_5 \gamma_{\nu} (\sla{q}+ \sla{k})^{-1} \gamma_{\mu} (\sla{q}-
\sla{p}_2)^{-1}\right] .& \eqnn\cr}$$  
We then shift
$q\mapsto q+p_2$ and interchange $(p_1,\mu)$ and $(p_2,\nu)$,   $$\eqalignno{C^{(2)}_2(k) & = - e^2 \int \d^4 p_1\,
\d^4 p_2\, A_{\mu} ( p_1 ) A_{\nu} ( p_2 ) \int{ \d^4 q
\over (2\pi)^4} \rho\bigl(\varepsilon (q-p_2)^2\bigr) \rho \bigl(\varepsilon
 q ^2\bigr)\rho\bigl(\varepsilon (q+p_1)^2 \bigr)& \cr & \qquad\times \tr\left[ \gamma_5
 \gamma_{\mu} (\sla{q}- \sla{p_2})^{-1} \gamma_{\nu}  \sla{q} ^{-1} \right] .& \eqnd\econtrib\cr}$$  
We  see that the two terms $C^{(2)}_1$ and $C^{(2)}_2$ would cancel in the absence of
regulators. This corresponds to the formal proof of current conservation. However  without regularization the integrals diverge and these manipulations are not  legitimate.\par
Instead here we find a non-vanishing sum due to the difference in  regulating factors:
$$\eqalignno{C^{(2)}(k) & = e^2 \int \d^4 p_1\,
\d^4 p_2\, A_{\mu} ( p_1 ) A_{\nu} ( p_2 ) \int{ \d^4 q
\over (2\pi)^4} 
\rho\bigl(\varepsilon (q-p_2)^2\bigr)\rho\bigl(\varepsilon q^2 \bigr) 
& \cr & \quad\times \tr\left[ \gamma_5
 \gamma_{\mu} (\sla{q}- \sla{p}_2)^{-1} \gamma_{\nu}\sla{q}^{-1}
\right] \left[\rho\bigl(\varepsilon (q+k )^2 \bigr)- \rho\bigl(\varepsilon(q+p_1 )^2\bigr)
\right].\qquad\quad&  \eqnn\cr}$$   
After evaluation of the trace  $C^{(2)}$ becomes (using \egamv): 
$$\eqalignno{C^{(2)}(k) & =- 4e^2 \int \d^4 p_1\,
\d^4 p_2\, A_{\mu} ( p_1 ) A_{\nu} ( p_2 ) \int{ \d^4 q
\over (2\pi)^4}
\rho\bigl(\varepsilon (q-p_2)^2\bigr)\rho\bigl(\varepsilon q^2 \bigr) 
& \cr & \quad\times
\epsilon_{\mu\nu\rho\sigma}{p_{2\rho} q_{\sigma} \over q^2 (
q-p_2)^2} 
\left[ \rho\bigl(\varepsilon(q+k)^2\bigr) -
\rho\bigl(\varepsilon (q+ p_1 )^2 \bigr) \right]
.& \eqnn \cr}$$   
Contributions coming from finite values of $q$ cancel in the $\varepsilon\to 0$ limit.
Due the  cut-off the relevant values of $q$ are of order $\varepsilon^{-1/2}$.
We therefore rescale $q$ accordingly $q\varepsilon^{ 1/2}\mapsto q$ and find
$$\eqalign {C^{(2)}(k) & =- 4e^2 \int \d^4 p_1\,
\d^4 p_2\, A_{\mu} ( p_1 ) A_{\nu} ( p_2 ) \int{ \d^4 q
\over (2\pi)^4}
\rho\bigl(  (q-p_2\sqrt{\varepsilon }
)^2\bigr)\rho\bigl(  q^2 \bigr) 
  \cr & \quad\times
\epsilon_{\mu\nu\rho\sigma}{p_{2\rho} q_{\sigma} \over q^2 (
q-p_2\sqrt{\varepsilon })^2} 
{ \rho\bigl( (q+k\sqrt{\varepsilon }
)^2\bigr) -
\rho\bigl( (q+ p_1\sqrt{\varepsilon }
 )^2 \bigr) \over \sqrt{\varepsilon }}
. \cr}$$   
Taking the $\varepsilon\to0$ limit we obtain a finite result: 
$$C^{(2)}(k)   =- 4e^2\epsilon_{\mu\nu\rho\sigma} \int \d^4 p_1\,
\d^4 p_2\, A_{\mu} ( p_1 ) A_{\nu} ( p_2 )I_{\rho \sigma }(p_1,p_2), $$
with
$$I_{\rho \sigma }(p_1,p_2)\sim \int{ \d^4 q \over(2\pi)^4 q^4 }p_{2\rho} q_{\sigma}
\rho^2\bigl(\varepsilon q^2\bigr) \rho'\bigl(\varepsilon q^2\bigr)\left[2\varepsilon q_\lambda  ( k-p _1)_\lambda  \right].\eqnn $$
The identity:
$$\int\d^4 q\, q_{\alpha}q_{\beta}f( q^2 )={\textstyle{ 1 \over 4}}
\delta_{\alpha\beta}\int\d^4 q \,q^2 f( q^2 ), $$
transforms the integral into:
$$I_{\rho \sigma }(p_1,p_2)\sim-\ud p_{2\rho} (2p_{1}+p_2{} )_{\sigma} \int{\varepsilon \d^4 q
\over(2\pi)^4 q^2 }\rho^2\bigl(\varepsilon q^2\bigr) \rho'\bigl(\varepsilon q^2\bigr) .\eqnn $$
The remaining integral can be calculated explicitly  (we recall $\rho(0)=1$)
$$\int{\varepsilon \d^4 q
\over(2\pi)^4 q^2 }\rho^2\bigl(\varepsilon q^2\bigr) \rho'\bigl(\varepsilon q^2\bigr) =
{1\over 8\pi^2}\int_0^\infty  \varepsilon q\d q\,\rho^2\bigl(\varepsilon q^2\bigr) \rho'\bigl(\varepsilon q^2\bigr)=- {1\over 48 \pi^2}, $$
and yields a result independent of the function $\rho$. We finally obtain:
$$\left<k_{\lambda}J_{\lambda}^5(k)\right>= -{ e^2 \over 12 \pi^2}
\epsilon_{\mu\nu\rho\sigma} \int \d^4 p_1\, \d^4 p_2\, p_{1\mu} A_{\nu}
( p_1 ) p_{2\rho} A_{\sigma} ( p_2 )  ,\eqnd{\eanoma} $$
and therefore from the definition \ethreept\ 
$$k_{\lambda}\Gamma^{(3)}_{\lambda\mu\nu}(k;p_1,p_2)={ e^2 \over 6
\pi^2}\epsilon_{\mu\nu\rho\sigma}p_{1\rho}p_{2\sigma}\,.\eqnn $$
This non-vanishing result implies that any definition of the
determinant $\det \Dbar$ breaks at least either current conservation or gauge
invariance. Since gauge invariance is essential to the consistency of a gauge theory we choose  to break current conservation. Exchanging arguments, we
obtain the value of
$p_{1\mu}\Gamma^{(3)}_{\lambda\mu\nu}(k;p_1,p_2)$:  
$$p_{1\mu}\Gamma^{(3)}_{\lambda\mu\nu}(k;p_1,p_2)={ e^2 \over 6
\pi^2}\epsilon_{\lambda\nu\rho\sigma}k_{\rho}p_{2\sigma}\,.\eqnn $$
If instead we had used a gauge invariant regularization, the result
for $\Gamma^{(3)}$ would have differed by a term 
$\delta\Gamma^{(3)}$ proportional to  \eambigu: 
$$\delta\Gamma^{(3)}_{\lambda\mu\nu}(k;p_1,p_2)=K
\epsilon_{\lambda\mu\nu\rho}(p_1-p_2)_{\rho}\,. \eqnn $$
The constant $K$ then is determined by the condition of gauge invariance
$$p_{1\mu}\left[ \Gamma^{(3)}_{\lambda\mu\nu}(k;p_1,p_2)+
 \delta\Gamma^{(3)}_{\lambda\mu\nu}(k;p_1,p_2)\right]=0\,,$$
which yields
$$p_{1\mu}\delta\Gamma^{(3)}_{\lambda\mu\nu}(k;p_1,p_2)=-{ e^2
\over 6\pi^2}\epsilon_{\lambda\nu\rho\sigma}k_{\rho}p_{2\sigma} \ \Rightarrow\ K= e^2 / (6\pi^2).\eqnn $$
This gives an additional contribution to
the divergence of the current
$$k_{\lambda}\delta\Gamma^{(3)}_{\lambda\mu\nu}(k;p_1,p_2)={ e^2
\over 3\pi^2}\epsilon_{\mu\lambda\rho\sigma}p_{1\rho}p_{2\sigma}\, .
\eqnn $$
Therefore in a QED-like gauge invariant field theory with massless fermions
the axial current is not conserved: this is called the chiral anomaly. For
any gauge invariant regularization one finds 
$$k_{\lambda}\Gamma^{(3)}_{\lambda\mu\nu}(k;p_1,p_2)=\left( { e^2
\over 2 \pi^2}\equiv {2\alpha \over \pi}\right)
\epsilon_{\mu\nu\rho\sigma}p_{1\rho}p_{2\sigma}\,.\eqnd \eanomchi  $$
The equation \eanomchi~can be rewritten after Fourier transformation as an
axial current non-conservation equation:
$$\partial_{\lambda}J^5_{\lambda}(x)=-i{\alpha \over
4\pi}\epsilon_{\mu\nu\rho\sigma} F_{\mu\nu}(x)F_{\rho\sigma}(x)\,. \eqnd{\eaxanom}
$$ 
Since global
chiral symmetry is not broken, the integral over the whole space of the
anomalous term must vanish. This condition is indeed verified since the
anomaly can immediately be written as a total derivative:
$$\epsilon_{\mu\nu\rho\sigma}
F_{\mu\nu}F_{\rho\sigma}=4 \partial_{\mu}(\epsilon_{\mu\nu\rho\sigma}
A_{\nu}\partial_{\rho}A_{\sigma} ). \eqnn $$ 
The space integral of the anomalous term depends only on the behaviour
of the gauge field at boundaries, and this property already indicates a connection between {\it topology
and anomalies}.\par 
The equation \eaxanom~also implies:
$$\ln\det\left[\e^{i\gamma_5 \theta(x)}\Dbar\e^{i\gamma_5 \theta(x)}\right]
= \ln \det\Dbar -i {\alpha \over 4\pi}\int \d^4 x\,
\theta(x)\epsilon_{\mu\nu\rho\sigma} F_{\mu\nu}(x)F_{\rho\sigma}(x). \eqnn $$
\smallskip
{\it Remark.} One might be surprised that in the calculation
the divergence of the axial current does not vanish, though the regularization of the fermion propagator seems to be consistent with chiral symmetry. The reason is simple: if we add for example higher derivative terms to the action, the form of the axial current is modified and the additional contributions cancel the term we have found.\par
In the form we have presented it the calculation generalizes without
difficulty to general even dimensions $2n$. Note simply that the permutation
 $({\bf p}_1, \mu)\leftrightarrow ({\bf p}_2, \nu)$ in  
equation \econtrib~is replaced by a cyclic permutation. If gauge invariance is maintained the anomaly in the divergence of the axial current $J^S_\lambda (x)$ in general is
$$\partial_{\lambda}J^S_{\lambda}(x)=-2i {e^n \over
(4\pi)^n n!}\epsilon_{\mu_1\nu_1\ldots\mu_n\nu_n} F_{\mu_1\nu_1}\ldots
F_{\mu_n\nu_n}\,,\eqnd \eanomAbgen  $$
where $\epsilon_{\mu_1\nu_1\ldots\mu_n\nu_n}$ is the completely antisymmetric tensor,
and $J^S_{\lambda}\equiv J^{(2n+1)}_\lambda $ is the axial current.

\medskip
{\it Boson regulator.} We have seen that we could also regularize by adding massive fermions and bosons with spin, the unpaired boson affecting transformation properties under space-dependent chiral transformations.
Denoting by $\phi$ the boson field and by $M$ its mass, we perform in the regularized functional integral a change of variables of the form of a space-dependent chiral transformation acting in the same way on the fermion and boson field. The variation $\delta {\cal S}$ of the action at first order in $\theta $ is
$$\delta {\cal S} =\int\d^4 x\left[\partial_{\mu}\theta(x)J_{\mu}^5(x) 
+2i M\theta (x)\bar\phi(x)\gamma_5 \phi(x)\right],$$
  with 
$$J_{\mu}^5(x)=i \bar\psi(x)\gamma_5 \gamma_{\mu}\psi(x)+i \bar\phi(x)\gamma_5 \gamma_{\mu}\phi(x).  $$
Expanding in $\theta $ and identifying the coefficient of $\theta (x)$ we therefore obtain the equation
$$\left<\partial _\mu J_{\mu}^5(x)\right>=2iM\left< \bar\phi(x)\gamma_5 \phi(x)\right>=-2iM\tr\gamma_5\left<x\right|\Dbar^{-1}\left|x\right> .\eqnd\eAbannce  $$
The divergence of the axial current comes here from the boson contribution. We know that in the large $M$ limit  it becomes quadratic in $A$. Expanding the r.h.s.~in powers of $A$, keeping the quadratic term we find after Fourier transformation:
$$\eqalignno{C^{(2)}(k) & = -2i M e^2 \int \d^4 p_1\,
\d^4 p_2\, A_{\mu} ( p_1 ) A_{\nu} ( p_2 ) \int{ \d^4 q
\over (2\pi)^4}  \cr&\quad\times  \tr\left[ \gamma_5
 (\sla{q}+ \sla{k}-iM)^{-1} \gamma_{\mu} (\sla{q}-
\sla{p}_2-iM )^{-1} \gamma_{\nu}(\sla{q}-iM)^{-1}\right] .\hskip5mm & \eqnd\eabanbose \cr}$$ 
The apparent divergence of this contribution is regularized  by formally vanishing diagrams that we do not write, but which justify the following formal manipulations. \par  
In the trace   the formal divergences cancel and one obtains 
$$\eqalign {C^{(2)}(k) & \mathop{\sim}_{M\to \infty }  8 M^2 e^2\epsilon _{\mu \nu \rho \sigma } \int \d^4 p_1\,
\d^4 p_2\,  p_{1\rho }p_{2\sigma }A_{\mu} ( p_1 ) A_{\nu} ( p_2 ) \cr
&\quad\times {1 \over (2\pi)^4} \int {\d^4 q  \over (q^2+M^2)^3}.\cr}$$
The limit  $M\to\infty$ corresponds to remove the regulator. The limit is finite because after rescaling of $q$ the mass can be eliminated. One finds:
$$C^{(2)}(k)   \mathop{\sim}_{M\to \infty }  {e^2\over 4\pi^2}   \epsilon _{\mu \nu \rho \sigma } \int \d^4 p_1\,
\d^4 p_2\, ( p_2 ) p_{1\rho }p_{2\sigma }  A_{\mu} ( p_1 ) A_{\nu}(p_2) \,,$$
in agreement with equation \eanomchi.
\medskip
{\it Point-splitting regularization.} Another calculation, based on regularization by point splitting, gives further insight into the mechanism that generates the anomaly. We thus consider the non-local operator
$$J^5_\mu (x,a)=i\bar\psi(x-a/2)\gamma _5\gamma _\mu\psi(x+a/2)\exp\left[ie
\int_{x-a/2}^{x+a/2}A_\lambda  (s)\d s_\lambda \right],\eqnn $$
in the limit $|a|\to 0$. To avoid a breaking of rotation symmetry by the regularization, before taking the limit   $|a|\to 0$ we will average over all orientations of the vector $a$.
The multiplicative gauge factor (parallel transporter) ensures gauge invariance of the
regularized operator (transformations \eAbgautr). 
The divergence of the operator for $|a|\to 0$ then becomes
$$\eqalignno{\partial^x _\mu  J^5_\mu (x,a)&\sim-e a_\lambda  \bar\psi(x-a/2)\gamma _5 \gamma _\mu F_{\mu\lambda }(x) \psi(x+a/2)\cr
&\quad\times \exp\left[ie \int_{x-a/2}^{x+a/2}A_\lambda  (s)\d s_\lambda \right],&\eqnn \cr}$$
where the $\psi,\bar\psi$ field equations have been used. We now expand the expectation value of the equation in powers of $A$. The first term vanishes. The second term is
quadratic in $A$ and yields 
$$  \left<\partial^x _\mu  J^5_\mu (x,a)\right>  \sim ie^2 a_\lambda F_{\mu\lambda }(x)   \int\d^4 y\,A_\nu(y+x)\tr \gamma _5\Delta_{\rm F}( y -a/2)\gamma _\nu
\Delta_{\rm F}(  -y-a/2)\gamma _\mu\,,  $$
where $\Delta_{\rm F}( y)$ is the fermion propagator:
$$\Delta_{\rm F}( y)=-{ i\over (2\pi)^4}\int\d^4 k{   \sla{k}\over k^2}={1\over 2\pi^2}{\sla{y}\over y^4} \,.\eqnn $$
We now take the trace. The propagator is singular for $|y|=O(|a|)$ and therefore we can expand $A_\nu(x+y)$ in powers of $y$. The first term vanishes for symmetry reasons ($y\mapsto -y$), and we obtain
$$ \left<\partial^x _\mu  J^5_\mu (x,a)\right>  \sim {ie^2 \over \pi^4}\epsilon _{\mu\nu\tau \sigma } a_\lambda F_{\mu\lambda }(x)  \partial _\rho A_\nu(x)  \int\d^4 y {y_\rho 
y_\sigma a_\tau \over |y+a/2|^4|y-a/2|^4}\,.$$
The integral over $y$ gives a linear combination of $\delta _{\rho \sigma }$ and
$a_\rho a_\sigma $ but the second term gives a vanishing contribution due to $\epsilon $ symbol. It follows
$$ \left<\partial^x _\mu  J^5_\mu (x,a)\right>  \sim {ie^2 \over 3\pi^4}\epsilon _{\mu\nu\tau \rho } a_\lambda a_\tau F_{\mu\lambda }(x)  \partial _\rho A_\nu(x)  \int\d^4 y {y^2-(y\cdot a)^2 
/a^2 \over |y+a/2|^4|y-a/2|^4}\,.$$
After integration we then find
$$ \left<\partial^x _\mu  J^5_\mu (x,a)\right>  \sim {ie^2 \over 4\pi^2}\epsilon _{\mu\nu\tau \rho }{ a_\lambda a_\tau \over a^2} F_{\mu\lambda }(x)  F_{\rho\nu}(x) . $$
Averaging over the $a$ directions we see that the divergence is finite for $|a|\to 0$,
and thus finally
$$\lim_{|a|\to 0} \left<\partial^x _\mu \overline{ J^5_\mu (x,a)}\right> ={ie^2 \over 16\pi^2}\epsilon _{\mu\nu \lambda  \rho }  F_{\mu\lambda }(x)  F_{\rho\nu}(x) , $$
and we recover the result \eaxanom.\par
On the lattice an averaging over $a_\mu$ is produced by summing over all lattice directions. Because the only expression  quadratic in $a_\mu $ that has the symmetry of the lattice is $a^2$ the same result is found: the anomaly is lattice-independent.
\medskip
{\it A direct physical application.} In a phenomenological model of Strong Interaction physics, where the a $SU(2)\times SU(2)$ chiral symmetry is softly 
broken by the pion mass, in the absence of anomalies the divergence of the neutral axial current is proportional to the $\pi_0$ field (corresponding to the neutral pion). A short calculation then shows the decay of $\pi_0$ into two photons would vanish a zero momentum. The axial anomaly \eaxanom\ gives instead a non-vanishing contribution to the decay, in good agreement with experimental data.
\smallskip
{\it Chiral gauge theory.} A gauge theory is consistent only if the gauge field is coupled to a conserved 
current. An anomaly that affects the current  destroys gauge invariance at the quantum level.
Therefore the theory with  axial gauge symmetry, where the action in the fermion sector  reads 
 $$ {\cal S} (\bar \psi ,\psi ;B ) = - \int \d  ^{4}x\, \bar
\psi(x) (\sla{\partial }+  ig \gamma_5 \Bbar) 
\psi(x),\eqnn$$
is inconsistent. Indeed current conservation  applies to the $BBB$ vertex at one-loop order.
Because now the three point vertex is symmetric the divergence is given by the expression \eanoma,
and thus does not vanish. 
 \par
More generally the anomaly   prevents the construction of a theory that would
have both an abelian gauge vector and axial symmetry, where the action in the fermion sector would read 
 $$ {\cal S} (\bar \psi ,\psi;A,B  ) = - \int \d  ^{4}x\, \bar
\psi(x) (\sla{\partial }+ie \Abar + i \gamma_5g \Bbar) 
\psi(x).\eqnn$$
A way to solve  both problems is to cancel the anomaly by introducing another fermion of opposite chiral coupling.  With more fermions other combinations of couplings are possible. Note, however that   in the purely axial gauge theory  it is easy to verify that a theory
with two fermions of opposite chiral charges can be rewritten as a vector theory
by combining differently the chiral components of both fermions.
\subsection Two dimensions 

As an exercise we   verify by explicit calculation the general expression \eanomAbgen\ in the special example of dimension two
$$\partial _\mu J_\mu ^3=- i {e \over 2\pi}\epsilon _{\mu \nu }F_{\mu \nu }\,.\eqnd\eAbaniiD $$
The general form of the l.h.s.~is again dictated by locality and power counting:
the anomaly must have canonical dimension two. The explicit calculation requires some care since massless fields may lead to IR divergences in two dimensions.
One thus gives a mass $m$ to fermions, which breaks chiral symmetry explicitly,
and takes the massless limit at the end of the calculation. The calculation involves only one diagram
$$\eqalign {\Gamma^{(2)}_{ \mu\nu} (k ,-k ) & =\left.
{\delta \over \delta A_\nu  (-k )} \left< J_ \mu ^3(k)\right>\right\vert_{A=0} 
=\left. {\delta \over \delta A_ \nu  (-k
 )}  i\tr\left[\gamma_3
\gamma_ \mu \Dbar^{-1}(k)\right] \right\vert_{A=0}  \cr &={e\over(2\pi)^2}
\tr\gamma_3 \gamma_ \mu\int\d^2 q {1\over i\sla{q}+m}\gamma _\nu 
 {1\over i\sla{q}+i\sla{k}+m}
.   \cr}$$ 
Here the $\gamma $-matrices are simply the ordinary Pauli matrices. 
Then
$$k_\mu \Gamma^{(2)}_{ \mu\nu} (k ,-k )={e\over(2\pi)^2}
\tr\gamma_3\sla{k} \int\d^2 q {1\over i\sla{q}+m}\gamma _\nu 
 {1\over i\sla{q}+i\sla{k}+m}
.  $$  
We use the method of the boson regulator, which leads to the two-dimensional analogue of equation \eAbannce.  It here leads to the calculation
of the difference between two diagrams (analogues of
equation \eabanbose) due to the explicit chiral symmetry breaking
$$\eqalign{C_\mu(k)&=2 m{e\over(2\pi)^2}
\tr \gamma_3\int\d^2 q {1\over i\sla{q}+m}\gamma _\nu 
 {1\over i\sla{q}+i\sla{k}+m}-\ (m\mapsto M)\cr
&=2 m{e\over(2\pi)^2}
\tr \gamma_3\int\d^2 q{\left(m-i\sla{q}\right)\gamma _\mu \left(m-i\sla{q}-i\sla{k}\right)
\over (q^2+m^2)[(k+q)^2+m^2]}-\ (m\mapsto M).\cr}$$
 In the trace again the divergent terms cancel 
$$C_\mu(k)= 4  e m^2\epsilon _{\mu \nu }k_\nu {1\over(2\pi)^2}\int{\d^2 q
\over (q^2+m^2)[(k+q)^2+m^2]}-\ (m\mapsto M). $$
The two contributions are now separately convergent. When $m\to0$
the $m^2$ factor dominates the logarithmic IR divergence and the contribution vanishes. In the second term in the limit $M\to\infty $ one obtains
$$\left. C_\mu(k)\right|_{m\to0\,, M\to\infty }\sim
-4  e M^2\epsilon _{\mu \nu }k_\nu {1\over(2\pi)^2}\int{\d^2 q
\over (q^2+M^2) ^2}=-{  e \over\pi} \epsilon _{\mu \nu }k_\nu\,,$$
in agreement with equation \eAbaniiD.
\subsection  Non-abelian vector gauge fields and abelian axial current

We still consider an abelian axial current but now in the framework of a non-abelian
gauge theory. The fermion fields transform non-trivially under a gauge group
$G$ and ${\bf A}_{\mu}$ is the corresponding gauge field. The action
is:\sslbl\sssAnonab 
$$ {\cal S} (\bar \psi ,\psi;A ) = - \int \d  ^{4}x\, \bar
\psi(x) \Ds \psi (x), \eqnn $$
with the convention \enonabcov\ and:
$$\Ds = \sla{\partial} + \Abbar\, .\eqnd \eDcova  $$
In a gauge transformation of unitary matrix ${\bf g}(x)$ the gauge field ${\bf A}_\mu$ and the  Dirac operator become
$${\bf A}_\mu(x) \mapsto  {\bf g}  (x )\partial_ \mu {\bf g}
^{-1}  (x)+  {\bf g}  (x ){\bf A}_\mu(x) {\bf g}
^{-1}  (x)  \ \Rightarrow \Ds\mapsto {\bf g}^{-1}(x)\Ds{\bf g} (x)\  .\eqnd\eAnogaugDir $$
The axial current $J^5_\mu$
$$ J_{\mu}^5(x)=i \bar\psi(x)\gamma_5 \gamma_{\mu}\psi(x),$$
is still gauge invariant. Therefore no new calculation is needed; the result is completely determined by dimensional analysis, gauge invariance and the preceding abelian calculation that yields the term of order ${\bf A}^2$, 
$$\partial_{\lambda}J^5_{\lambda}(x)=-{i \over
16\pi^2}\epsilon_{\mu\nu\rho\sigma}\tr{\bf F}_{\mu\nu}{\bf F}_{\rho\sigma}\,,
\eqnd \enabanom  $$
in which ${\bf F}_{\mu\nu}$  now is the corresponding curvature \enonabcur.
Again this
expression must be a total derivative. One indeed verifies:
$$\epsilon_{\mu\nu\rho\sigma}\tr{\bf F}_{\mu\nu}{\bf F}_{\rho\sigma}=4\,
 \epsilon_{\mu\nu\rho\sigma}\partial_{\mu}\tr( {\bf A}_ \nu \partial_ \rho  {\bf A}_ \sigma + \frac{2}{3}  {\bf 
A}_\nu {\bf A}_ \rho{\bf A}_\sigma). \eqnd \etotder  $$
\subsection Anomaly and eigenvalues of the Dirac operator

We assume that the spectrum of $\Ds$, the Dirac operator  in a
non-abelian gauge field (equation \eDcova), is discrete (putting temporarily the fermions in a box if necessary) and  call 
$d_n$ and $\varphi_n(x)$ the corresponding eigenvalues and eigenvectors:%
\sslbl\sssSManf
$$\Ds \varphi_n = d_n \varphi_n\, .\eqnn $$
For a unitary or orthogonal group the massless Dirac operator is anti-hermitian;
therefore the eigenvalues are imaginary and the eigenvectors orthogonal. 
In addition we choose them with unit norm. \par
The eigenvalues are gauge invariant, because in a gauge transformation of unitary matrix ${\bf g}(x)$ the Dirac operator transforms like in \eAnogaugDir, 
and thus simply
$$  \varphi_n(x)\mapsto {\bf g} (x) \varphi_n(x).
$$
The anticommutation $\Ds\gamma_5+\gamma_5\Ds=0$ implies
$$\Ds\gamma_5 \varphi_n =- d_n\gamma_5 \varphi_n\, .\eqnn $$
Therefore either $d_n$ is different from zero, and $\gamma_5\varphi_n$ is an
eigenvector of $\Ds$ with eigenvalue $-d_n$, or $d_n$ vanishes. The eigenspace corresponding to the eigenvalue $0$ then is invariant under $\gamma _5$, which can be diagonalized: the eigenvectors of $\Ds$ can be chosen eigenvectors of definite
chirality, i.e.~eigenvectors of $\gamma_5$ with eigenvalue $\pm 1$,
$$\Ds\varphi_n=0\,,\quad \gamma_5 \varphi_n = \pm \varphi_n \,. $$
We call $n_+$ and $n_-$ the dimensions of the eigenspace of positive and negative chirality respectively. \par
We now  consider the determinant of  the operator $\Ds +m$ regularized by mode truncation (mode regularization):
$$ \det_N(\Ds+m) =\prod_{n\le N}(d_n+m), \eqnn $$
keeping the $N$ lowest eigenvalues of $\Ds$ (in modulus), with $N-n_+-n_-$ even, in such a way that the corresponding subspace is $\gamma _5$ invariant.\par
The regularization is gauge invariant because the eigenvalues of $\Ds$ are gauge invariant.\par
Note that in the truncated space 
$$\tr \gamma _5=n_+-n_-\,. \eqnd\eAnomtrgs $$
The trace of $\gamma _5$  equals $n_+- n_-$, the {\it index}\/ of the Dirac operator $\Ds$.  A non-vanishing  index  thus endangers axial current conservation.\par
In  a chiral transformation \espachir~with constant $\theta $ the determinant of $(\Ds +m)$  becomes
$$ \det_ N(\Ds+m)  \mapsto  \det_ N\bigl(\e^{i\theta \gamma _5}(\Ds+m)\e^{i\theta \gamma _5}\bigr)  .$$
We now consider the various eigenspaces.\par
If $d_n\ne 0$ the matrix $\gamma _5$ is represented by the Pauli matrix $\sigma _1$ 
in the sum of eigenspaces corresponding to the two eigenvalues $\pm d_n$
and $\Ds+m$ by $d_n\sigma _3+m$. The determinant in the subspace then is
$$\det\bigl(\e^{i\theta \sigma _1}(d_n\sigma _3+m)\e^{i\theta \sigma _1}\bigr)=
\det \e^{2i\theta \sigma _1}\det(d_n\sigma _3+m)= m^2-d_n^2 ,$$
because $\sigma _1$ is traceless.\par
In the eigenspace of dimension $n_+$ of vanishing eigenvalues $d_n$ with eigenvectors with positive chirality, $\gamma _5$ is diagonal with eigenvalue $1$  and thus
$$m^{n_+}\mapsto m^{n_+}\e^{2i\theta n_+}.$$
Similarly  in the eigenspace of chirality $-1$ and dimension $n_-$
$$m^{n_-}\mapsto m^{n_-}\e^{-2i\theta n_-}.$$
We conclude 
$$\det_ N\bigl(\e^{i\theta \gamma _5}(\Ds+m)\e^{i\theta \gamma _5}\bigr) =
 \e^{2i\theta (n_+-n_-)}\det_ N(\Ds+m),$$
The ratio of both determinants is independent of $N$. Taking the limit $N\to\infty $ we find
$$\det \left[\left(\e^{i\gamma_5 \theta}(\Ds +m)\e^{i\gamma_5
\theta}\right) \left( \Ds +m \right)^{-1}\right]= \e^{2i\theta (n_+-n_-)}. \eqnd \enumzer$$
Note   that the l.h.s.\ of equation \enumzer~is obviously $1$ when
$\theta= n\pi$, which implies that the coefficient of $2\theta $ in the r.h.s.~must indeed be an integer.\par 
The variation of  $\ln\det ( \Ds +m  )$
$$ \ln \det \left[\left(\e^{i\gamma_5 \theta}(\Ds +m)\e^{i\gamma_5
\theta}\right) \left( \Ds +m \right)^{-1}\right] =2i\theta\left( n_+ - n_- \right)  , $$ 
at first order in $\theta $ 
is related  to the variation of the action \eQEDfer~and  thus to the expectation value of the integral of the divergence of the axial current, $\int\d^4x \left<\partial _\mu J^5_\mu(x)\right>$ in four dimensions. In the limit $m=0$ it is thus 
related to the space integral of the
chiral anomaly \enabanom. \par
We have thus found a local expression giving the index of the Dirac operator:
$$ -{ 1\over 32\pi^2}\epsilon_{\mu\nu\rho\sigma}\int \d^4 x\, \tr{\bf
F}_{\mu\nu}{\bf F}_{\rho\sigma}=   n_+ - n_-  \, .
\eqnd \eFFnumz  $$
Concerning this result several comments can be made:
\smallskip
(i) At first order in $\theta $ in the absence of regularization we have calculated ($\ln\det=\tr\ln$)
$$\ln\det\left[1+i\theta \left(\gamma _5+(\Ds+m)\gamma _5(\Ds+m)^{-1}\right)\right]
\sim 2i\theta \tr\gamma _5\,,$$
where we have used the cyclic property of the trace. Since the trace of the matrix $\gamma _5$ vanishes we could expect  naively a vanishing result. But trace here means trace in $\gamma $ space and in coordinate space, and  $\gamma _5$ really stands here for $\gamma _5\delta (x-y)$. The mode regularization  yields a well-defined finite result for the ill defined product $0\times \delta ^d(0) $.\par 
(ii) The property that the integral \eFFnumz\ is quantized
shows that the form of the anomaly is related to topological properties  of
the gauge field since the integral does not change when the gauge field is
deformed continuously. The integral of the anomaly over the whole
space thus depends only on the behaviour at large distances of the curvature tensor ${\bf F_{\mu\nu}}$ and the anomaly must a total derivative as equation \etotder~confirms. \par  
\par
 (iii) One might be surprised   that $\det \Ds$ is not invariant
under global chiral transformations. However we have just established that
when the integral of the anomaly does not vanish, $\det \Ds$ vanishes. This
explains that to  give a meaning to the r.h.s.\ of equation \enumzer~we have
been forced to introduce a mass   to find a non-trivial result. The
determinant of $  \Ds $ in the subspace orthogonal to eigenvectors with vanishing
eigenvalue, even in presence of a mass, is chiral invariant by parity doubling, but for $n_+\ne n_-$ not the determinant in the eigenspace of eigenvalue zero because the trace of $\gamma _5$ does not vanish in the eigenspace (equation \eAnomtrgs).
In the limit $m\to0$ the complete determinant vanishes but not the ratio of determinants for different values
of $\theta $ because the powers of $m$ cancel.\par
(iv) The discussion of the index of the Dirac operator is valid in any even dimension. Therefore the topological character and the quantization of the space integral of the anomaly are general.
\def\sigmab{\sigma}
\section Instantons,  anomalies and $\theta $-vacua

We now discuss the role of instantons in several examples  where the classical potential has a periodic structure with an infinite set of degenerate minima. We  exhibit their topological character, and in the presence of gauge fields relate them to anomalies and the index of the Dirac operator. Instantons imply that the eigenstates of the hamiltonian
depend on an angle $\theta $. In the quantum field theory the notion of $\theta $-vacuum emerges.  
\subsection The periodic cosine potential

As a first example of the role of instantons when topology is involved we
consider a  simple hamiltonian with a periodic potential:
$$ H=-{g\over 2}\left(\d \left/ \d x \right. \right)^2+ {1\over 2g} \sin^2 x\,. \eqnd\ecosham $$ 
The potential   has  an infinite
number of degenerate minima $x=n\pi$, $n\in{\Bbb Z}$. Each minimum is an equivalent starting point for a perturbative calculation of the eigenvalues of $H$. Periodicity implies that the perturbative expansions are identical to all orders in $g$, a property that seems to imply that the quantum
  hamiltonian   has an infinite number of degenerate eigenstates. In reality we know that the exact spectrum of the hamiltonian $ H $ is not degenerate, as a result
of  barrier penetration. Instead it is continuous and has, at least for $ g $ small enough, a band structure. 
\medskip
{\it The structure of the ground state.} 
To characterize more precisely the structure of the spectrum of the hamiltonian \ecosham\ we introduce the operator $
T$ that generates an elementary translation of one period $  \pi$
$$T\psi(x)=\psi(x+\pi).$$
Since $T$ commutes with the hamiltonian,
$$ \left[ T,H \right] =0\,, \eqnn $$
both operators can be diagonalized simultaneously. Because the eigenfunctions of $H$ must  be bounded at infinity, the eigenvalues of $T$
are pure phases. Each eigenfunction of $H$ thus is characterized by an
angle $ \theta $ (pseudo-momentum) eigenvalue of $T$:
$$ T \left| \theta \right> = \e^{i\theta} \left|
\theta \right> . \eqnn $$
The corresponding eigenvalues $E_n(\theta )$ are periodic functions of $\theta $ and for $g\to0$ are close to the eigenvalues of the harmonic oscillator
$$E_n(\theta )=n+1/2+O(g). $$
To all orders in powers of $g$ $E_n(\theta )$ is independent of $\theta $ and
the spectrum of  $H$ is infinitely degenerate. Exponentially small contributions due to instantons lift the degeneracy and introduce a $\theta $ dependence. To each  value of $n$ then corresponds a band when $\theta $ varies in $[0,2\pi]$.
\medskip
{\it Path integral representation.}  The spectrum of $H$ can be extracted from the calculation of   the quantity ${\cal Z}_\ell$ 
$${\cal Z}_\ell(\beta )=\tr T^\ell \e^{-\beta H}={1\over 2\pi}\sum_{n=0}^\infty \int\d\theta \,
\e^{-i\ell\theta }\e^{-\beta E_n(\theta )}. $$
Indeed
$${\cal Z}(\theta,\beta  )\equiv \sum_\ell \e^{i\ell\theta  }{\cal Z}_\ell(\beta )=\sum_n \e^{-\beta E_n(\theta )}, \eqnd\ecosZthet $$
where ${\cal Z}(\theta,\beta  )$ is the partition function restricted to states with a fixed
$\theta $ angle.\par
The path integral representation  of ${\cal Z}_\ell(\beta )$ differs from the representation of the partition function  ${\cal Z}_0(\beta )$ only by the boundary conditions.
The operator $T$ has the effect of   translating the argument $x$ in the
matrix element $\langle x'| \tr\e^{-\beta H}|x\rangle$ before taking the trace:
 $$ \eqalignno{ {\cal Z}_\ell(\beta ) & = \int_{x(\beta /2)=x(-\beta
/2)+\ell \pi}  \left[ \d x (t) \right]
\exp\left[-{\cal S}(x)\right] , & \eqnd\eZpathcosel  \cr   {\cal S}  ( x  ) & ={1\over 2g} \int^{\beta
/2}_{-\beta /2} \left[ \dot x^2(t)+   \sin^2\bigl(x(t)\bigr) \right] \d  t\,. & \eqnn  \cr} $$ 
A careful study of the trace operation in the case of periodic potentials shows  that $ x  (-\beta  /2  ) $ varies over only  one period (see Appendix \label{\ssDegMtr}). \par
Therefore from \ecosZthet\ we derive the path integral representation of $  {\cal Z}(\theta,\beta  )$:
$$\eqalignno{ {\cal Z}(\theta,\beta  )&=\sum_\ell  \int_{x(\beta /2)=x(-\beta
/2)+\ell \pi}  \left[ \d x (t) \right]
\exp\left[-{\cal S}(x)  +i\ell \theta \right]\cr
&=\int_{x(\beta /2)=x(-\beta/2) \hskip-1.5mm\pmod \pi}  \left[ \d x (t) \right]
\exp\left[-{\cal S}(x)+i{\theta\over \pi}\int_{-\beta /2}^{\beta /2}\d t\,\dot x(t) \right]. \hskip7mm &
\eqnd\ecosZpath \cr} $$
Note that $\ell $ is a topological number since two trajectories with
different values of $\ell$ cannot be related continuously. In the same way
$$Q={1\over\pi}\int_{-\beta /2}^{\beta /2}\d t\,\dot x(t),\eqnn $$
is a topological charge; it depends on the trajectory only through the boundary conditions.\par
For $ \beta $ large and $ g\to 0 $ the path integral  is dominated by the
 constant solutions $x_c(t)=0\ {\rm mod}\ \pi$ corresponding to the $\ell=0$ sector. 
A non-trivial $\theta $ dependence can  come only from 
  instanton (non-constant finite action saddle points) contributions  corresponding  to quantum tunnelling. 
Note that quite generally
$$\int\d t\left[\dot x(t)\pm \sin\bigl(x(t)\bigr)\right]^2\ge 0\ 
\Rightarrow \ {\cal S}\ge\left| \cos\bigl(x(+\infty )\bigr) -\cos\bigl(x(-\infty )\bigr)\right|  /g.\eqnd\eBPScos $$
The action is finite only if $x(\pm \infty )\in\{0,\pi\}$. The non-vanishing value of the l.h.s.\ is $2$. The minimum   is reached for trajectories $x_c$ solution of
$$\dot x_c=\pm\sin x_c\ \Rightarrow x_c(t)=2\tan^{-1} \e^{\pm (t-t_0)}, \eqnn $$
 and the corresponding classical action then is:
$$ {\cal S} ( x_{c}  ) =2/g\,  . \eqnn $$
The instanton solutions belong to the  $\ell=\pm 1$ sector
and  connect two consecutive minima of the potential. They yield the leading contribution  to barrier penetration for $g\to0$. An explicit calculation yields
$$E_0(g)=E_{\rm pert.}(g)-{4\over\sqrt{\pi g}}\e^{-2/g}\cos \theta [1+O(g)]. $$
\def\varphib{\varphi}
\subsection Instantons and anomaly:  $ CP(N-1)$ models

We now consider   field theories, the two-dimensional   $CP(N-1)$  models, where again instantons and topology play a role and the semi-classical vacuum has a similar periodic structure. The new feature is the relation between the  topological charge and the two-dimensional chiral anomaly.\par
We here
mainly describe the nature of the instanton solutions and refer the reader to the
literature for a   more detailed analysis. Note that the explicit calculation 
of  instanton contributions in the small coupling limit in the 
$CP(N-1)$  models, as well as in the non-abelian gauge theories we discuss in section \label{\ssdmSUii}, remains to large extent an unsolved problem.
Due to the scale invariance of the classical theory, instantons depend on a scale (or size) parameter. Instanton contributions then involves the running coupling constant at the instanton size. Both families of theories are UV asymptotically free. Therefore the running coupling is small for small instantons and the semi-classical approximation is justified. However, in the absence of any IR cut-off,
the running coupling becomes large for large instantons, and it is unclear whether a  semi-classical approximation remains valid.
\sslbl\ssdmCPN\par 
\smallskip
{\it The $CP(N-1)$  manifolds.} We consider a $N$-component complex vector $\varphib$ of unit length,
$$ \bar \varphib \cdot\varphib =1\,. \eqnd \ezzbar  $$
This condition characterizes a space isomorphic to the quotient space $U(N)/U(N-1)$.
In addition two vectors $ \varphib  $ and $ \varphib' $ are
considered   equivalent if 
$$\varphib'\equiv \varphib\ \Leftrightarrow \  \varphi_\alpha'  =\e^{i\Lambda }\varphi_\alpha  \,.
\eqnd \eCPNequiv  $$ 
This condition  characterizes the symmetric space and complex Grassmannian manifold $U(N)/U(1)/U(N-1) $.  It is isomorphic to the manifold  $CP(N-1)$ (for $N-1$-dimensional Complex Projective), which is obtained from ${\Bbb C}^N$ by the equivalence relation
$$ z_\alpha \equiv z'_\alpha \quad {\rm if}\quad z'_\alpha= \lambda z_\alpha\,$$
where $\lambda $ belongs to the Riemann sphere (compactified complex plane).
\smallskip
{\it The $CP(N-1)$  models.} A symmetric space admits a unique metric, up to a multiplicative factor, and this leads to a unique action with two derivatives.
One form of the unique symmetric classical
action is: 
$$ {\cal S}(\varphib, A_\mu) ={1 \over g} \int \d^{2}x\,
\overline{ {\rm D}_{\mu}\varphib}\cdot {\rm D}_{\mu}\varphib\,, \eqnn $$
in which $g$ is a coupling constant and $ {\rm D}_{\mu}$   the covariant derivative:
$$ {\rm D}_{\mu}=\partial_{\mu}+i A_\mu\,. \eqnn $$
The field $A_\mu$ is a gauge field for the $U(1)$ transformations 
$$ \varphib'(x)=\e^{i\Lambda(x)}\varphib (x)\,, \quad A'_\mu(x)=A_\mu(x)-\partial _\mu \Lambda (x).\eqnd{\egaugequi} $$ 
The action is obviously $U(N)$ symmetric and the gauge symmetry ensures the equivalence  \eCPNequiv. \par
Since the action contains no kinetic term for   $A_\mu$ the gauge field is not a dynamical field but only an auxiliary field
that can be integrated out. The action is quadratic in $A$ and the gaussian  integration  results in   replacing in the action $A_\mu$ by the solution of the $A $-field equation:
$$A_\mu=i\bar \varphib \cdot \partial_{\mu}\varphib\,,\eqnd\eCPNAphi $$
where the equation \ezzbar\ has been used.
After this substitution the field $\bar \varphib \cdot \partial_{\mu}\varphib$ acts as a composite gauge field. \par
For what follows however we find more convenient to keep $A_\mu$ as an independent  field.
\smallskip
{\it Instantons.} To prove the existence of locally stable non-trivial minima of the action   the following Bogomolnyi inequality can be used (note the analogy with \eqns{\eBPScos}):
$$ \int \d^2x \left| {\rm D}_{\mu}\varphib\mp i\epsilon_{\mu\nu}
{\rm D}_{\nu}\varphib \right|^2\geq 0\,, \eqnn $$ 
($\epsilon_{\mu\nu}$ being the antisymmetric tensor, $\epsilon_{12}=1$). After expansion the inequality can be cast in the form
$$ {\cal S}(\varphib)\geq   2\pi |Q(\varphib)|/g\,, \eqnn $$
with
$$Q(\varphib)=-  {i\over 2\pi} \epsilon_{\mu \nu} \int \d^2x\, {\rm D}_{\mu}\varphib
\cdot \overline{ {\rm D}_{\nu}\varphib} = { i\over2\pi}  \int \d^2x\, \epsilon_{\mu \nu}{\rm D}_{\nu}{\rm D}_{\mu}\varphib
\cdot \bar \varphib \,. \eqnd{\etopineq} $$ 
Then
$$ i \epsilon_{\mu \nu}{\rm D}_{\nu}{\rm D}_{\mu}=\ud i \epsilon_{\mu \nu}[{\rm D}_{\nu},{\rm D}_{\mu}]=
\ud F_{\mu \nu }\,,\eqnd\eDGmFmn $$
where $F_{\mu \nu }$ is the curvature
$$F_{\mu \nu }=\partial _\mu A_\nu -\partial _\nu A_\mu \,.$$
Therefore, using \ezzbar,
$$Q(\varphib)={1\over 4\pi} \int \d^2x\, \epsilon_{\mu \nu}F_{\mu \nu } \, .\eqnd\eDGCPtop $$
The integrand is proportional to the two-dimensional abelian chiral anomaly  \eAbaniiD, and thus is a total divergence 
 $$\ud\epsilon_{\mu \nu}F_{\mu \nu }=\partial _\mu \epsilon _{\mu \nu } A_\nu \,.$$
Substituting this form into equation \eDGCPtop\  and integrating in a large disk of radius $R$  one obtains:
$$Q(\varphib)={1\over 2\pi} \lim_{R\to \infty }\oint_{|x|=R}  \d x_\mu \, A_\mu(x).\eqnd\eCPNQiform $$
$Q(\varphib)$ thus depends only on the
behaviour of the classical solution for $|x|$ large and  is a topological charge. Finiteness 
of the action demands that at large distances $ {\rm D}_\mu\varphib$ vanishes
and therefore
$${\rm D}_\mu\varphib=0\ \Rightarrow \ [{\rm D}_\mu,{\rm D}_\nu]\varphib=F_{\mu\nu}
\varphib=0
\,, $$
Since $\varphib\ne0$ this equation implies  that $F_{\mu\nu}$ vanishes and thus $A_\mu$ is a pure gauge (and $\varphib$ a gauge transform of a constant vector)
$$A_\mu  =\partial_\mu  \Lambda (x)\ \Rightarrow\ Q(\varphib)   ={1\over 2\pi}\lim_{R\to \infty }\oint_{|x|=R}  \d x_{\mu}\partial_{\mu}\Lambda(x) \,.   \eqnd{\eQtopol}  $$
The topological charge measures the variation of the angle $
\Lambda(x) $ on a large circle, which is a multiple of $ 2\pi $  because $\varphib$ is regular. One is thus led to the consideration of the homotopy classes of
mappings from $ U(1)$, i.e.\ $ S_{1} $ to $S_{1} $, which are characterized by an integer $n$, the winding number. This is equivalent to the statement  that the homotopy group $\pi_1(S_1)$ is isomorphic to the additive group of integers ${\Bbb Z}$.\par
Then
$$  Q(\varphib)   =  n \ \Longrightarrow  {\cal S}(\varphib)
\geq 2\pi |n| /g\, .   \eqnn   $$
The equality
 $ {\cal S}(\varphib)=2\pi |n| /g $ corresponds to a local minimum and implies
that the classical solutions  satisfy first order
partial differential (self-duality) equations:
$$ {\rm D}_{\mu}\varphib=\pm i\epsilon_{\mu \nu}{\rm D}_{\nu}\varphib\,. \eqnd{\eCPNsd}
$$
For each sign there is really only one equation for instance $\mu=1,\nu=2$. 
It is simple to verify that both equations imply the $\varphi$-field equations, and combined with the constraint \ezzbar\  the $A$-field equation \eCPNAphi. In the complex coordinates $z=x_1+i x_2$, $\bar z=x_1-i x_2$ they can be written 
$$\eqalign{\partial_ z \varphi_\alpha(z,\bar z)  &=-i A_z (z,\bar z) \varphi_\alpha(z,\bar z) , \cr
\partial_ {\bar z} \varphi_\alpha(z,\bar z)  &=-i A_{\bar z}(z,\bar z)  \varphi_\alpha(z,\bar z)   . \cr} $$
Exchanging the two equations just amounts to exchange $\varphib $ and
$\bar \varphib$. We   therefore solve only the second equation
$$ \varphi_\alpha(z,\bar z) =\kappa (z,\bar z)P_{\alpha}(z), $$
where $\kappa (z,\bar z)$ is a particular solution of
$$\partial_ {\bar z} \kappa (z,\bar z) =-i A_{\bar z}  (z,\bar z)\kappa (z,\bar z) .$$
Vectors solutions of the equations \eCPNsd\
are proportional to holomorphic or anti-holo\-mor\-phic (depending on the sign)
vectors  (this reflects the conformal invariance of the classical field
theory).
The function $  \kappa (z,\bar z)$, which  gauge invariance allows to  choose  real (this corresponds to the $\partial _\mu A_\mu=0$ gauge),  then is constrained by the condition \ezzbar
$$  \kappa^2 (z,\bar z) \,  P\cdot\bar P=1  \,. $$
The asymptotic conditions constrain the functions $P_\alpha (z)$ to be polynomials. Common roots to all $P_\alpha $  would correspond to non-integrable singularities for $ \varphi_{\alpha}$,  and therefore are excluded by the  condition of finiteness of the action. Finally if the polynomials have maximal degree $n$, asymptotically
 $$P_\alpha (z)\sim c_\alpha z^n\ \Rightarrow \ \varphi_\alpha\sim {c_\alpha \over
\sqrt{{\bf c}\cdot \bar{  \bf c}}}(z/\bar z)^{n/2}.$$
When the phase of $z$ varies by $2\pi $, the phase of $\varphi_\alpha $ varies
by $2n\pi $, showing that the corresponding winding number is $n$.
 
\medskip
{\it The structure of the semi-classical vacuum.} In contrast to our analysis
of  periodic potentials in quantum mechanics, we have here discussed the existence of
instantons without reference to the structure of the classical vacuum. To
find an interpretation of   instantons in gauge theories, it is useful to express the results in the temporal gauge. Then 
classical minima of the potential correspond to fields $\varphib(x_1)$, where $x_1$ is only the space variable, gauge transforms of a constant vector: $$\varphib(x_1)=\e^{i\Lambda (x_1)}{\bf v}\,,\quad \bar{\bf v}\cdot {\bf v}=1\,.$$
Moreover if the vacuum state is invariant under space reflection
$\varphib(+\infty )=\varphib(-\infty )$ and thus
$$\Lambda (+\infty )-\Lambda (-\infty )=2\nu \pi \quad \nu \in {\Bbb Z}\,.$$
Again $\nu$ is a topological number that classifies degenerate classical minima, and the semi-classical vacuum thus has a periodic structure. This analysis is consistent with Gauss's law   that 
only implies that  states are invariant under infinitesimal gauge transformations,
and therefore under gauge transformations of the class $\nu =0$ that are continuously 
connected to the identity.\par 
We now consider a large rectangle with extension $R$ in the space direction and $T$ in the euclidean time direction and by a smooth gauge transformation continue the instanton solution to the temporal gauge. Then the variation of the pure gauge comes entirely from the sides at fixed time. One
finds for $R\to\infty $,
$$\Lambda (+\infty ,0)- \Lambda (-\infty ,0)-\left[\Lambda (+\infty ,T)- \Lambda (-\infty ,T)\right]=2n\pi\,.$$
Therefore instantons interpolate between different  classical minima. Like in the case of the cosine potential, one projects onto a proper quantum  eigenstate, the ``$\theta$-vacuum" corresponding to an angle $\theta $ by adding, in analogy with the expression \ecosZpath, a topological term to the classical action 
$${\cal S}(\varphib)\mapsto {\cal S}(\varphib)+i{\theta \over 4\pi}\int\d^2x \,\epsilon _{\mu\nu}F_{\mu\nu}. $$
\smallskip
{\it Remark.} Replacing in the topological charge $Q$ the gauge field by the explicit expression \eCPNAphi\ we find
$$Q(\varphi)={i\over 2\pi}\int\d^2 x\, \epsilon _{\mu \nu }\partial _\mu\bar\varphib\cdot\partial _\nu \varphib={i\over 2\pi}\int\d^2 x\,\d\bar\varphi_\alpha \wedge \d \varphi_\alpha\,, $$
where the notation of exterior differential calculus has been used.
We recognize the integral of a two-form, a symplectic form and
$4\pi Q$ is the area of a 2-surface embedded in $CP(N-1)$.
A symplectic form is always closed, here it is also exact, so that  $Q$ is the integral of a one-form
(equation \eCPNQiform)
$$Q(\varphi)={i\over 2\pi}\int\bar\varphi_\alpha  \d \varphi_\alpha={i\over 4\pi}
 \int\left( \bar\varphi_\alpha  \d \varphi_\alpha-\varphi_\alpha  \d \bar\varphi_\alpha
\right). $$
\medskip
{\it The $O(3)$ non-linear $\sigma $-model.} The $CP(1)$ model is locally isomorphic to the $ O (3)$ non-linear
$ \sigma$-model, with the identification  
$$\phi^i(x)=\bar \varphi_{\alpha}(x) \sigma^i_{\alpha \beta}\varphi_{\beta}(x)\,. \eqnn
$$ 
where $\sigma^i$ are the three Pauli matrices. \par
Using for example an explicit representation of Pauli matrices, one verifies indeed
$$\phi^i(x) \phi^i(x) =1\,,\quad \partial _\mu \phi^i(x) \partial _\mu \phi^i(x) 
=4\overline{ {\rm D}_{\mu}\varphib}\cdot {\rm D}_{\mu}\varphib\,.$$
Therefore the field theory can be expressed in terms of the field $\phi^i$, and takes the form of the non-linear $\sigma $-model.
The fields   $\phi$ are gauge invariant, and the whole physical picture is a picture of confinement of the charged scalar ``quarks"  $\varphi_{\alpha}(x)$
and the propagation of the $\phi^i$ neutral bound states. \par
Instantons in the $\phi $ description take the form of $\phi$ configurations 
with uniform limit for $|x|\to \infty $. They thus define a mapping from the compactified plane topologically equivalent to $S_2$ to the sphere $S_2$ (the $\phi^i$ configurations). Since $\pi_2(S_2)={\Bbb Z}$ the $\varphi$ and $\phi $ pictures are consistent.\par
In the example of $CP(1)$ a solution of winding number 1 is
$$\varphi_1={1\over\sqrt{1+z\bar z}}\,,\quad \varphi_2={z\over\sqrt{1+z\bar z}}\,.$$
Translating the $CP(1)$ minimal solution  into
the $O(3)$ $\sigma$-model language one finds $$\phi_1={z+\bar z\over 1+\bar z z}\,,\quad \phi_2={1\over i}{z-\bar z\over 1+\bar z z}\,,
\quad \phi_3={1-\bar z z \over  1+\bar z z}\,.$$
This defines a stereographic mapping of  the plane onto the sphere $S_2$, as one verifies by setting $z=\tan(\eta /2)\e^{i\theta }$, $\eta \in[0,\pi]$. \par
In the $O(3)$ representation the topological charge $4\pi Q$ has  the interpretation of an area  of surface in $S_2$, itself embedded in ${\Bbb R}^3$:
$$Q={i\over 2\pi}\int\d\bar \varphi_\alpha \wedge\d   \varphi_\alpha=
{1\over 8\pi}\epsilon _{ijk}\int \phi _i \d \phi _j\wedge \phi _k\equiv
{1\over 8\pi}\epsilon _{\mu \nu }\epsilon _{ijk}\int\d^2 x\, \phi _i \partial _\mu \phi _j \partial _\nu \phi _k\,.$$
The result is a multiple of the area of the sphere $S_2$, which in this interpretation explains the quantization.
\subsection  Instantons and anomaly: non-abelian gauge theories

We now consider non-abelian gauge theories  in four dimensions. Again gauge field configurations can be found that contribute to the chiral anomaly and 
for which therefore the r.h.s.\ of equation \eFFnumz~does not vanish. A specially interesting example is provided  by instantons, i.e.~finite action solutions of euclidean field equations. \par
To discuss this problem  it is sufficient to consider  pure gauge theories and the gauge group $SU(2)$
since a general  theorem states that for a Lie group containing $SU(2)$ as a
subgroup the instantons are those of the $ SU(2) $
subgroup. \sslbl\ssdmSUii
\par  
In the absence of matter fields it is convenient  to use a $SO(3)$ notation. The gauge field  ${\bf A}_\mu$ is a  vector that is related to the element ${\goth A}_\mu$ of the Lie algebra used previously as gauge field by
$${\goth A}_\mu=- \ud i{\bf A}_{\mu}\cdot \sigmab \,,\eqnn $$
where $ \sigma_i  $ are the three Pauli matrices. The gauge action then reads:
$$ {\cal S} ( {\bf A}_ \mu ) = {1 \over 4g^2} \int \left[ {\bf F}_{\mu
\nu}(x) \right]^2 \d^{4}x\,, \eqnn $$
($g$ is the gauge coupling constant)
where the curvature ${\bf F}_{\mu \nu}$ is also a vector:
$$ {\bf F}_{\mu \nu}=\partial_{\mu} {\bf A}_{\nu}-\partial_{\nu}
{\bf A}_{\mu}+  {\bf A}_{\mu}\times {\bf A}_{\nu}\,. \eqnn $$
The corresponding classical field equations are
$${\bf D}_\nu{\bf F}_{\nu\mu}= \partial _\nu {\bf F}_{\nu\mu}+ {\bf A}_\nu\times 
{\bf F}_{\nu\mu}=0\,.\eqnd \eSUiifledeq$$
The existence and some properties of instantons in this theory
follow from  considerations analogous to those presented for the $
CP (N-1 ) $ model. \par
We define the dual of the tensor $ {\bf F}_{\mu \nu} $ by
$$ \tilde{\bf F}_{\mu \nu}=\ud\epsilon_{\mu \nu \rho \sigma}
{\bf F}_{\rho \sigma}\,. \eqnn $$
Then the  Bogomolnyi inequality
$$ \int \d^{4}x \left[ {\bf F}_{\mu \nu}(x)\pm {\bf \tilde F}
_{\mu \nu}(x) \right]^2\geq 0\,, \eqnn $$
implies:
$$ {\cal S}({\bf A}_\mu)  \ge 8\pi^2 |Q({\bf A}_\mu)|/g^2\,, \eqnn $$
with
$$Q({\bf A}_\mu)= {1\over 32\pi^2}  \int \d^{4}x\, {\bf F}_{\mu \nu}\cdot {\bf \tilde F}_{\mu \nu}\,  . \eqnd{\eineqFFd} $$ 
The expression $Q({\bf A}_\mu)$ is proportional to the integral of the chiral anomaly \enabanom,  here written in $SO(3)$ notation. 
 We have shown that the quantity $ {\bf F}_{\mu \nu}\cdot {\bf \tilde F}_{\mu \nu}
$  is a pure divergence (equation \etotder):
$$ {\bf F}_{\mu \nu}\cdot {\bf \tilde F}_{\mu \nu}= \partial_{\mu}  V_{\mu}\,, $$
with
\eqna \etotderii 
$$\eqalignno{ V_\mu&=  -4\,
 \epsilon_{\mu\nu\rho\sigma} \tr\left( {\goth A}_ \nu \partial_ \rho  {\goth A}_ \sigma + \frac{2}{3}  {\goth A}_\nu {\goth A}_ \rho{\goth A}_\sigma\right) &  \etotderii{a}\cr
&=2\epsilon_{\mu \nu \rho \sigma} \left[ {\bf A}
_{\nu}\cdot \partial_{\rho} {\bf A}_{\sigma}+{\textstyle{1 \over 3}}
{\bf A}_{\nu}\cdot \left( {\bf A}_{\rho}\times {\bf A}_{\sigma}
\right) \right] .&  \etotderii{b} \cr} $$
The integral thus depends only on the behaviour of the
gauge field at large distances and its values are quantized (equation \eFFnumz). 
Here again, as in the $CP(N-1)$ model, the bound involves  a topological charge, $Q({\bf A}_\mu)$.
\par
We now use equation \etotder~and Stokes theorem 
$$\int_{\cal D}\d^4 x\,\partial _\mu V_\mu =\int_{\partial {\cal D}}\d  \Omega\ \hat  n_\mu   V _{\mu}\,, $$
where $ \d  \Omega $ is the measure  on the boundary $\partial  {\cal D} $  of the four-volume $\cal D$ and $\hat n_\mu $ the unit vector normal to $\partial  {\cal D} $. We take for $\cal D$  a sphere of large radius $R$ and finds for the topological charge $Q$,
$$Q({\bf A}_\mu)=  {1\over 32\pi^ 2 }\int \d^{4}x\,\tr {\bf F}_{\mu \nu} \cdot  \tilde{\bf F}_{\mu
\nu}=  {1\over 32\pi^ 2 }R^3\int_{r=R} \d  \Omega\  \hat   n_\mu   V _{\mu}\,,
\eqnd\eSUiitop $$
The finiteness of the action  implies that the classical solution
must asymptotically become a pure gauge, i.e.\ with our conventions,
$${\goth A}_\mu= -\ud i{\bf A}_{\mu}\cdot \sigmab ={\bf g}  (x )\partial_{\mu}{\bf g}
^{-1}  (x)+O\left( | x |^{-2} \right)\  \ | x|\to \infty\, . \eqnd\eAnopgau $$
The element $\bf g$ of the $SU(2)$ group  can be parametrized  in terms of Pauli  matrices
$${\bf g}=u_4+i{\bf u}\cdot \sigmab\,, \eqnd\eSUiigas $$
where $(u_4,\bf u)$ is a four-component real vector belonging the unit sphere $S_3$:
$$u_4^2+{\bf u}^2=1\,.$$
Since $SU(2)$ is topologically equivalent
to the sphere $S_ 3  $,  the pure gauge configurations on a sphere of large radius $|x|=R$
define a mapping from $  S_ 3  $ to $ S_ 3  $. Such mappings belong to different homotopy classes that are characterized by an integer
called the {\it winding number}. 
Here we identify the homotopy group $\pi_3(S_3)$, which again is isomorphic to the additive group of integers ${\Bbb Z}$.\par
The simplest one to one mapping corresponds to an element ${\bf g}(x) $ of the form
$$ {\bf g}  (x  )={x_{4}+i {\bf x} \cdot \sigmab \over r},\quad
 r=  (x^2_{4}+ {\bf x}^2 )^{1/2}, \eqnn $$
and thus
$$ A^{i}_{m}\mathop{\sim}_{r\to \infty }2 \left(x_{4}\delta_{im}+\epsilon_{imk}x_{k}
\right)r^{-2},\quad A^{i}_{4}=-2x_{i}r^{-2}. \eqnn $$
Note that the transformation
$${\bf g}(x)\mapsto {\bf U}_1{\bf g}(x){\bf U}_2^\dagger={\bf g}({\bf R}x), $$
where ${\bf U}_1$ and ${\bf U}_2$ are two constant $SU(2)$ matrices, induces a $SO(4)$ rotation of matrix ${\bf R}$ of the vector $x_\mu$.
Then 
$$ {\bf U}_2\partial _\mu{\bf g}^\dagger(x) {\bf U}_1^\dagger =R_{\mu\nu}\partial _\nu
{\bf g}^\dagger({\bf R}x),\quad  {\bf U}_1{\bf g}(x)\partial _\mu{\bf g}^\dagger(x) {\bf U}_1^\dagger={\bf g}({\bf R}x) R_{\mu\nu}\partial _\nu
{\bf g}^\dagger({\bf R}x), $$
and therefore
$${\bf U}_1{\goth A}_\mu(x)  {\bf U}_1^\dagger = R_{\mu\nu} {\goth A}_\nu({\bf R}x).$$
Introducing this relation into the definition \etotderii{a} of $V_\mu$ we verify  that the dependence in the matrix $ {\bf U}_1$ cancels in the trace and  
thus $V_\mu$ transforms like a four-vector. Since only one vector is available, and taking into account dimensional analysis we conclude
$$V_\mu\propto x_\mu /r^4\,.$$
\par
For $R\to \infty $  ${\bf A}_\mu$ approaches a pure gauge (equation \eAnopgau), 
and therefore $V_\mu$ can be transformed into
$$V_\mu\sim-\frac{1}{3}\epsilon _{\mu \nu \rho \sigma } {\bf A}_\nu\cdot( {\bf A}_ \rho\times {\bf A}_\sigma).$$
It is sufficient to calculate $V_1$. We can choose $\rho =3,\sigma =4$ and multiply by a factor six to take into account all other choices. Then
 $$V_1= 16 \epsilon _{ijk} (x_4\delta _{2i}+\epsilon _{i2l}x_l)(x_4\delta _{3j}+\epsilon _{j3m}x_m)x_k/r^6= 16 x_1/r^4 ,$$
and thus
$$V_\mu\sim  16x_\mu/ r^4= 16\hat n_\mu/ R^3\,. $$
The powers of $R$ in equation \eSUiitop\ cancel and since $\int\d\Omega =2\pi^2$ the value of the topological charge is simply  
$$Q({\bf A}_\mu)= 1\, .\eqnn $$
If we compare this result with equation \eFFnumz\ we see that we have indeed
found the minimal action solution. \par
Without explicit calculation we know already from the analysis of the index of the Dirac operator that the topological charge is an integer: 
$$Q({\bf A}_\mu)={1\over 32\pi^2} \int \d^{4}x\, {\bf F}_{\mu \nu}\cdot {\bf \tilde F}_{\mu
\nu}= n\,.\eqnn $$
As in the case of the $CP(N-1)$ model this result has a geometric interpretation.
In general in the parametrization \eSUiigas, 
$$V_\mu\mathop{\sim}_{r\to \infty }\frac{8}{3}\epsilon _{\mu \nu \rho \sigma }\epsilon _{\alpha \beta \gamma \delta }
u_\alpha \partial _\nu u_\beta \partial _\rho u_\gamma \partial _\sigma u_\delta \,.$$
A few algebraic manipulations starting from
$$\int_{S_3} R^3\d  \Omega\  \hat   n_\mu   V _{\mu}=
\frac{1}{6}\epsilon _{\mu \nu \rho \sigma }\int V_\mu \d u_\nu \wedge \d u_\rho\wedge \d u_\sigma \,, $$
then yield 
$$Q={1\over 12\pi^2}\epsilon _{\mu \nu \rho \sigma }\int
u_\mu \d u_\nu \wedge \d u_\rho\wedge \d u_\sigma \,, \eqnd\eSUiiQSiii $$
where the notation of exterior differential calculus again has been used. 
The area $\Sigma _p$ of the sphere $S_{p-1}$ in the same notation can be written
$$\Sigma _p={2\pi^{p/2}\over \Gamma (p/2)}={1\over (p-1)!}
\epsilon _{\mu_1\ldots \mu_p}\int\, u_{\mu_1}\d u_{\mu_2}\wedge\ldots\wedge
 \d u_{\mu_p}\,,$$
when the vector $u_\mu$ describes the sphere $S_{p-1}$ only once. In the l.h.s.~of equation \eSUiiQSiii\ one thus recognizes an expression proportional to the area of the sphere $S_3$.  Because in general   $u_\mu$ describes $S_3$ $n$ times when $x_\mu$ describes $S_3$ only once a  factor $n$ is generated
\par
The inequality \eineqFFd\ then implies
$$ {\cal S} ({\bf A}_{\mu} ) \geq 8\pi^2|n| /g^2\, . \eqnn $$
The equality, which corresponds to a local minimum of the action, is
obtained for fields satisfying the {\it self-duality equations}
$${\bf F}_{\mu \nu}= \pm\tilde{\bf F}_{\mu \nu}\,. \eqnn $$
These equations, unlike the general classical field equations \eSUiifledeq, are first order partial differential equations and therefore easier to solve. The one-instanton
solution, which depends on an arbitrary scale parameter $\lambda$, is
$$   A^{i}_{m}   = {2 \over r^2+\lambda
^2} \left(x_{4}\delta_{im}+\epsilon_{imk}x_{k} \right),\
\ m=1,2,3\,, \quad  A^{i}_{4}   =- {2x_{i} \over r^2+\lambda
^2}.   \eqnd{\euninstan} $$
\smallskip
{\it The semi-classical vacuum.}  We now proceed in analogy with the analysis of the $CP(N-1)$ model. In the temporal gauge ${\bf A}_4=0$  the
classical minima of the potential correspond to gauge field components ${\bf A}_i$, $i=1,2,3$,
which are pure gauge functions of the three space variables $x_i$: 
$${\goth A}_m=- \ud i{\bf A}_{m}\cdot\sigmab = {\bf g}(x_{i})\partial_{m}{\bf g}^{-1}(x_{i})
\,. \eqnn $$
 The structure
of the classical minima is related to the homotopy classes of mappings of
the group elements $ {\bf g}$ into compactified ${\Bbb R}^3$ (because ${\bf g}(x)$ goes to a constant for $|x|\to \infty $), i.e.\ again of
$ S_3$ into $S_3$ and thus the  semi-classical vacuum has a periodic
structure. One   verifies that
the  instanton solution \euninstan, transported into the temporal gauge by a
gauge transformation, connects minima with  different winding numbers. 
Therefore, as in the case of the  $CP(N-1)$ model,  to project
onto a  $\theta$-vacuum,  one adds a term
to the classical action of gauge theories: 
$$ {\cal S}_{\theta}  ( {\bf A}_{\mu} )={\cal S}  ( {\bf A}_{\mu}
 )+{i\theta \over 32\pi^2} \int d^{4}x\, {\bf F}_{\mu
\nu}\cdot {\bf \tilde F}_{\mu \nu}\,, \eqnn $$
and   then integrates over all fields $ {\bf A}_{\mu} $
without restriction.   At least in the semi-classical approximation, the gauge theory depends on
an additional parameter, the angle $ \theta $. For non-vanishing values of $
\theta  $ the additional term violates CP conservation, and is at the
origin of the strong CP violation problem, because if  $\theta $ does not vanish experimental bounds are consistent only with unnaturally small values.
\subsection Fermions in an instanton background 

 We now apply this analysis to QCD, the  theory of Strong Interactions, where  $N_F$ Dirac fermions  ${\bf Q}$, $\bar{\bf Q}$, the quark fields,   are coupled 
to  non-abelian gauge fields ${\bf A}_\mu$ corresponding to the $SU(3)$ colour group. The action can be written (returning here to   standard $SU(3)$ notation) 
$${\cal S} ({\bf A}_{\mu},\bar{\bf Q},{\bf Q} )=- \int \d^4x\left[ {1 \over
4 g^2} \tr {\bf F}_{\mu\nu}^2 + \sum_{f=1}^{N_f}\bar{\bf Q}_f \left(\Ds +
m_f \right) {\bf Q}_f \right] .$$
The existence of abelian anomalies and instantons has several physical consequences. We mention here two of them.
\smallskip
{\it The strong CP problem.}
 According to the analysis of Section \sssSManf\ only configurations with a non-vanishing index of the Dirac operator contribute to the $\theta $-term. Then the Dirac operator has at least one vanishing eigenvalue. If one fermion field is massless, the determinant resulting
from the fermion integration thus vanishes,
 the instantons do not contribute to the functional
integral and the strong CP violation problem  is solved. However
such an hypothesis  seems to be  inconsistent  with experimental data on  quark masses. Another scheme was based on a scalar field, the axion, which unfortunately has remained experimentally invisible.
\smallskip
{\it The solution of the $U(1)$ problem.} Experimentally it is observed that
masses of pseudo-scalar mesons are smaller or even much smaller
(in the case of pions) than the masses of the corresponding scalar mesons.
This strongly suggests an approximate chiral symmetry
corresponding to small quark $\bf u$ and $\bf d$ masses, and a more badly violated chiral symmetry corresponding to the strange quark, realized in a phase of spontaneous symmetry breaking. This picture is confirmed by its many other physical consequences.
\par
In a  theory 
in which the quarks would be massless the action would have a chiral $U(N_{\rm F})\times U(N_{\rm F})$ symmetry, in which
$N_{\rm F}$ is the number of flavours.  The spontaneous breaking of chiral
symmetry to its diagonal subgroup $U(N_{\rm F})$ leads to expect $N_{\rm F}^2$
Goldstone bosons associated with all axial currents (corresponding to the generators of $U(N)\times U(N)$ that do not belong to the remaining $U(N)$ symmetry group). If the masses of quarks are non-vanishing but small one expects this picture to survive approximately, with instead of Goldstone bosons light  pseudo-scalar mesons.
From the preceding analysis we know that the axial
current corresponding to the $U(1)$ abelian subgroup has an anomaly.  
The WT identities which imply  the existence of Goldstone bosons
correspond to constant group transformations and therefore involve only the
space integral of the divergence of the current. Since the anomaly is a total
derivative one might have expected the integral to vanish. However  non-abelian gauge theories have configurations that give  non-vanishing values  of the form \eFFnumz\ to the space integral of the anomaly \enabanom. For small couplings these configurations are in the neighbourhood of instanton 
solutions  (as discussed in section \ssdmSUii).   This indicates (but no satisfactory calculation of the instanton
contribution has been performed) that for small, but non-vanishing, quark
masses the $U(1)$ axial current is far from being conserved and therefore no
corresponding light would-be Goldstone boson is generated. This argument resolves a long
standing puzzle since experimentally no corresponding light pseudoscalar boson
is indeed observed for $N_F=2,3$. \par
Note that the usual derivation of WT identities involves only global chiral transformations, and therefore there is no need to introduce axial currents. In the case of massive quarks  chiral symmetry is explicitly broken by soft mass terms and WT identities involve insertions of the operators
$${\cal M}_f =m_f\int\d^4 x \, \bar{\bf Q}_f(x)\gamma _5
  {\bf Q}_f(x),$$
which are the variations of the mass terms in an infinitesimal chiral transformation.
If the contributions of ${\cal M}_f$ vanish when $m_f\to 0$, as one would normally expect, then a situation of approximate chiral symmetry is realized (in a symmetric or spontaneously broken phase).
However if we integrate over fermions first, at fixed gauge fields
we find (disconnected) contributions proportional to
$$\left<{\cal M}_f \right> = m_f \tr\gamma _5\left(\Ds +
m_f \right)^{-1}.  $$
We have shown in section \sssSManf\ that for topologically non-trivial gauge field configurations $\Ds$ has zero eigenmodes, which for $m_f\to 0$ give the leading contributions
$$\eqalign{\left<{\cal M}_f \right>&= m_f\sum_n \int\d^4 x\,\varphi_n^*(x)\gamma _5 
\varphi_n (x){1\over m_f}+O(m_f)\cr
&=(n_+-n_-)+O(m_f).\cr}$$
These contributions do not vanish for $m_f\to0$ and are responsible, after integration over gauge fields, for a violation of chiral symmetry. 
\section Non-abelian anomaly

We first consider the problem of conservation of a general axial current in a non-abelian vector gauge theory, and then the issue of obstruction to gauge invariance in chiral gauge theories.
\subsection General axial current

We now discuss  the problem of the conservation of a general
axial  current in the example of an  action with $N$ massless Dirac fermions in the background of non-abelian vector gauge fields
$$ {\cal S} (\psi , \bar{ \psi};A )=- \int \d ^{4}x\, \bar{ \psi}
_{i} (x) \Ds \psi_{i} (x). \eqnn $$
In the absence of gauge fields the action $ {\cal S} (\psi , \bar{ \psi};0 )$ has a 
$U(N)\times U(N)$ symmetry
corresponding to the transformations:
$$\eqalignno{ \psib ' &= \left[ \ud ( {\bf 1} + 
\gamma_5 ) {\bf U}_+ +\ud ( {\bf 1} - 
\gamma_5 ) {\bf U}_- \right]\psib\, ,&
\eqnd \eAnomchirtra \cr
\bar{ \psib} ' &= \bar{ \psib} \left[\ud (
{\bf 1} +   \gamma_5)  {\bf U}^{\dagger}_-  +\ud (
{\bf 1} -  \gamma_5 ) {\bf U}^{\dagger}_+ \right],
&\eqnd \eAnomchirtrb \cr} $$ 
where $ {\bf U}_\pm $ are  $ N\times N $ unitary matrices. We denote by   ${\bf t}^{\alpha}$ the anti-hermitian generators of $U(N)$,
$${\bf U}={\bf 1}+\theta_\alpha  {\bf t}^{\alpha}+O(\theta ^2).$$ 
We consider gauge fields coupled to all vector currents, corresponding to
diagonal subgroup of $U(N)\times U(N)$ corresponding to ${ \bf U}_+= {\bf U}_-$,
$${\bf A}_\mu = {\bf t}^{\alpha}A_\mu^\alpha .$$
We define axial currents in terms of infinitesimal
space-dependent chiral transformation:
$$ {\bf U}_\pm={\bf 1}\pm \theta _\alpha (x) {\bf t}^{\alpha}+O(\theta ^2)\ \Rightarrow 
\delta \psi =\theta _\alpha (x)\gamma _5{\bf t}^{\alpha}\psi ,\quad
\delta \bar \psi =\theta _\alpha (x)\bar\psi \gamma _5{\bf t}^{\alpha}
.$$
The variation of the action then reads
$$\delta {\cal S}=\int \d^4 x \left\{J^{5\alpha}_{\mu}(x) \partial _\mu  \theta _\alpha (x)
+\theta _\alpha (x) \bar\psi(x) \gamma_5 \gamma_{\mu}[ {\bf A}_\mu, {\bf t}^{\alpha}] 
\psi(x)\right\}  , \eqnn $$ 
where $J^{5\alpha}_{\mu}(x)$ is the axial current: 
$$J^{5\alpha}_{\mu}(x)=  \bar\psi \gamma_5 \gamma_{\mu}
{\bf t}^{\alpha} \psi\, . \eqnn $$ 
Since the gauge group has a non-trivial intersection with the chiral group, 
the commutator $[ {\bf A}_\mu, {\bf t}^{\alpha}]$   no longer vanishes
$$[ {\bf A}_\mu, {\bf t}^{\alpha}]=A_\mu^\beta f_{\beta \alpha \gamma }{\bf t}^  \gamma \,,$$
where the $ f_{\beta \alpha \gamma }$ are the totally antisymmetric structure constants of the Lie algebra of $U(N)$. Thus
$$\delta {\cal S}=\int \d^4 x\,\theta _\alpha (x)  \left\{-\partial _\mu  J^{5\alpha}_{\mu}(x) 
+ f_{ \beta \alpha\gamma }  A _\mu^\beta (x)  J^{5 \gamma }_{\mu}(x)   \right\}  . \eqnn $$ 
The classical current
conservation equation is replaced by a gauge covariant conservation equation:
$${\bf D}_{\mu}{\bf J}^5_{\mu}=0\,, \eqnn $$
where we have defined the covariant divergence of the current by
$$\left({\bf D}_{\mu}  J^{5 }_{\mu}\right)_\alpha \equiv \partial _\mu J^{5\alpha }_{\mu}+f_{\alpha \beta \gamma }A_\mu^\beta J^{5 \gamma  }_{\mu}.$$
In  the contribution to the anomaly the terms quadratic in the gauge fields are modified, compared to   the expression \etotder,   only by the appearance of 
a new geometric  factor. Then the complete form of the anomaly is
dictated by gauge covariance. One finds:
$${{\bf D}_{\lambda}J^{5\alpha}_{\lambda}(x)} =-{1 \over
16\pi^2}\epsilon_{\mu\nu\rho\sigma}\tr {\bf t}^{\alpha} {\bf F}_{\mu\nu}{\bf
F}_{\rho\sigma}\,. \eqnn $$
This is result for the most general chiral and gauge transformations. If we restrict
both groups in such a way that the gauge group has an empty intersection with the chiral group the anomaly becomes proportional to $\tr {\bf t}^{\alpha}$, where ${\bf t}^{\alpha}$
are the generators of the chiral group $G\times G$, and is therefore different from
zero only for the abelian factors of $G$.
\subsection Obstruction to gauge invariance

We now  consider a non-abelian gauge field coupled to left or
right-handed fermions for instance:\sslbl\sssAnomR
$$ {\cal S} (\bar \psi ,\psi;A ) = - \int \d  ^{4}x\, \bar
\psi (x)\ud \left( 1 + \gamma_5 \right) \Ds \psi(x ),
\eqnn $$
(the discussion with $\ud(1-\gamma_5)$ is  similar).\par
The gauge theory can be consistent  only if the partition function
$$ {\cal Z} ({\bf A}_{\mu})=\int \left[\d \psi \d \bar\psi\right]
\exp\left[ -{\cal S}(\psi ,\bar\psi;A)\right] \eqnn $$
is gauge invariant. \par
We introduce the generators ${\bf t}^{\alpha}$ of the gauge group in the fermion
representation and define the corresponding current ${\bf J}_{\mu}$ as: 
$$J^{\alpha}_{\mu}(x)= \bar\psi \ud \left(1+\gamma_5\right) \gamma_{\mu}
{\bf t}^{\alpha} \psi\, . \eqnn $$
The invariance of $ {\cal Z}({\bf A}_{\mu})$ under an infinitesimal gauge
transformation again leads to a covariant conservation equation for the
current: 
$$\left< {\bf D}_{\mu}{\bf J}_{\mu}\right>=0\,, $$
with
$${\bf D}_\mu =\partial _\mu + [{\bf A}_\mu,\bullet]. $$
The calculation of the quadratic contribution to the anomaly is
simple: the first regularization adopted for the calculation in section
\sssAnomc\ is also suited to the present situation since the current-gauge field
three-point function is symmetric in the external arguments. The group structure
is reflected by a simple geometric  factor. The global factor can be taken
from the abelian calculation. It differs from result \eanoma\ by a factor
$1/2$ that comes from the projector $\ud (1+\gamma_5 )$. The general
form of the term of degree three in the gauge field can also  easily be found, but
the calculation of the global factor is somewhat tedious. We  argue in the
 section \label{\ssAnoWZ} that it can be obtained from consistency conditions. The complete
expression reads:
$$\left({\bf D}_{\mu}{\bf J}_{\mu}(x)\right)^{\alpha}=-{1 \over 24\pi^2}
\partial_{\mu}\epsilon_{\mu\nu\rho\sigma}  \tr\left[
{\bf t}^{\alpha} \left( {\bf A}_{\nu} \partial_{\rho} {\bf A}_{\sigma}+ \ud {\bf
A}_{\nu}{\bf A}_{\rho}{\bf A}_{\sigma}\right) \right]. \eqnd{\egauamon} $$
If the projector $\ud(1+\gamma_5)$ is replaced by $\ud (1-\gamma_5)$ the sign
of the anomaly changes.\par
Unless this term vanishes identically there is an obstruction to the
construction of the gauge theory. The first term is proportional to $d_{\alpha\beta\gamma}$:
$$d_{\alpha\beta\gamma}= \ud\tr\left[ {\bf t}^{\alpha}\left(
{\bf t}^{\beta}{\bf t}^{\gamma}+{\bf t}^{\gamma} {\bf t}^{\beta}\right)\right]. \eqnn $$
The second term involves the product of four generators, but taking into account
the antisymmetry of the $\epsilon$ tensor, one product of two consecutive can be replaced by a commutator. Therefore the term is also proportional to $d_{\alpha\beta\gamma}$.\par
For a unitary representation the generators ${\bf t}^{\alpha}$ are, with our
conventions, antihermitian. Therefore the coefficients
$d_{\alpha\beta\gamma}$ are purely imaginary:
$$d_{\alpha\beta\gamma}^{*}= \ud\tr \left[{\bf t}^{\alpha}\left(
{\bf t}^{\beta}{\bf t}^{\gamma}+{\bf t}^{\gamma} {\bf t}^{\beta}\right)\right]^\dagger= -
d_{\alpha\beta\gamma}\, . \eqnn $$
For all real (the ${\bf t}^{\alpha}$ antisymmetric) or ``pseudo-real"
(${\bf t}^{\alpha}=-S\,\hskip2pt{}^{T}\hskip-2pt {\bf t}^{\alpha}S^{-1}$) representations
these coefficients 
vanish. It follows that the only non-abelian groups that can lead to
anomalies in four dimensions are:
$SU(N)$ for $N\geq 3$, $SO(6)$ and $E_6$. 
\subsection Wess--Zumino consistency conditions
 
In section \sssAnomR~we have calculated the part of the
anomaly that is quadratic in the
gauge field and asserted that the remaining part could be obtained from
geometric arguments. The method is based on BRS transformations. The anomaly is the
variation of a functional under an infinitesimal gauge transformation. 
This implies compatibility
conditions, which here are constraints on the form of the anomaly. \par
One
convenient way to express these constraints is to express the nilpotency of BRS
transformations. In a BRS transformation the variation of the gauge field ${\bf A}_{\mu}$ takes the form:\sslbl\ssAnoWZ
$$\delta_{\rm BRS} {\bf A}_{\mu}(x) = {\bf D}_{\mu}{\bf C}(x)\bar\varepsilon\,,
\eqnd\eBRSAmu $$
where ${\bf C}$ is a (fermion) ghost field and $\bar \varepsilon $ an anticommuting constant.   
The corresponding variation of $\ln {\cal Z} ( {\bf A}_ \mu  )$ is:
$$\delta_{\rm BRS} \ln{\cal Z} ( {\bf A}_ \mu )= -\int \d^4 x \,\left< {\bf
J}_{\mu}(x) \right>  {\bf D}_{\mu}{\bf C}(x)\bar\varepsilon\,.
\eqnd \eBRSlnZ  $$
The anomaly equation has the general form:
$$\left<{\bf D}_{\mu}{\bf J}_{\mu}(x)\right>= {\cal A}\left({\bf A}_{\mu};x
\right).\eqnn $$
In terms of $\cal A$ the equation \eBRSlnZ, after an integration by parts, can be rewritten:
$$\delta_{\rm BRS} \ln {\cal Z} ( {\bf A}_ \mu )= \int \d^4 x \, {\cal A}
\left({\bf A}_{\mu};x \right) {\bf C}(x)\bar\varepsilon\,.\eqnn $$
We then calculate the BRS variation of ${\cal A}{\bf C }$. We therefore need
also  the BRS transformation of the fermion ghost ${\bf C}(x)$: 
$$\delta_{\rm BRS} {\bf C}(x)= \bar\varepsilon\, {\bf C}^2(x).\eqnd\eBRSghost  $$
We express that the square of the BRS operator $\delta_{\rm BRS}$
vanishes (like a cohomology operator), and thus that ${\cal A} C$ is BRS invariant
$$\delta_{\rm BRS}^2=0\ \Rightarrow \ \delta_{\rm BRS}\int \d^4 x \, {\cal A}
\left({\bf A}_{\mu};x \right) {\bf C}(x)=0\,.$$
 This condition   yields   a constraint on  the possible form  of anomalies that   determines the term cubic
in ${\bf A}$ in the r.h.s.~of equation \egauamon\  completely. One can verify 
$$ \delta_{\rm BRS}\,\epsilon_{\mu\nu\rho\sigma}\int \d^4 x\,
  \tr\left[
{\bf C}(x)\partial_{\mu} \left( {\bf A}_{\nu} \partial_{\rho} {\bf A}_{\sigma}+ \ud {\bf
A}_{\nu}{\bf A}_{\rho}{\bf A}_{\sigma}\right) \right]=0\,.$$
Explicitly, after integration by parts, the equation takes the form
$$\eqalign{ &\epsilon_{\mu\nu\rho\sigma} \tr\int \d^4 x\,\left\{
 \partial_{\mu}{\bf C}^2(x)  {\bf A}_{\nu} \partial_{\rho} {\bf A}_{\sigma} +
\partial_{\mu}{\bf C}{\bf D}_{\nu}{\bf C}\partial_{\rho} {\bf A}_{\sigma}+\partial_{\mu}{\bf C} {\bf A}_{\nu} \partial_{\rho}{\bf D}_{\sigma}{\bf C} \right.\cr
&\left.
+\ud\partial_{\mu}{\bf C}^2(x)  {\bf
A}_{\nu}{\bf A}_{\rho}{\bf A}_{\sigma}
+\ud \partial_{\mu}{\bf C}\left(
{\bf D}_{\nu}{\bf C}{\bf A}_{\rho}{\bf A}_{\sigma}+
{\bf A}_{\nu}{\bf D}_{\rho}{\bf C}{\bf A}_{\sigma}+
{\bf A}_{\nu}{\bf A}_{\rho}{\bf D}_{\sigma}{\bf C}\right)
\right\}=0\,.\cr}$$
The terms linear   in $A$ cancels automatically 
$$ \epsilon_{\mu\nu\rho\sigma} \tr\int \d^4 x\left( \partial_{\mu}{\bf C} \partial _{\nu}{\bf C}\partial_{\rho} {\bf A}_{\sigma}+\partial_{\mu}{\bf C} {\bf A}_{\nu} \partial_{\rho} \partial _{\sigma}{\bf C}\right) = 0\,,$$
after integrating by parts the first term and using the antisymmetry of the $\epsilon $ symbol.\par
In the same way the cubic terms cancel (one has to remember the anticommuting properties of ${\bf C}$)
$$\eqalign{&\epsilon_{\mu\nu\rho\sigma} \tr\int \d^4 x\left\{\left(\partial_{\mu}{\bf C} {\bf C} +{\bf C} \partial_{\mu}{\bf C}\right){\bf
A}_{\nu}{\bf A}_{\rho}{\bf A}_{\sigma}
+\partial_{\mu}{\bf C}\left(
[ {\bf A}_{\nu},{\bf C}] {\bf C}{\bf A}_{\rho}{\bf A}_{\sigma}\right.\right. \cr&\left.\left.
\qquad+
{\bf A}_{\nu}[ {\bf A}_{ \rho },{\bf C}]  {\bf A}_{\sigma}+
{\bf A}_{\nu}{\bf A}_{\rho}[ {\bf A}_ \sigma ,{\bf C}]  \right)
\right\}=0\,.\cr}$$
It is only the quadratic terms that gives a relation between the quadratic and cubic terms in the anomaly, both contributions being proportional to
$$\epsilon_{\mu\nu\rho\sigma} \tr\int \d^4 x\,\partial _\mu {\bf C}\partial _\nu {\bf C}
{\bf A}_\rho {\bf A}_\sigma \,.$$
\section Lattice fermions: Ginsparg--Wilson relation

{\it Notation.} We now return to the problem of lattice fermions discussed in section \sslattfer. For convenience we   set the lattice spacing $a=1$ and
use for the fields the notation $\psi(x)\equiv \psi_x$.\sslbl\ssrotGW
\medskip
{\it Ginsparg--Wilson relation.} It had been noted, many years ago, that a potential way to avoid the doubling
problem while still retaining chiral properties in the continuum limit was to look for  lattice Dirac operators $\bf D$ that, instead of anticommuting with $\gamma _5$, would satisfy  the relation
$${\bf D}^{-1} \gamma_5+  \gamma_5 {\bf D}^{-1}  =\gamma_5{\bf 1}\,\eqnd\eGinWil $$ 
where $ \bf 1$ stands for the identity  both  for lattice sites and in the algebra of $\gamma $-matrices. More explicitly
$$({\bf D}^{-1})_{xy}\gamma_5+\gamma_5({\bf D}^{-1})_{xy}=\gamma_5\delta_{xy}\,.$$
More generally the r.h.s.~can be replaced by any local positive operator on the lattice. Locality on the lattice is defined by a decrease that is at least exponential when the point $x,y $ are separated.
The anti-commutator being local, it  is expected that it does not affect
correlation functions at large distance  and that chiral properties are recovered in the continuum limit. Note that when $\bf D$ is the Dirac operator in a gauge background the condition \eGinWil\ is gauge invariant.\par
However, only recently have lattice Dirac operators solutions to the
Ginsparg--Wilson relation \eGinWil~been discovered, because the demands that
both ${\bf D}$ and the anticommutator  $\{{\bf D}^{-1},\gamma_5\}$ should be
local, seemed difficult to satisfy, specially in the most interesting 
case of gauge theories. \par  
Note that while the relation \eGinWil\ implies some generalized form of chirality on the lattice, as we now show, it does not guarantee the absence of doublers, as examples illustrate. But the important point is that in this class solutions can be found without doublers.
\subsection Chiral symmetry and index

We first discuss the main properties of a Dirac operator satisfying relation \eGinWil, and exhibit a generalized form of chiral transformations on the lattice. \par
Using the relation, quite generally
true for an euclidean Dirac operator (consequence of hermiticity and reflection symmetry), 
$${\bf D}^\dagger =\gamma_5 {\bf D}\gamma_5\,, \eqnd\eGivsym $$
one can  rewrite the relation \eGinWil, after multiplication by $\gamma _5$,
$${\bf D}^{-1}+\left({\bf D}^{-1}\right)^\dagger ={\bf 1}\,, $$
and therefore
$${\bf D}+{\bf D}^\dagger={\bf D}{\bf D}^\dagger= {\bf D}^\dagger{\bf D}\,.\eqnd\eDirnorm $$
This implies that the lattice operator $\bf D$ has an index, and in addition
$${\bf S}={\bf 1}-{\bf D}\,, \eqnd\erotlatun $$
is unitary 
$${\bf S}{\bf S}^\dagger ={\bf 1}\,. \eqnn $$
The eigenvalues of $\bf S$ lie on the unit circle. The eigenvalue one
corresponds to the pole of the Dirac propagator.\par
Note also the relations
$$\gamma _5{\bf S}={\bf S}^\dagger \gamma _5\,,\quad (\gamma _5{\bf S})^2={\bf 1}\,. \eqnd\erotSgamv $$
The matrix $\gamma _5  {\bf S} $ is  hermitian and $\ud({\bf 1}\pm \gamma _5  {\bf S} $) are two orthogonal projectors. If ${\bf D}$ is a Dirac operator in a gauge background these projectors depend on the gauge field.
\par
It is then possible to construct lattice actions that have a chiral symmetry
that corresponds to local but non point-like transformations. In the abelian
example, 
$$\eqalignno{ \psi'_x&= \sum_y\left( \e^{i\theta\gamma_5 {\bf S}} \right)_{xy} \psi_y\,, &\eqnn \cr
 \bar \psi'_x&= \bar \psi_x  \e^{i\theta \gamma_5} \,. &\eqnn \cr} $$
Indeed, 
$$\bar \psi'_x{\bf D}  \psi'_x= \bar \psi _x{\bf D}  \psi _x \ \Leftrightarrow 
\ \e^{i\theta \gamma_5}{\bf D}\e^{i\theta\gamma_5 {\bf S}}={\bf D}
\ \Leftrightarrow 
\ {\bf D}\e^{i\theta\gamma_5 {\bf S}}=\e^{-i\theta \gamma_5} {\bf D}\,.$$
Using the second relation \erotSgamv\ we expand the exponentials
and reduce the equation to
$${\bf D}\gamma _5{\bf S}=-\gamma _5{\bf D}\,,\eqnd\eGinWilii $$
which is another form of relation \eGinWil.\par
These transformations, however, no longer leave the integration
measure over the fermion fields, 
$$\prod_x \d\psi_x \d\bar\psi_x\,,$$
automatically invariant. The jacobian $J$ of the change of variables $\psi \mapsto \psi'$ is
$$J=\det \e^{i\theta \gamma_5}\e^{i\theta\gamma_5 {\bf S}} =1+2i\theta \tr \gamma_5  ({\bf 1}-{\bf D}/2)+O(\theta ^2),\eqnd\erotGWjac $$
where trace means trace in the space of $\gamma $ matrices and in the lattice indices.
This leaves open the possibility of generating the expected anomalies, when the
Dirac operator of the free theory is replaced by the covariant
operator in the background of a gauge field, as we now show. 
\smallskip
{\it Eigenvalues of Dirac operator in a gauge background.} We briefly discuss the index of a  lattice Dirac operator ${\bf D}$  satisfying the relation \eGinWil, in a gauge background.
We assume that its spectrum is discrete (this is certainly true on a finite lattice where ${\bf D}$ is a matrix). The operator ${\bf D}$ is related by \erotlatun\ to a unitary operator
${\bf S}$ whose eigenvalues have modulus one. Therefore if we denote by $ |n\rangle$ its $n^{\rm th}$
eigenvector, 
$${\bf D}\left|n\right>=(1-{\bf S})\left|n\right>= (1-\e^{i\theta _n})\left|n\right>\ \Rightarrow {\bf D}^\dagger\left|n\right>=(1-\e^{-i\theta _n})\left|n\right>\,.$$
Then using equation \eGivsym, we infer
$${\bf D}\gamma_5\left|n\right>=(1-\e^{-i\theta _n})\gamma_5\left|n\right>. $$
The discussion that follows then is analogous  to  the discussion of Section \sssSManf~to which we refer for details.
We note that when the eigenvalues are not real, $\theta _n\ne 0\pmod\pi$, 
$\gamma_5\left|n\right>$ is an eigenvector different from $\left|n\right>$ because the eigenvalues are different. Instead in the two subspaces corresponding to the eigenvalues $0$ and $2$, we can choose  eigenvectors with definite chirality
$$\gamma_5\left|n\right>=\pm \left|n\right>.$$
We call below $n_\pm$ the number of  eigenvalues $0$,  and $\nu _\pm$  the number of eigenvalues $2$ with chirality $\pm1$. \par
Note that on a finite lattice  $\delta _{xy} $ is a finite matrix, and thus 
$$\tr\gamma _5\delta _{xy}=0\,.$$
Therefore
$$ \tr\gamma_5(2-{\bf D})=-\tr\gamma _5{\bf D}\,,$$
which implies
$$\sum_n\left<n\right|\gamma_5(2-{\bf D} )\left|n\right>= -\sum_n\left<n\right|\gamma_5 {\bf D} \left|n\right>.\eqnn $$
In the equation all complex eigenvalues cancel because the vector
$\left|n\right>$ and $\gamma_5\left|n\right>$ are orthogonal. The sum reduces to the subspace of real eigenvalues, where the eigenvectors have definite chirality.
In the l.h.s.~only the eigenvalue $0$ contributes, and in the r.h.s.~only the eigenvalue $2$.   We find
$$ n_+-n_- =- (\nu _+-\nu _-).$$
The equation tells us that the  difference between the number of states of different chirality in the zero eigenvalue sector is cancelled by the difference in the massive sector of  eigenvalue two.
\smallskip
{\it Remark.} It is interesting to note the relation between the spectrum of ${\bf D}$ and $\gamma _5 {\bf D}$, which from relation \eGivsym~is a hermitian matrix,
$$\gamma _5 {\bf D}={\bf D}^\dagger\gamma _5=(\gamma _5 {\bf D})^\dagger, $$
and thus is diagonalizable with real eigenvalues. It is easy to verify the two equations, the second being obtained by changing $\theta $ in $\theta +2\pi$,
$$\eqalign{\gamma _5 {\bf D} (1-i\e^{i\theta _n/2}\gamma _5) \left|n\right>&=2\sin(\theta _n/2)
(1-i\e^{i\theta _n/2}\gamma _5) \left|n\right>, \cr
\gamma _5 {\bf D} (1+i\e^{i\theta _n/2}\gamma _5) \left|n\right>&=-2\sin(\theta _n/2)
(1+i\e^{i\theta _n/2}\gamma _5) \left|n\right>, \cr}
$$
which implies that the eigenvalues are paired $\pm \sin(\theta _n/2)$ except for $ \theta _n=0$  (mod $\pi$) where $\left|n\right>$ and $\gamma _5 \left|n\right>$
are proportional. For $\theta _n=0$ $\gamma _5 {\bf D}$ has also eigenvalue $0$.
For $ \theta _n=\pi$ $\gamma _5 {\bf D}$ has eigenvalue $\pm2$ depending on the
chirality of  $ |n\rangle$.
 \par
In the same way
$$\eqalign{\gamma _5(2- {\bf D}) (1+\e^{i\theta _n/2}\gamma _5) \left|n\right>&=2\cos(\theta _n/2)
(1+\e^{i\theta _n/2}\gamma _5) \left|n\right>, \cr
\gamma _5(2- {\bf D}) (1-\e^{i\theta _n/2}\gamma _5) \left|n\right>&=-2\cos(\theta _n/2)
(1-\e^{i\theta _n/2}\gamma _5) \left|n\right>. \cr}
$$
\smallskip
{\it Jacobian and lattice anomaly.}
The variation of the jacobian \erotGWjac\ can now be evaluated. Opposite eigenvalues
of $  \gamma _5(2- {\bf D}) $ cancel. The eigenvalues $\theta _n=\pi$ give factors one.
Only $\theta _n=0$ gives a non-trivial contribution 
$$J=\det \e^{i\theta\gamma_5 (2-{\bf D})}=\e^{2i\theta (n_+-n_-)}. $$
The quantity $\tr\gamma_5(2-{\bf D})$, coefficient of the term of order $\theta $,  is a sum of terms that are local, gauge invariant, pseudoscalar and topological as the continuum anomaly \enabanom\ since
$$ \tr\gamma_5(2-{\bf D})= \sum_n\left<n\right|\gamma_5(2-{\bf D} )\left|n\right>  =2(n_+ - n_-) .$$
\smallskip
{\it Non-abelian generalization.} We now consider  the non-abelian chiral transformations
$$\eqalignno{ \psib _U &= \left[ \ud ( {\bf 1} + 
\gamma_5 {\bf S}) {\bf U}_+ +\ud ( {\bf 1} - 
\gamma_5{\bf S} ) {\bf U}_- \right]\psib\, ,&
\eqnd \elattchirtra \cr
\bar{ \psib}_U &= \bar{ \psib} \left[\ud (
{\bf 1} +   \gamma_5)  {\bf U}^{\dagger}_-  +\ud (
{\bf 1} -  \gamma_5 ) {\bf U}^{\dagger}_+ \right],
&\eqnd \elattchirtrb \cr} $$ 
where $ {\bf U}_\pm $ are matrices belonging  to some unitary group $G$
$${\bf U}={\bf 1}+\Theta +O(\Theta ^2).$$
We note that this amounts to define differently chiral components of
$\bar\psi $ and $\psi $, for $\psi $ the definition being even gauge field dependent.\par
We assume that $G$ is a vector symmetry of the fermion action,
and thus the Dirac operator commutes with the elements $\Theta$ of the 
Lie algebra
$$[{\bf D},\Theta ]=0\,.$$
Then again the relation \eGinWil\ in the form \eGinWilii\  
implies the invariance of the fermion action
$$ \bar{ \psib}_U\, {\bf D}\,{ \psib}_U= \bar{ \psib}\,  {\bf D}\,  \psib\,.$$

The jacobian of an infinitesimal chiral transformation $\Theta=\Theta _+=-\Theta _-$
is
$$J=1+\tr\gamma _5\Theta ({\bf 2}-{\bf D})+O(\Theta ^2).$$
\smallskip
{\it  Wess--Zumino consistency conditions.} To determine anomalies in the case of gauge fields coupling differently to fermion chiral components one can on the lattice also play with the nilpotency of BRS transformations, which then take the form
$$\eqalign{\delta {\bf U}_{xy}&=\bar\varepsilon \left({\bf C}_x {\bf U}_{xy}- {\bf U}_{xy}{\bf C}_y\right), \cr
\delta{\bf  C}_x &=\bar\varepsilon {\bf C}^2_x\,, \cr}$$
instead of \eqns{\eBRSAmu,\eBRSghost}. Moreover the matrix elements  ${\bf D}_{xy}$ of the gauge covariant Dirac operator  transform  like ${\bf U}_{xy}$.
\subsection Explicit construction: Overlap fermions

An explicit solution to the Ginsparg--Wilson relation without doublers can be derived from operators ${\bf D}_{\rm W}$ that share the properties of the Wilson--Dirac operator
  of equation \eWferDop, i.e.~which  avoid doublers at the price of breaking chiral symmetry explicitly. Setting
$${\bf A}={\bf 1}-{\bf D}_{\rm W}/M\,,$$
where $M>0$ is a mass parameter, one takes (the idea of overlap fermions)
$${\bf S}={\bf A}\left({\bf A}^\dagger{\bf A}\right)^{-1/2}\ \Rightarrow\
{\bf D}={\bf 1}-{\bf A}\left({\bf A}^\dagger{\bf A}\right)^{-1/2}. \eqnn $$
With this ansatz $\bf D$ has a zero eigenmode when 
${\bf A}\left({\bf A}^\dagger{\bf A}\right)^{-1/2}$ has the eigenvalue one.
This can happen when ${\bf A}$ and ${\bf A}^\dagger$ have the same eigenvector
with a {\it positive}\/ eigenvalue. 
\smallskip
{\it Free fermions.} We now verify the absence of doublers in the absence of gauge fields.
The Fourier representation has the general form
$$D_W(p)=\alpha (p)+i\gamma _\mu \beta _\mu(p), \eqnd\elatWDgen  $$
where $\alpha (p)$ and $ \beta _\mu(p)$ are real, periodic, smooth functions such that
$$ \beta _\mu(p)\mathop{\sim}_{|p|\to 0} p_\mu\,,\quad \alpha (p)\ge 0\mathop{\propto}_{|p|\to 0} p^2, $$
and $ \alpha (p) >0$ for all values of $p_\mu$ that correspond to doublers,
i.e.~such that $\beta _\mu(p)=0$ for $|p|\ne 0$.\par 
In the case of the  operator \elatWDgen~a short calculation shows that the determinant of $\bf D$ vanishes when
$$\left[\sqrt{\bigl(M-\alpha (p)\bigr)^2+\beta ^2_\mu(p)}-M+\alpha (p)\right]^2
+\beta ^2_\mu(p)=0\,.$$
This implies $\beta  _\mu(p) =0$, an equation that necessarily admits doubler solutions, and 
$$|M-\alpha (p)|=M-\alpha (p). $$
The solutions to this equation depend on the value of $\alpha (p)$ with respect to $M$ for the doubler modes, i.e.\ for the values of $p$ such that $\beta _\mu(p)=0$. If $\alpha (p)\le M$ the equation is automatically satisfied 
and the corresponding doubler survives. As mentioned in the introduction, the relation \eGinWil\ alone does not  guarantee the absence of doublers.
If instead $\alpha (p)>M$ the equation implies $\alpha (p)=M$, which is impossible. Therefore by rescaling $\alpha (p)$, if necessary,   we can keep  the wanted $p_\mu=0$ mode, while eliminating all doublers, which then correspond to the eigenvalue two for $\bf D$, and the doubling problem is  solved, at least in a free theory.\par
In presence of a gauge field the construction can be generalized and works provided the plaquette action is sufficiently close to one.
\smallskip
{\it Remark}. 
Let us stress that, if it seems that the doubling problem has been solved from the formal point of view, from the numerical point of view the calculation of the operator
$ ({\bf A}^\dagger{\bf A} )^{-1/2}$ in a gauge background represents a major challenge.
\section Supersymmetric quantum mechanics and domain wall fermions

Because the construction of  lattice fermions without doublers we have just described is somewhat artificial, one may wonder whether there is a context in which they  would appear more naturally. Therefore we now briefly outline how
a similar lattice Dirac operator can be generated by  embedding first four-dimensional space in a larger five dimensional space. This is the method of domain wall fermions.
\par 
Because the general idea behind domain wall fermions has emerged first in another context, as a preparation, we first 
recall a few  properties of the spectrum of the  hamiltonian in supersymmetric quantum mechanics,
a topic also related to the index of the Dirac operator (section \sssSManf), and
 stochastic dynamics or Fokker--Planck equation.
\subsection Supersymmetric quantum mechanics

We consider  a first order differential operator  $\goth D $ that is a $2\times2$ matrix ($\sigma _i$ still are the Pauli matrices):\sslbl\ssSUSYQM
$$ {\goth D} \equiv  \sigma _1 \d_x - i\sigma _2 A(x) .\eqnd\eSUSYDx $$
The function $A(x)$ is real, and thus the operator ${\goth D}$ is anti-hermitian. 
\par
The operator $\goth D$ shares several properties of the Dirac operator of section
\sssSManf. In particular it satisfies
$$\sigma _3 {\goth D} +{\goth D} \sigma _3=0\, ,$$
and thus has an index.
We introduce the operator ${\rm D}$
$${\rm D}= \d_x +A(x)\ \Rightarrow \ {\rm D}^\dagger= -\d_x +A (x), \eqnn $$
and
$$Q={\rm D}  \pmatrix{ 0 & 0\cr 1 & 0 \cr} \Rightarrow  Q^\dagger
={\rm D}^\dagger \pmatrix{ 0 & 1\cr 0 & 0 \cr}.$$
Then
$${\goth D}=Q-Q^\dagger\,,\quad Q^2=(Q^\dagger)^2=0\,.$$
Thus $\{Q,Q^\dagger\}$ are generators of the simplest form of supersymmetry, and the
supersymmetric hamiltonian $H$ is given by
$$H=QQ^\dagger +Q^\dagger Q=-{\goth D}^2=\pmatrix{{\rm D}^\dagger{\rm D}  &0 \cr 0
 & {\rm D} {\rm D}^\dagger} . \eqnn $$
We note that $H$ is a positive operator. The  eigenvectors  of $H$ have the form $ \psi_+(x)(1,0)$
and $\psi_-(x) (0,1)$ and 
satisfy
$${\rm D}^\dagger{\rm D} \left|\psi_+\right>  =\varepsilon_+ \left|\psi_+\right> , \quad {\rm or}\quad {\rm D} {\rm D}^\dagger \left|\psi_- \right>=\varepsilon_- \left| \psi_- \right>, \quad \varepsilon_\pm\ge 0\,,\eqnd\eSUSYQMspec $$
where
$$ {\rm D}^\dagger {\rm D}=-\d_x^2+A ^2(x)-A'(x) ,\quad       {\rm D}{\rm D}^\dagger=-\d_x^2+A ^2(x)+A'(x).$$
Moreover if  $x$ belongs to a bounded interval or
$A (x)\to \infty $ for $|x|\to \infty $ then the spectrum of $H$ is discrete.\par
Multiplying the first equation \eSUSYQMspec\ by $\rm D$, we conclude that if ${\rm D} \left|\psi_+\right>\ne 0$, and thus $\epsilon _+$ does not vanish, it is an eigenvector of $ {\rm D}{\rm D}^\dagger$ with eigenvalue $\varepsilon _+$, and conversely. Therefore except for a possible ground state with vanishing eigenvalue, the spectrum of $H$ is doubly degenerate.
\par This observation  is consistent with the analysis of section \sssSManf\ applied to 
  the operator $\goth D$. We know from that either eigenvectors are paired
$ | \psi  \rangle,  \sigma _3| \psi  \rangle$ with opposite eigenvalues
$\pm i\sqrt{\varepsilon}$, or they correspond to the eigenvalue zero and can be chosen with  definite chirality
$${\goth D}\left| \psi  \right>=0   \,,\quad   \sigma _3\left| \psi  \right>= \pm \left| \psi   \right>.$$
It is convenient to now  introduce the function $S(x)$:
$$S'(x)=A(x),\eqnn $$
and for simplicity discuss only the situation of operators on the entire real line.\par
We   assume that 
$$ S(x)/|x|\mathop{\ge}_{ x \to \pm \infty }\ell>0\,.$$
Then the function   $S(x)$  is such that  $\e^{-S(x)}$ is a normalizable wave function: $\int\d x\, \e^{-2S(x)}<\infty $. 
In the stochastic interpretation $\e^{-2S(x)}$ is the equilibrium distribution.

When $\e^{-S(x)}$ is  normalizable we know one  eigenvector with vanishing eigenvalue and chirality $+ 1$,
which corresponds to the isolated ground state of ${\rm D}^\dagger{\rm D}$
 $${\goth D}\left|  \psi_+,0 \right>=0 \ \Leftrightarrow\ {\rm D}\left| \psi_+  \right>=0 \,,\quad   \sigma _3\left|\psi_ +, 0 \right>=  \left| \psi_+,0  \right>,$$
with
$$\psi_+(x)=\e^{ -S(x)}.$$
On the other hand the formal solution of ${\rm D}^\dagger | \psi_-   \rangle=0$,
$$\psi_-(x)=\e^{ S(x)}$$
is not normalizable, and therefore no eigenvector with negative chirality is found.
\par
We conclude that the operator ${\goth D}$ has only one eigenvector with zero eigenvalue corresponding to positive chirality: the index of  ${\goth D}$ is one. Note that  expressions for the index of the Dirac operator in a general background have been derived. In the present example they yield
$${\rm Index}= \ud\left[ \sgn A(+\infty)  - \sgn  A(-\infty)\right] ,$$
in agreement with the explicit calculation.
\smallskip
{\it The resolvent.} For later purpose it is useful to exhibit some properties of the resolvent
${\goth G}$:
$${\goth G}=\left({\goth D}-k\right)^{-1},$$
for real values of the parameter $k$.
Parametrizing ${\goth G}$ as a $2\times2$ matrix
$${\goth G}=\pmatrix{G_{11} & G_{12} \cr G_{21} & G_{22} \cr},$$
one obtains
$$\eqalign{ G_{11}=-k\left({\bf D}^\dagger{\bf D}+k^2\right)^{-1} &\quad G_{21} =-{\rm D}\left({\bf D}^\dagger{\bf D}+k^2\right)^{-1}
 \cr G_{22}=-k\left({\bf D}{\bf D}^\dagger+k^2\right)^{-1} &\quad  G_{12}= {\bf D}^\dagger
\left({\bf D}{\bf D}^\dagger+k^2\right)^{-1}.\cr}$$
For $k^2$ real one verifies $G_{21}=-G_{12}^\dagger$. \par
A number of  properties then follow directly from the analysis of \label{\sshamresolv}. 
\par
When $k\to 0$ only $G_{11}$ has a pole,
$G_{11}=O(1/k)$, $G_{22}$ vanishes as $k$ and $G_{12}(x,y)=-G_{21}(y,x)$ have finite limits:
$${\goth G}(x,y)\mathop{\sim}_{k\to 0}\pmatrix {-{1\over k}\psi _+(x)\psi _+(y)/\|\psi _+\|^2 &
-G_{21}(y,x) \cr G_{12}(x,y) & 0\,\cr}\sim -{1\over 2k}{\psi _+(x)\psi _+(y)\over\|\psi _+\|^2}(1+\sigma _3) . $$
Another limit of interest is the limit $y\to x$. The non-diagonal elements are discontinuous
but the limit of interest for domain wall fermions is the average of the two limits
$$\overline{{\goth G}(x,x)}=\ud(1+\sigma _3)G_{11}(x,x)+\ud(1-\sigma _3)G_{22}(x,x)
+i \sigma _2\overline{ G_{12}(x,x)}\,.$$
When the function $A(x)$ is odd, $A(-x)=-A(x)$, in the limit $x=0$ the matrix $\overline{{\goth G}(x,x)}$ reduces to
$$\overline{{\goth G}(0,0)}=\ud(1+\sigma _3)G_{11}(0,0)+\ud(1-\sigma _3)G_{22}(0,0).$$

\smallskip
{\it Examples.} \par
(i) In the example of the function $S(x)=\ud x^2$, the two components of the hamiltonian $H$ become
$${\rm D}{\rm D}^\dagger=-\d_x^2+ x^2+ 1\,,\quad {\rm D}^\dagger {\rm D}=-\d_x^2+x^2-1\,.$$
We recognize two shifted harmonic oscillators and the spectrum of $\goth D$
contains one eigenvalue zero, and a spectrum of opposite eigenvalues $\pm i\sqrt{2n}$, $n\ge 1$.\par
(ii) Another example useful for later purpose is
$S(x)= |x|$. Then $A(x)= \epsilon (x)$ ($ \epsilon (x)$ is the sign function), and
$A'(x)=2\delta (x)$. The two components of the hamiltonian $H$ become
$${\rm D}{\rm D}^\dagger=-\d_x^2+   1 +2\delta (x) ,\quad {\rm D}^\dagger {\rm D}=-\d_x^2+1-2 \delta (x).
\eqnd\eSUSYdelx $$
Here one finds one isolated eigenvalue zero, and  a continuous spectrum $\varepsilon \ge 1$.
\par
(iii) A less singular but similar example that can be solved analytically corresponds to
$A(x)=\mu\tanh(x)$, where $\mu$ is for instance a positive constant. It leads to the potentials
$$V(x)=A^2(x)\pm A'(x)=\mu ^2-{\mu (\mu \mp1)\over \cosh^2(x)}.$$
The two operators have a continuous spectrum starting at $\mu ^2$ and a discrete spectrum
$$\mu ^2-(\mu -n)^2, \quad n\in{\Bbb N}\le \mu\,,\qquad \mu ^2-(\mu -n-1)^2, \quad n\in{\Bbb N}\le \mu-1\,.$$
\subsection Field theory in two dimensions 

A natural  realization in quantum field theory of such a situation corresponds to a two-dimension model of a Dirac fermion
in the background of a static soliton (finite energy solution of field equations).\par
We consider the action ${\cal S}(\bar \psi ,\psi,\varphi )$, $\psi,\bar\psi$ being Dirac fermions, and $\varphi$ a scalar boson:
$${\cal S}(\bar \psi ,\psi,\varphi)=\int\d  x\,\d t \left[-\bar \psi\left(\sla{\partial }+m+M \varphi\right)\psi+\ud \left(\partial _\mu \varphi\right)^2+V(\varphi)\right].$$
We assume that $V(\varphi)$ has degenerate minima, like $(\varphi^2-1)^2$
or $\cos\varphi$ and field equations therefore admit  soliton solutions $\varphi(x)$,  static solitons being the instantons of the one-dimension quantum $\varphi$ model. \par
Let us now study the spectrum of the corresponding   Dirac operator  
$${\cal D}= \sigma  _1 \partial _x+  \sigma  _2 \partial _t+m + M\varphi(x). $$
We assume for  definiteness that $\varphi(x)$ go from $-1$ for $x=-\infty $ to $+1$ for $x=+\infty $,  a typical example being
$$ \varphi(x)= \tanh (x).$$
Since time translation symmetry remains, we can introduce the (euclidean) time
Fourier components and study
$${\cal D}= \sigma  _1 \d _x+ i\omega  \sigma  _2 +m + M\varphi(x). $$
The zero eigenmodes of $\cal D$ are  also the  solutions of the eigenvalue equation
$${\goth D}\left|\psi \right>=\omega \left|\psi \right>, \quad
{\goth D}=\omega+i \sigma _2{\cal D}= \sigma _3 \d_x+ i\sigma _2\bigl( m + M\varphi(x)
\bigr), $$
which differs from equation \eSUSYDx\ by an exchange between the matrices $\sigma _3$ and $\sigma _1$. The possible zero eigenmodes of $\goth D$ ($\omega =0$) thus satisfy
$$\sigma _1\left|\psi \right>=\epsilon \left|\psi \right>, \quad \epsilon=\pm 1\,,$$
and therefore are proportional to $\psi _\epsilon (x)$ solution of
$$\epsilon \psi _\epsilon'+ \bigl( m + M\varphi(x)
\bigr)\psi _\epsilon=0\,.$$
This equation has a normalizable solution only if $|m|<|M|$ and $\epsilon =+1$.
Then we find  one  fermion zero-mode. \par
A soliton solution breaks space translation symmetry, but leads to a zero-mode
would generate IR divergences. The zero-mode is removed by taking the position of the soliton as a collective coordinate and translation symmetry then is restored.
The implications of the fermion zero-mode then require further analysis.
It is found that it is associated with a double degeneracy of the soliton state, which carries $1/2$ fermion number.
\subsection Domain wall fermions

{\it Continuum formulation.} One now considers four-dimensional space
(but the strategy applies to all even dimensional spaces) as a surface embedded
in five-dimensional space. We denote by $x_\mu$ the usual four coordinates,
and $t$ the  coordinate in the fifth dimension. Physical space corresponds to 
$t=0$. We then study the five-dimensional Dirac operator $\cal D$ in the background of a classical scalar field $\varphi (t)$ that depends only on $t$. The fermion action reads
$${\cal S}(\bar\psi ,\psi )=-\int\d t\,\d^4 x\,\bar\psi(t,x) {\cal D}\psi(t,x),$$
with
$${\cal D}=\sla{\partial }+\gamma _5  \d_t +M \varphi (Mt).$$
Since translation symmetry in four-space is not broken, we can work in Fourier representation
$${\cal D}=ip_\mu \gamma _\mu +\gamma _5 \d_t +M \varphi (Mt),$$
where the parameter $M$ is a mass large with respect to the mass of all
physical particles.\par
To find the mass spectrum corresponding to $\cal D$, it is convenient to write it
$$ {\cal D}=\gamma _{\bf p}\left[i|{\bf p}| +\gamma _{\bf p}\gamma _5 \d_t + \gamma _{\bf p}M\varphi (Mt)\right],$$
where $\gamma _{\bf p}=p_\mu \gamma _\mu/|{\bf p}|$ and thus  $\gamma _{\bf p}^2=1$.
The eigenvectors with vanishing eigenvalue of ${\cal D}$ are also those
of the operator $\goth D$ 
$${\goth D}=i\gamma _{\bf p}{\cal D}+|{\bf p}|=i\gamma _{\bf p}\gamma _5 \d_t +i \gamma _{\bf p}M\varphi (Mt),$$
with eigenvalue $|{\bf p}|$. \par
We then note that $ i\gamma _{\bf p}\gamma _5$, $\gamma _{\bf p}$ and $-\gamma _5$ are hermitian matrices that form a representation of the algebra of Pauli matrices. The operator $\goth D$ can then be compared with the
operator \eSUSYDx, and $M\varphi (Mt)$ corresponds to $A(x)$. Under the same conditions $\goth D$ has an eigenvector with an isolated vanishing eigenvalue
corresponding to an eigenvector with positive chirality. All other eigenvalues,
for dimensional reasons are proportional to $M$ and thus correspond to fermions of large masses. Moreover
the eigenfunction with eigenvalue zero decays on a scale $t=O(1/M)$.
Therefore for $M$ large one is left with a fermion that has a single chiral component, 
confined on the $t=0$ surface.  \par
One possible  interpretation of the function $\varphi(t)$  is that $\varphi(t)$ is a  solution of  field 
equations and connects two minima  $\varphi=\pm 1$ of the $\varphi$ potential.
In the limit of very sharp transition one is led to the hamiltonian \eSUSYdelx.
Note that this interpretation is possible only for an even dimension $d\ge 4$; in dimension two like in the two-dimensional field theory, zero-modes forbid a static wall. \par
More precise results follow from the study of section \ssSUSYQM.
We have noticed that   ${\goth G}(t_1,t_2;p)$, the inverse of the Dirac operator in  Fourier representation, has a short distance singularity for $t_2\to t_1$ in the form of a discontinuity. This is here an artifact of treating the fifth dimension differently from the four others. In real space for the function ${\goth G}(t_1,t_2;x_1-x_2)$ with separate points on the surface, $x_1\ne x_2$, the limit $t_1=t_2$ corresponds to points in five dimensions that do not coincide and  this singularity is absent. A short analysis shows that this amounts in   Fourier representation to  take the average of the limiting values (a property that can easily be verified for the free propagator). 
Then if $\varphi(t)$ is an odd function one finds for $t_1=t_2=0$
$${\cal D}^{-1}(p)={i\over 2\sla{p }}\left[d_1(p^2)(1+\gamma _5)+(1-\gamma _5) p^2d_2(p^2)\right],$$
where $d_1,d_2$ are regular functions of $p^2$. Therefore ${\cal D}^{-1}$ anticommutes with 
$\gamma _5$ and chiral symmetry is realized in the usual way. If however $\varphi(t)$ is of more general type one finds
$${\cal D}^{-1}={i\over 2\sla{p }}\left[d_1(p^2)(1+\gamma _5)+(1-\gamma _5) p^2d_2(p^2)\right]+d_3(p^2),$$
where $d_3$ is   regular.
As a consequence
$$\gamma _5{\cal D}^{-1}+{\cal D}^{-1}\gamma _5=2d_3(p^2)\gamma _5\,,$$
which is a form of Wilson--Ginsparg's relation because the r.h.s.~is local.
\smallskip
{\it Domain wall fermions: lattice.} We now replace four-dimensional continuum space by a lattice but  keep the fifth dimension continuous.
We replace the Dirac operator by the Wilson--Dirac operator \elatWDgen\ to avoid doublers. In Fourier representation we  find
$${\cal D}= \alpha ({\bf p})+i\beta _\mu({\bf p}) \gamma _\mu +\gamma _5 \d_t +M \varphi (Mt).$$
This has the effect of replacing $p_\mu$ by $\beta _\mu({\bf p})$ and shifting
$M \varphi (Mt)\mapsto M \varphi (Mt)+   \alpha ({\bf p})$.
To ensure the absence of doublers we require that  for the values
for which $\beta _\mu({\bf p})=0$ and ${\bf p}\ne 0$ none of the solutions
to the zero eigenvalue equation is normalizable. This is realizes if 
for $|t|\to\infty $  $\varphi(t)$ is bounded, for instance
$$ |\varphi(t)|\le 1\,$$
and $M<|\alpha ({\bf p})|$. \par
The inverse Dirac operator on the surface $t=0$  takes the general form
$${\cal D}^{-1}=i\sla{\beta }\left[\delta _1(p^2)(1+\gamma _5)+(1-\gamma _5) \delta _2(p^2)\right]+\delta _3(p^2),$$
where $\delta _1$ is the only function that has a pole for $p=0$, and where $\delta _2,\delta _3$ are regular.
The function $d_3$ does not vanish even if $\varphi(t) $ is odd because the addition of $\alpha (p^2)$ breaks the symmetry. We then always find Wilson--Ginsparg's relation
$$\gamma _5{\cal D}^{-1}+{\cal D}^{-1}\gamma _5=2\delta _3(p^2)\gamma _5\,$$
More explicit expressions can be obtained in the limit $\varphi(t)=\sgn(t)$
(a situation analogous to \eSUSYdelx), using the analysis of  section \label{\sshamresolv}.
\par
Of course to simulate domain walls on the computer one has also to discretize the fifth dimension. 
\medskip
{\bf Acknowledgments.} Useful discussions with T.W.~Chiu, P.~Hasenfratz and H.~Neuberger are gratefully acknowledged. 
\beginbib

These lectures are an expansion of several sections of \rf
J.~Zinn-Justin, {\it Quantum Field Theory and Critical Phenomena},
Clarendon Press (Oxford 1989, fourth ed.~2002),
\nrf to which the reader is referred for background in particular about    euclidean field theory and general gauge theories,  and references.
\par
For an early reference on momentum cut-off regularization see\rf
W. Pauli and F. Villars, {\it Rev. Mod. Phys.} 21 (1949) 434.
\nrf
Renormalizability of gauge theories has been proven using momentum regularization in\rf
B.W. Lee and J. Zinn-Justin, {\it Phys. Rev.} D5 (1972) 3121, 3137, 3155; D7
 (1973) 1049.
\nrf The proof has been generalized using BRS symmetry and the master equation in\rf
J. Zinn-Justin in {\it Trends in Elementary Particle Physics (Lectures Notes
in Physics} 37),  H. Rollnik and K. Dietz eds. (Springer-Verlag,
Berlin 1975);  in {\it Proc. of the 12th School of Theoretical Physics,
Karpacz 1975}, Acta Universitatis   Wratislaviensis 368.
\nrf A short summary can be found in \rf
J. Zinn-Justin {\it Mod. Phys.Lett.} A19 (1999) 1227.
\nrf Dimensional regularization has been introduced by:\rf
J. Ashmore, {\it Lett. Nuovo Cimento} 4 (1972) 289; 
G.~'t~Hooft and M.~Veltman, {\it Nucl. Phys.} B44 (1972) 189;
C.G.~Bollini and J.J.~Giambiaggi, {\it Phys. Lett.} 40B (1972) 566, {\it Nuovo
Cimento} 12B (1972).
\nrf See also \rf
E.R. Speer, {\it J. Math. Phys.} 15 (1974) 1; M.C. Berg\`ere and F. David,   {\it J. Math. Phys.} 20 (19779 1244.
\nrf Its use in problems with chiral anomalies has been proposed in\rf
D.A. Akyeampong and R. Delbourgo, {\it Nuovo Cimento} 17A (1973) 578.
\nrf For an early review see\rf
G. Leibbrandt, {\it Rev. Mod. Phys.} 47 (1975) 849.
\nrf For dimensional regularization and other schemes see also\rf
E.R. Speer in {\it Renormalization Theory},  Erice 1975, G. Velo and
A.S.~Wightman eds. (D. Reidel, Dordrecht, Holland 1976).
\nrf The consistency of the lattice regularization is rigorously established (except for
theories with chiral fermions) in\rf
T. Reisz, {\it Commun. Math. Phys.} 117 (1988) 79, 639.
\nrf The generality of the doubling phenomenon for lattice fermions has been proven   by\rf
H.B. Nielsen and M. Ninomiya, {\it Nucl. Phys.} B185 (1981) 20.
\nrf Wilson's solution to the fermion doubling problem is described in\rf
K.G. Wilson in {\it New Phenomena in Subnuclear Physics}, Erice 1975,
\goodbreak A.~Zichichi ed. (Plenum, New York 1977).  
\nrf Staggered fermions have been proposed in\rf
T. Banks, L. Susskind and J. Kogut, {\it Phys. Rev.} D13 (1976) 1043.
\nrf The problem of chiral anomalies is discussed in\rf
J.S. Bell and R. Jackiw, {\it Nuovo Cimento} A60 (1969) 47;
S.L. Adler, {\it Phys. Rev.} 177 (1969) 2426;
W.A. Bardeen, {\it Phys. Rev.} 184 (1969) 1848;  
D.J. Gross and R. Jackiw, {\it Phys. Rev.} D6 (1972) 477;
H. Georgi and S.L. Glashow, {\it Phys. Rev.} D6 (1972) 429;
C. Bouchiat, J. Iliopoulos and Ph. Meyer, {\it Phys. Lett.} 38B (1972) 519.
\nrf See also the lectures\rf
S.L. Adler, in {\it Lectures on Elementary Particles and Quantum Field Theory},
S. Deser {\it et al}\/ eds. (MIT Press, Cambridge 1970); 
M. Peskin, in {\it Recent Advances in Field Theory and Statistical Mechanics},
Les Houches 1982, R. Stora and J.-B. Zuber eds. (North-Holland, Amsterdam
1984);
L. Alvarez-Gaume, in   {\it Fundamental problems of gauge theory}, Erice 1985
G. Velo and A.S. Wightman eds. (Plenum Press, New-York 1986).
\nrf The index of the Dirac operator in a gauge background is related to Atiyah--Singer's theorem\rf
M. Atiyah, R. Bott and V. Patodi, {\it Invent. Math.} 19 (1973) 279.
\nrf It is at the basis of the analysis relating anomalies to the regularization of  the fermion measure\rf
K. Fujikawa, {\it Phys. Rev.} D21 (1980) 2848; D22 (1980) 1499(E).
\nrf The same strategy has been applied to the conformal anomaly\rf
K. Fujikawa, {\it Phys. Rev. Lett.} 44 (1980) 1733.
\nrf For non-perturbative global gauge anomalies see\rf
E. Witten, {\it Phys. Lett.} B117 (1982) 324; {\it Nucl. Phys.} B223 (1983) 422;
S. Elitzur, V.P. Nair, {\it Nucl. Phys.} B243 (1984) 205.
\nrf The gravitational anomaly is discussed in \rf
L. Alvarez-Gaume and E. Witten, {\it Nucl. Phys.} B234 (1983) 269.
\nrf See also the volumes \rf
S.B. Treiman, R. Jackiw, B. Zumino and E. Witten, {\it Current Algebra and
Anomalies} (World Scientific, Singapore 1985) and references therein;
R.A. Bertlman, {\it Anomalies in Quantum Field theory}, Oxford Univ.~Press, Oxford 1996.
\nrf Instanton contributions to the cosine potential have been calculated 
with increasing accuracy in \rf
E. Br\'ezin, G. Parisi and J. Zinn-Justin, {\it Phys. Rev.} D16 (1977) 408;
E.B. Bogomolny, {\it Phys. Lett.} 91B (1980) 431; J. Zinn-Justin, {\it Nucl. Phys.} B192 (1981) 125; B218 (1983) 333; {\it J.
Math. Phys.} 22 (1981) 511; 25 (1984) 549.
\nrf Classical references on instantons in the $CP(N-1)$ models include \rf
A. Jevicki {\it Nucl. Phys.} B127 (1977) 125;
D. F\"orster, {\it Nucl. Phys.} B130 (1977) 38;
M. L\"uscher, {\it Phys. Lett.} 78B (1978) 465;
A. D'Adda, P. Di Vecchia and M. L\"uscher, {\it Nucl. Phys.} B146 (1978) 63;
B152 (1979) 125;
H. Eichenherr, {\it Nucl. Phys.} B146 (1978) 215;
V.L. Golo and A. Perelemov, {\it Phys. Lett.} 79B (1978) 112;
A.M. Perelemov, {\it Phys. Rep.} 146 (1987) 135.
\nrf For instantons in gauge theories see\rf
A.A. Belavin, A.M. Polyakov, A.S. Schwartz and Yu S. Tyupkin, {\it Phys.
Lett.} 59B (1975) 85;
G. 't Hooft, {\it Phys. Rev. Lett.} 37 (1976) 8; {\it Phys. Rev.} D14 (1976)
3432 (Erratum {\it Phys. Rev.} D18 (1978) 2199);
R. Jackiw and C. Rebbi, {\it Phys. Rev. Lett.} 37 (1976) 172;
C.G. Callan, R.F. Dashen and D.J. Gross, {\it Phys. Lett.} 63B (1976) 334;
A.A. Belavin and A.M. Polyakov, {\it Nucl. Phys.} B123 (1977) 429;
F.R. Ore, {\it Phys. Rev.} D16 (1977) 2577;
S. Chadha,  P. Di Vecchia, A. D'Adda and F. Nicodemi, {\it Phys. Lett.} 72B
(1977) 103; 
T. Yoneya, {\it Phys. Lett.} 71B (1977) 407;
I.V. Frolov and A.S. Schwarz, {\it Phys. Lett.} 80B (1979) 406;
E. Corrigan, P. Goddard and S. Templeton, {\it Nucl. Phys.} B151 (1979) 93.
\nrf For a  solution of the $U(1)$ problem based on anomalies and instantons see\rf
G. 't Hooft, {\it Phys. Rep.} 142 (1986) 357.
\nrf The Bogomolnyi bound is discussed in \rf 
E.B. Bogomolnyi, {\it Sov. J. Nucl. Phys.} 24 (1976) 449; M.K. Prasad and C.M. Sommerfeld,
{\it Phys. Rev. Lett.} 35 (1975) 760.
\nrf For more details see Coleman lectures in\rf
S. Coleman, {\it Aspects of symmetry}, Cambridge Univ.~Press (Cambridge 1985). 
\nrf BRS symmetry has been introduced in \rf
C. Becchi, A. Rouet and R. Stora,   {\it Comm. Math. Phys.} 42 (1975) 127; {\it Ann. Phys.} (NY) 98 (1976) 287. 
\nrf It has been used to determine the non-abelian anomaly by\rf
J. Wess and B. Zumino, {\it Phys. Lett.} 37B (1971) 95.
\nrf The overlap Dirac operators for chiral fermions
is constructed explicitly in \rf
 H. Neuberger, {\it Phys. Lett.} B417 (1998) 141,
hep-lat/9707022, {\it ibidem}\/ B427 (1998) 353, hep-lat/9801031.
\nrf The index theorem in lattice gauge theory is discussed in\rf
P. Hasenfratz, V. Laliena, F. Niedermayer,
{\it Phys. Lett.} B427 (1998) 125, hep-lat/9801021.
\nrf A modified exact chiral symmetry on the lattice was exhibited in\rf
M. L\"uscher, {\it Phys. Lett.} B428 (1998) 342, hep-lat/9802011,
hep-lat/9811032.
\nrf The overlap Dirac operator  was found to provide solutions to  the Ginsparg--Wilson relation\rf
P.H. Ginsparg and K.G. Wilson, {\it Phys. Rev.} D25 (1982) 2649.
\nrf Supersymmetric quantum mechanics is studied in \rf
E. Witten, {\it Nucl. Phys.} B188 (1981) 513.
\nrf
General determinations of the index of the Dirac operator can be found in \rf
C. Callias, {\it Com. Math. Phys.} 62 (1978) 213.
\nrf The fermion zero-mode in a soliton background in two dimensions is investigated in \rf
R. Jackiw and C. Rebbi, {\it Phys. Rev.} D13 (1976) 3398.
\nrf Special properties of fermions in presence of domain walls were noticed in \rf
C.G. Callan and J.A. Harvey, {\it Nucl. Phys.} B250 (1985) 427. 
\nrf Domain wall fermions on the lattice were discussed in \rf
D.B. Kaplan, {\it Phys. Lett.} B288 (1992) 342. 
\nrf For more discussions and references \rf
  Proceedings of the workshop ``Chiral 99", {\it Chinese Journal of Physics}, 38 (2000) 521--743;
K. Fujikawa, {\it Int. J. Mod. Phys.} A16 (2001) 331; T.-W. Chiu, {\it Phys. Rev.} D58(1998) 074511,  hep-lat/9804016;
M. L\"uscher, Lectures given at International School of Subnuclear Physics {\it Theory and Experiment Heading for New Physics}, Erice,  27 Aug - 5 Sep 2000, hep-th/0102028 and references therein; P. Hasenfratz, Proceedings of ``Lattice 2001",  
{\it Nucl. Proc. Suppl.} 106 (2002) 159, hep-lat/0111023. 
\nrf In particular the $U(1)$ problem  has been discussed analytically and studied numerically on the lattice. For a recent reference see for instance \rf
L. Giusti, G.C. Rossi, M. Testa, G. Veneziano, hep-lat/0108009.
\nrf A number of simulations use overlap or domain wall fermions; a few examples are \rf
R.G. Edwards, U. Heller, J. Kiskis, R. Narayanan, Phys. Rev D61 (2000) 074504, hep-lat/9910041; 
P. Hern\'andez, K. Jansen, L. Lellouch, {\it Nucl. Phys. Proc. Suppl.} 83 (2000) 633, hep-lat/9909026; 
S.J. Dong, F.X. Lee, K.F. Liu, J.B. Zhang, {\it  Phys. Rev. Lett.} 85 (2000) 5051,  hep-lat/0006004; 
T. DeGrand, {\it Phys.Rev.} D63 (2001) 03450, hep-lat/0007046; 
R.V. Gavai, S. Gupta, R. Lacaze, hep-lat/0107022; 
L. Giusti,C. Hoelbling, C. Rebbi, hep-lat/0110184; {\it ibidem} hep-lat/0108007.
\endbib
\vfill\eject
\def\xxt{X}
\appendix{} 

\vskip-33pt
\section Trace formula for periodic potentials

We consider a hamiltonian $ H $ corresponding to a real periodic
potential $ V (x) $ with period $ \xxt $:\sslbl\ssDegMtr
$$ V (x+\xxt )=V (x). \eqnn $$
Eigenfunctions $ \psi_{\theta} (x) $ are then also
eigenfunctions of the translation operator $ T $:
$$ T\psi_{\theta} (x)\equiv \psi_{\theta}  (x+\xxt
 )= \e^{i\theta}\psi_{\theta} (x). \eqnd{\etransop}
$$
We first restrict space to a box of size $ N\xxt $ with periodic
boundary conditions. This implies a quantization of the angle $ \theta $
$$ \e^{iN\theta}=1 \Rightarrow \theta =\theta_{p}\equiv  2\pi p
/N\,,\quad  0\leq p<N\,. \eqnn $$
We call $ \psi_{p,n} $ the normalized eigenfunctions of $ H $ corresponding
to the band $ n $ and the pseudo-momentum $ \theta_{p} $
$$\int_0^{N\xxt }\d x\,  \psi^* _{p,m} (x)\psi_{q,n}(x)=\delta _{mn}\delta _{pq}\,,$$
and $E_n( \theta_{p}) $ the corresponding eigenvalues. 
Reality implies:
$$ E_{n}  (\theta  )=E_{n}  (-\theta  ). \eqnn $$
This leads to a decomposition of the identity operator in $[0,N\xxt]$
$$\delta (x-y)=\sum_{p,n}\psi_{p,n}(x)\psi^*_{p,n}(y).$$
We now consider an operator $O$ that commutes with $T$
$$[T,O]=0\ \Rightarrow \ \left<x\right|O\left|y\right>=\left<x+\xxt \right|O\left|y+\xxt \right> .$$
Then
$$\left<q,n\right|O\left|p,m\right>=\int_0^{N\xxt}\d x \,\d y\,\psi^* _{q,n} (x)\left<x\right|O\left|y\right>\psi_{p,m}(y)=\delta _{pq}O_{mn}(\theta _p). $$
Its trace  can be written
$$\tr O=\int_0^{N\xxt} \d x\,\left<x\right|O\left|x\right>=N  \int_0^{ \xxt} \d x\,\left<x\right|O\left|x\right>=\sum_{p,n} O_{nn}(\theta _p).$$
We then take the infinite box limit $N\to\infty $. Then
$${1\over N}\sum_p\to {1\over 2\pi}\int_0^{2\pi} \d\theta\,,$$
and thus we find
$$\int_0^{ \xxt} \d x\,\left<x\right|O\left|x\right>= \sum_n{1\over 2\pi}\int_0^{2\pi} 
O_{nn}(\theta )\d\theta\, .\eqnd\etrperiod $$
We now apply this general result to the operator
$$O=T^\ell \e^{-\beta  H}\,.$$
Then
$$ \int^{\xxt}_0\left< x \right| T^\ell \e^{-\beta H}
\left| x \right> \d  x={1 \over 2\pi} \sum_{n} \int
^{2\pi}_0 \e^{i\ell \theta -\beta E_{n}  (\theta  )}
\d  \theta\, ,\eqnn $$
which using the definition of $T$ can be rewritten
$$ \int^{\xxt}_0\left< x+\ell \xxt  \right| \e^{-\beta H}
\left| x \right> \d  x={1 \over 2\pi} \sum_{n} \int
^{2\pi}_0 \e^{i\ell \theta -\beta E_{n}  (\theta  )}
\d  \theta\, .\eqnn $$
In the path integral formulation this leads to a representation of the form
$$\int_{x(\beta /2)=x(-\beta /2)+\ell \xxt }[\d x(t)]\exp\left[-{\cal S}(x)\right]=
{1 \over 2\pi} \sum_{n} \int
^{2\pi}_0 \e^{i\ell \theta -\beta E_{n}  (\theta  )}
\d  \theta\, ,\eqnn $$
where $x(-\beta /2)$ varies only in $[0,\xxt]$, justifying the representation
\eZpathcosel.
\section Resolvent of the hamiltonian in supersymmetric QM 

The resolvent $G(z)=(H+z)^{-1}$ of the hermitian operator $H$,\sslbl\sshamresolv
$$ d H=-\d_x^2+V(x),$$
where $-z$ is below the spectrum of $H$, satisfies the differential  equation:
$$\bigl(-\d^2+V(x)+z\bigr)G(z;x,y)=\delta(x-y)\,. \eqnd\eresol $$
We recall how $G(z;x,y)$ can be expressed in terms of two independent
solutions of the homogeneous equation
$$\bigl(-\d^2+V(x)+z\bigr)\varphi_{1,2}(x)=0\ .\eqnd\ehom $$
If we partially normalize by the wronskian $W$,
$$W(\varphi_1,\varphi_2)\equiv\varphi'_{1}(x)\varphi_2(x)-\varphi_1(x) \varphi'_2(x)=1\ ,$$
and moreover impose the boundary conditions
$$\varphi_1(x)\to 0\ {\rm for}\ x\to -\infty,\quad
\varphi_2(x)\to 0\ {\rm for}\ x\to +\infty\ ,$$
then it is easily verified that $G(z;x,y)$ is given by
$$G(z;x,y)=\varphi_1(y)\varphi_2(x)\,\theta(x-y)+
\varphi_1(x)\varphi_2(y)\,\theta(y-x)\ .\eqnd\eHresol $$
If the potential is an even function $V(-x)=V(x)$,
$$\varphi_2(x)\propto \varphi_1(-x).$$
We now apply this result to the operator
$$H={\bf D}{\bf D}^\dagger\quad{\rm with}\quad z=-k^2.$$
The functions $\varphi_i$ then satisfy
$$\left({\bf D}{\bf D}^\dagger+k^2\right)\varphi_i(x)=\equiv \left[-\d^2+A^2(x)+A'(x)+k^2\right]\varphi_i(x)=0\,,$$
and the equation \eHresol\ yields the resolvent $G_-(k^2;x,y)$.\par
The corresponding solutions for the operator ${\bf D}^\dagger {\bf D}+k^2$ follow
since
$${\bf D}^\dagger \left({\bf D}{\bf D}^\dagger+k^2\right)\varphi_i =0=\left({\bf D}^\dagger  {\bf D}+k^2\right){\bf D}^\dagger\varphi_i=0\,.$$
Setting 
$$\chi_i(x)= {\bf D}^\dagger\varphi_i(x), $$
we calculate the wronskian for normalization purpose
$$W(\chi_1,\chi_2)\equiv\chi'_{1}(x)\chi_2(x)-\chi_1(x) \chi'_2(x)=-k^2.$$
Thus the corresponding resolvent $G_+$ reads
$$G_+(k^2;x,y)=-{1\over k^2}\left[ \chi_1(y)\chi_2(x)\,\theta(x-y)+
\chi_1(x)\chi_2(y)\,\theta(y-x)\right].\eqnn $$
The limits $x=y$ are
$$G_-(k^2;x,x)=\varphi_1(x)\varphi_2(x),\quad G_+(k^2;x,x)=-{1\over k^2}\chi _1(x)\chi _2(x).$$
Note that these functions satisfy third order linear differential equations. 
If the potential is even, here this implies that $A(x)$ is odd, $ G_\pm(k^2;x,x)$ are even functions.
\par
Finally we also need ${\bf D}^\dagger G_-(k^2;x,y)$:
$${\bf D}^\dagger G_-(k^2;x,y)=\varphi_1(y){\bf D}^\dagger\varphi_2(x)\theta (x-y)+
\varphi_2(y){\bf D}^\dagger\varphi_1(x)\theta (y-x).$$
We note that  ${\bf D}^\dagger G_-(k^2;x,y)$ is not continuous at $x=y$:
$$\lim_{y\to x_+}{\bf D}^\dagger G_-(k^2;x,y)=\varphi_2(x){\bf D}^\dagger\varphi_1(x),\quad
\lim_{y\to x_-}{\bf D}^\dagger G_-(k^2;x,y)=\varphi_1(x){\bf D}^\dagger\varphi_2(x),$$
and therefore from the wronskian,
$$\lim_{y\to x_-}{\bf D}^\dagger G_-(k^2;x,y)-\lim_{y\to x_+}{\bf D}^\dagger G_-(k^2;x,y)=1\,.$$
The half sum is given by
$$\eqalign{\overline{{\bf D}^\dagger G_-(k^2;x,x)}&=\ud\lim_{y\to x_-}{\bf D}^\dagger G_-(k^2;x,y)+\ud\lim_{y\to x_+}{\bf D}^\dagger G_-(k^2;x,y)\cr
&=\ud{\bf D}^\dagger \varphi_1(x) \varphi_2(x) +\ud{\bf D}^\dagger \varphi_2(x) \varphi_1(x) \cr
&=\ud \left(\varphi_1\varphi_2\right)'(x)+A(x)\varphi_1(x)\varphi_2(x) .\cr}
 $$
This function is odd when $A(x)$ is odd.\par
Finally in the limit $k\to0$ one finds
$$\varphi_1(x)=N\e^{S(x)}\int_{-\infty }^x\d u\,\e^{-2S(u)},\quad \varphi_2(x)=N\e^{S(x)}\int_x ^{\infty }\d u\,\e^{-2S(u)}, $$
with 
$$N^2 \int_{-\infty }^{+\infty }\d u\,\e^{-2S(u)}=1\,.$$
Moreover
$${\bf D}^\dagger\varphi_1(x)=-N\e^{-S(x)},\quad 
 {\bf D}^\dagger\varphi_2(x)= N\e^{-S(x)}.$$ 
Therefore, as expected
$$G_+(k^2;x,y)\mathop{\sim}_{k\to0} {1\over k^2}N^2\e^{-S(x)-S(y)}.$$
Finally
$${\bf D}^\dagger G_-(0;x,y)=N^2\left[\theta (x-y) \e^{-S(x)+S(y)} \int_{-\infty }^y \d u\,\e^{-2S(u)}+
 (x\leftrightarrow y)\right],$$
and therefore
$$ \overline{{\bf D}^\dagger G_-(0;x,x)}=\ud N^2\int_{-\infty }^\infty \d t\,\sgn(x-t)\e^{-2S(t)}.   $$
\bye

%% file: zmacxxx.tex
\input hyperbasics
\catcode`\@=11
\def\unredoffs{\voffset=13mm \hoffset=6.5truemm} 
\def\redoffs{\voffset=-12.truemm\hoffset=-3truemm} 
\def\speclscape{}
%
\newbox\leftpage \newdimen\fullhsize \newdimen\hstitle \newdimen\hsbody
\newdimen\hdim
\hfuzz=1pt
\ifx\hyperdef\UNd@FiNeD\def\hyperdef#1#2#3#4{#4}\def\hyperref#1#2#3#4{#4}\fi
\def\newans{y }
\def\answb{y }
\ifx\answb\newans\message{(This uses normal fonts.)}%
%
\def\bigans{b }
\def\answ{b }
\ifx\answ\bigans\message{(Format simple colonne 12pts.}
\magnification=1200 \unredoffs\hsize=147truemm\vsize=219truemm
\hsbody=\hsize \hstitle=\hsize 
\else\message{(Format double colonne, 10pts.} \let\l@r=L
\magnification=1000 \vsize=182.5truemm
\redoffs%
\hstitle=122.5truemm\hsbody=122.5truemm\fullhsize=258truemm\hsize=\hsbody 
\output={
  \almostshipout{\leftline{\vbox{\makeheadline\pagebody\makefootline}}}
\advancepageno%
}
\def\almostshipout#1{\if L\l@r \count1=1 \message{[\the\count0.\the\count1]}
      \global\setbox\leftpage=#1 \global\let\l@r=R
 \else \count1=2
  \shipout\vbox{\speclscape{\hsize\fullhsize}
      \hbox to\fullhsize{\box\leftpage\hfil#1}}  \global\let\l@r=L\fi}
\fi

\input lfont
\def\sla#1{\mkern-1.5mu\raise0.4pt\hbox{$\not$}\mkern1.2mu #1\mkern 0.7mu}
\def\Dbar{\mkern-1.5mu\raise0.4pt\hbox{$\not$}\mkern-.1mu {\rm D}\mkern.1mu}
\def\Abar{\mkern1.mu\raise0.4pt\hbox{$\not$}\mkern-1.3mu A\mkern.1mu}
\nopagenumbers
\headline={\ifnum\pageno=1\hfill\else\draftdate\hfil{\headrm\folio}%
\hfil\fi}	 
\else\message{(This uses pseudo 12pts fonts.}
\hoffset=8mm
\voffset=16mm
\input lfont12 

\def\sla#1{\mkern-1.5mu\raise0.5pt\hbox{$\not$}\mkern1.2mu #1\mkern 0.7mu}
\def\Dbar{\mkern-1.5mu\raise0.5pt\hbox{$\not$}\mkern-.1mu {\rm D}\mkern.1mu}
\def\Abar{\mkern1.mu\raise0.5pt\hbox{$\not$}\mkern-1.3mu A\mkern.1mu}
\fi

\newcount\yearltd\yearltd=\year\advance\yearltd by -1900
\newif\ifdraftmode
\draftmodefalse
\def\draft{\draftmodetrue{\count255=\time\divide\count255 by 60
\xdef\hourmin{\number\count255} 
  \multiply\count255 by-60\advance\count255 by\time
  \xdef\hourmin{\hourmin:\ifnum\count255<10 0\fi\the\count255}}}
\def\draftdate{\ifdraftmode{\headrm\quad (\jobname,\ le
\number\day/\number\month/\number\yearltd\ \ \hourmin)}\else{}\fi} 
\newif\iffrancmode
\francmodefalse
\def\e{\mathop{\rm e}\nolimits}
\def\sgn{\mathop{\rm sgn}\nolimits}

\def\d{{\rm d}}
\def\ud{{\textstyle{1\over 2}}}

\def\tr{\mathop{\rm tr}\nolimits}
\def\det{\mathop{\rm det}\nolimits}

\chardef\sigmat=27
\def\n{\noindent}  
\def\rf{\par\item{}}
\def\nrf{\par\n}
\def\frac#1#2{{\textstyle{#1\over#2}}}

\def\leaderfill{\leaders\hbox to 1em{\hss.\hss}\hfill}
\catcode`\@=11
\def\deqalignno#1{\displ@y\tabskip\centering \halign to
\displaywidth{\hfil$\displaystyle{##}$\tabskip0pt&$\displaystyle{{}##}$
\hfil\tabskip0pt &\quad
\hfil$\displaystyle{##}$\tabskip0pt&$\displaystyle{{}##}$ 
\hfil\tabskip\centering& \llap{$##$}\tabskip0pt \crcr #1 \crcr}}
\def\deqalign#1{\null\,\vcenter{\openup\jot\m@th\ialign{
\strut\hfil$\displaystyle{##}$&$\displaystyle{{}##}$\hfil
&&\quad\strut\hfil$\displaystyle{##}$&$\displaystyle{{}##}$
\hfil\crcr#1\crcr}}\,}
\def\xlabel#1{\expandafter\xl@bel#1}\def\xl@bel#1{#1}
\def\label#1{\l@bel #1{\hbox{}}}
\def\l@bel#1{\ifx\UNd@FiNeD#1\message{label \string#1 is undefined.}%
\xdef#1{?.? }\fi{\let\hyperref=\relax\xdef\next{#1}}%
\ifx\next\em@rk\def\next{}%
\else\def\next{#1}\fi\next}
\def\DefWarn#1{\ifx\UNd@FiNeD#1\else
\immediate\write16{*** WARNING: the label \string#1 is already defined%
***}\fi}%
\newread\ch@ckfile
\def\cinput#1{\def\filen@me{#1 }
\immediate\openin\ch@ckfile=\filen@me
\ifeof\ch@ckfile\message{<< (\filen@me\ DOES NOT EXIST in this pass)>>}\else%
\closein \ch@ckfile\input\filen@me\fi}
\ifx\UNd@FiNeD\prefix\def\prefix{}\fi 
\newread\ch@ckfile
\immediate\openin\ch@ckfile=\jobname.def
\ifeof\ch@ckfile\message{<< (\jobname.def DOES NOT EXIST in this pass) >>}
\else
\def\DefWarn#1{}%
\closein \ch@ckfile%
\input\jobname.def\fi
\def\listcontent{
\immediate\openin\ch@ckfile=\jobname.tab 
\ifeof\ch@ckfile\message{no file \jobname.tab, no table of contents this
pass}%
\else\closein\ch@ckfile\centerline{\bf\iffrancmode Table des
mati\`eres \else Contents\fi}\nobreak\medskip%
{\baselineskip=12pt\parskip=0pt\catcode`\@=11\input\jobname.tab
\catcode`\@=12\bigbreak\bigskip}\fi}
\newcount\nosection
\newcount\nosubsection
\newcount\neqno
\newcount\notenumber
\newcount\nofigure
\newcount\notable
\newcount\noexerc
\newif\ifappmode
\def\equation{\jobname.equ}
\newwrite\equa

\newdimen\hulp
\def\maketitle#1{
\edef\oneliner##1{\centerline{##1}}
\edef\twoliner##1{\vbox{\parindent=0pt\leftskip=0pt plus 1fill\rightskip=0pt
plus 1fill 
                     \parfillskip=0pt\relax##1}} 
\setbox0=\vbox{#1}\hulp=0.5\hsize
                 \ifdim\wd0<\hulp\oneliner{#1}\else
                 \twoliner{#1}\fi}
\def\preprint#1{\ifdraftmode\gdef\prepname{\jobname/#1}\else%
\gdef\prepname{#1}\fi\hfill{
\expandafter{\prepname}}\vskip20mm} 
\def\title#1\par{\gdef\titlename{#1}
\maketitle{\ssbx\uppercase\expandafter{\titlename}}
\vskip20truemm
\nosection=0
\neqno=0
\notenumber=0
\nofigure=0
\notable=0
\def\prefix{}
\appmodefalse
\mark{\the\nosection}
\message{#1}
\immediate\openout\equa=\equation
}
\def\abstract{\vskip8mm\iffrancmode\centerline{R\'ESUM\'E}\else%
\centerline{ABSTRACT}\fi \smallskip \begingroup\narrower
\elevenpoint\baselineskip10pt} 
\def\endabstract{\par\endgroup \bigskip}
\def\section#1\par{\vskip0pt plus.1\vsize\penalty-100\vskip0pt plus-.1
\vsize\bigskip\vskip\parskip
\ifnum\nosection=0\ifappmode\relax\else\writetoc
\fi\fi
\advance\nosection by 1\global\nosubsection=0\global\neqno=0
\vbox{\noindent\bf{\hyperdef\hypernoname{section}{\prefix\the\nosection}%
{\prefix\the\nosection}\ #1}}
\writetoca{{\string\hyperref{}{section}{\prefix\the\nosection}%
{\prefix\the\nosection}} {#1}}
\message{\the\nosection\ #1}
\mark{\the\nosection}\bigskip\noindent
}

\def\appendix#1#2\par{\bigbreak\nosection=0
\appmodetrue
\notenumber=0
\neqno=0
\def\prefix{A}
\mark{\the\nosection}
\message{APPENDICES}
{\leftline{APPENDICES} \hyperdef\hypernoname{appendix}{\prefix}{ 
\leftline{\uppercase\expandafter{#1}}
\leftline{\uppercase\expandafter{#2}}}}
\bigskip\noindent\nonfrenchspacing
\writetoca{\string\hyperref{}{appendix}{\prefix}{Appendices}.\ #1 \ #2}%
}
\def\subsection#1\par {\vskip0pt plus.05\vsize\penalty-100\vskip0pt
plus-.05\vsize\bigskip\vskip\parskip\advance\nosubsection by 1
\vbox{\noindent\it{\hyperdef\hypernoname{subsection}{\prefix\the\nosection.%
\the\nosubsection}{\prefix\the\nosection.\the\nosubsection\ #1}}}%
\smallskip\noindent 
\writetoca{{\string\hyperref{}{subsection}{\prefix\the\nosection.%
\the\nosubsection}{\prefix\the\nosection.\the\nosubsection}} {#1}}
\message{\the\nosection.\the\nosubsection\ #1}
} 
\def\note #1{\advance\notenumber by 1
\footnote{$^{\the\notenumber}$}{\sevenrm #1}} 

\parindent=1em 
\newinsert\margin
\dimen\margin=\maxdimen
\count\margin=0 \skip\margin=0pt
\def\sslbl#1{\DefWarn#1%
\ifdraftmode{\hfill\escapechar-1{\rlap{\hskip-1mm%
\sevenrm\string#1}}}\fi%
\ifnum\nosection=0\if\prefix{}\xdef#1{}%
\edef\ewrite{\write\equa{{\string#1}}%
\write\equa{}}\ewrite%
\else
\xdef#1{\noexpand\hyperref{}{appendix}{\prefix}{\prefix}}%
\edef\ewrite{\write\equa{{\string#1},\prefix}%
\write\equa{}}\ewrite%
\writedef{#1\leftbracket#1}
\fi
\else%
\ifnum\nosubsection=0%
\xdef#1{\noexpand\hyperref{}{section}{\prefix\the\nosection}%
{\prefix\the\nosection}}%
\edef\ewrite{\write\equa{{\string#1},\prefix\the\nosection}%
\write\equa{}}\ewrite%
\writedef{#1\leftbracket#1}
\else%
\xdef#1{\noexpand\hyperref{}{subsection}{\prefix\the\nosection.%
\the\nosubsection}{\prefix\the\nosection.\the\nosubsection}}%
\writedef{#1\leftbracket#1}
\edef\ewrite{\write\equa{{\string#1},\prefix\the\nosection%
.\the\nosubsection}\write\equa{}}\ewrite\fi\fi}%

\newwrite\tfile \def\writetoca#1{}
\def\writetoc{\immediate\openout\tfile=\jobname.tab
\def\writetoca##1{{\edef\next{\write\tfile{\noindent ##1 \string\leaderfill%
\noexpand\number\pageno\par}}\next}}}

%
\def\nolabels{\def\wrlabeL##1{}\def\eqlabeL##1{}\def\reflabeL##1{}}
\def\writelabels{\def\wrlabeL##1{\leavevmode\vadjust{\rlap{\smash%
{\line{{\escapechar=` \hfill\rlap{\sevenrm\hskip.03in\string##1}}}}}}}%
\def\eqlabeL##1{{\escapechar-1\rlap{\sevenrm\hskip.05in\string##1}}}%
\def\reflabeL##1{\noexpand\llap{\noexpand\sevenrm\string\string\string##1}}}
\nolabels

\global\newcount\refno \global\refno=1
\newwrite\rfile
\def\ref{[\hyperref{}{reference}{\the\refno}{\the\refno}]\nref}
\def\nref#1{\DefWarn#1%
\xdef#1{[\noexpand\hyperref{}{reference}{\the\refno}{\the\refno}]}%
\writedef{#1\leftbracket#1}%
\ifnum\refno=1\immediate\openout\rfile=\jobname.ref\fi
\chardef\wfile=\rfile\immediate\write\rfile{\noexpand\item{[\noexpand\hyperdef%
\noexpand\hypernoname{reference}{\the\refno}{\the\refno}]\ }%
\reflabeL{#1\hskip.31in}\pctsign}\global\advance\refno by1\findarg}
\def\findarg#1#{\begingroup\obeylines\newlinechar=`\^^M\pass@rg}
{\obeylines\gdef\pass@rg#1{\writ@line\relax #1^^M\hbox{}^^M}%
\gdef\writ@line#1^^M{\expandafter\toks0\expandafter{\striprel@x #1}%
\edef\next{\the\toks0}\ifx\next\em@rk\let\next=\endgroup\else\ifx\next\empty%
\else\immediate\write\wfile{\the\toks0}\fi\let\next=\writ@line\fi\next\relax}}
\def\striprel@x#1{} \def\em@rk{\hbox{}}
\def\lref{\begingroup\obeylines\lr@f}
\def\lr@f#1#2{\DefWarn#1\gdef#1{\let#1=\UNd@FiNeD\ref#1{#2}}\endgroup\unskip}

\def\addref#1{\immediate\write\rfile{\noexpand\item{}#1}} 
\def\listrefs{{}\vfill\supereject\immediate\closeout\rfile\writestoppt
\baselineskip=14pt\centerline{{\bf\iffrancmode R\'eferences\else References%
\fi}}
\bigskip{\parindent=20pt%
\frenchspacing\escapechar=` \input \jobname.ref\vfill\eject}\nonfrenchspacing}
\def\startrefs#1{\immediate\openout\rfile=\jobname.ref\refno=#1}
\def\xref{\expandafter\xr@f}\def\xr@f[#1]{#1}
\def\refs#1{\count255=1[\r@fs #1{\hbox{}}]}
\def\r@fs#1{\ifx\UNd@FiNeD#1\message{reflabel \string#1 is undefined.}%
\nref#1{need to supply reference \string#1.}\fi%
\vphantom{\hphantom{#1}}{\let\hyperref=\relax\xdef\next{#1}}%
\ifx\next\em@rk\def\next{}%
\else\ifx\next#1\ifodd\count255\relax\xref#1\count255=0\fi%
\else#1\count255=1\fi\let\next=\r@fs\fi\next}
%
\newwrite\lfile
{\escapechar-1\xdef\pctsign{\string\%}\xdef\leftbracket{\string\{}
\xdef\rightbracket{\string\}}\xdef\numbersign{\string\#}}
\def\writedefs{\immediate\openout\lfile=\jobname.def \def\writedef##1{%
{\let\hyperref=\relax\let\hyperdef=\relax\let\hypernoname=\relax
 \immediate\write\lfile{\string\def\string##1\rightbracket}}}}%
\def\writestop{\def\writestoppt{\immediate\write\lfile{\string\pageno%
\the\pageno\string\startrefs\leftbracket\the\refno\rightbracket%
\string\def\string\secsym\leftbracket\secsym\rightbracket%
\string\secno\the\secno\string\meqno\the\meqno}\immediate\closeout\lfile}}
\def\writestoppt{}\def\writedef#1{}
\writedefs
\def\biblio\par{\vskip0pt plus.1\vsize\penalty-100\vskip0pt plus-.1
\vsize\bigskip\vskip\parskip
\message{Bibliographie}
{\leftline{\bf \hyperdef\hypernoname{biblio}{bib}{Bibliographical Notes}}}
\nobreak\medskip\noindent\frenchspacing
\writetoca{\string\hyperref{}{biblio}{bib}{Bibliographical Notes}}}%

\def\biblionote{\iffrancmode Notes Bibliographiques\else Bibliographical Notes
\fi}
\def\beginbib\par{\vskip0pt plus.1\vsize\penalty-100\vskip0pt plus-.1
\vsize\bigskip\vskip\parskip
\message{Bibliographie}
{\leftline{\bf \hyperdef\hypernoname{biblio}{\the\nosection}%
{\biblionote}}}
\nobreak\medskip\noindent\frenchspacing
\writetoca{\string\hyperref{}{biblio}{\the\nosection}%
{\biblionote}}}%
\def\endbib{\nonfrenchspacing}

\def\Exercises{\iffrancmode Exercices\else Exercises
\fi}
\def\exerc\par{\vskip0pt plus.1\vsize\penalty-100\vskip0pt plus-.1
\vsize\bigskip\vskip\parskip\global\noexerc=0
\message{Exercises}
{\leftline{\bf \hyperdef\hypernoname{exercise}{\the\nosection}{\Exercises}}}
\nobreak\medskip\noindent\frenchspacing
\writetoca{\string\hyperref{}{exercise}{\the\nosection}{\Exercises}}
}
\def\esubsec{\ifnum\noexerc=0\vskip-12pt\else\vskip0pt plus.05\vsize%
\penalty-100\vskip0pt plus-.05\vsize\bigskip\vskip\parskip\fi%
\global\advance\noexerc by 1
\hyperdef\hypernoname{exercise}{\the\nosection.\the\noexerc}{}%
\vbox{\noindent\it \iffrancmode Exercice\else Exercise\fi\ \the\nosection.\the\noexerc}\smallskip\noindent}
\def\exelbl#1{\ifdraftmode{\hfill\escapechar-1{\rlap{\hskip-1mm%
\sevenrm\string#1}}}\fi%
{\xdef#1{\noexpand\hyperref{}{exercise}{\the\nosection.\the\noexerc}%
{\the\nosection.\the\noexerc}}}%
\edef\ewrite{\write\equa{{\string#1}\the\nosection.\the\noexerc}%
\write\equa{}}\ewrite%
\writedef{#1\leftbracket#1}}

\def\eqnn{\global\advance\neqno by 1 \ifinner\relax\else%
\eqno\fi(\prefix\the\nosection.\the\neqno)}
%
\def\eqnd#1{\DefWarn#1%
\global\advance\neqno by 1 
{\xdef#1{($\noexpand\hyperref{}{equation}{\prefix\the\nosection.\the\neqno}%
{\prefix\the\nosection.\the\neqno}$)}}
\ifinner\relax\else\eqno\fi(\hyperdef\hypernoname{equation}{\prefix\the%
\nosection.\the\neqno}{\prefix\the\nosection.\the\neqno})
\writedef{#1\leftbracket#1}
\ifdraftmode{\escapechar-1{\rlap{\hskip.2mm\sevenrm\string#1}}}\fi
\edef\ewrite{\write\equa{{\string#1},(\prefix\the\nosection.\the\neqno)
{\noexpand\number\pageno}}\write\equa{}}\ewrite}
%
\def\checkm@de#1#2{\ifmmode{\def\f@rst##1{##1}\hyperdef\hypernoname{equation}%
{#1}{#2}}\else\hyperref{}{equation}{#1}{#2}\fi}
\def\f@rst#1{\c@t#1a\em@ark}\def\c@t#1#2\em@ark{#1}
\def\eqna#1{\DefWarn#1%
\global\advance\neqno by1\ifdraftmode{\hfill%
\escapechar-1{\rlap{\sevenrm\string#1}}}\fi%
\xdef #1##1{(\noexpand\relax\noexpand%
\checkm@de{\prefix\the\nosection.\the\neqno\noexpand\f@rst{##1}1}%
{\hbox{$\prefix\the\nosection.\the\neqno##1$}})}
\writedef{#1\numbersign1\leftbracket#1{\numbersign1}}%
} 
%

%
\def\em@rk{\hbox{}} 
\def\xeqn{\expandafter\xe@n}\def\xe@n(#1){#1}
\def\xeqna#1{\expandafter\xe@na#1}\def\xe@na\hbox#1{\xe@nap #1}
\def\xe@nap$(#1)${\hbox{$#1$}}
\def\eqns#1{(\e@ns #1{\hbox{}})}
\def\e@ns#1{\ifx\UNd@FiNeD#1\message{eqnlabel \string#1 is undefined.}%
\xdef#1{(?.?)}\fi{\let\hyperref=\relax\xdef\next{#1}}%
\ifx\next\em@rk\def\next{}%
\else\ifx\next#1\xeqn#1\else\def\n@xt{#1}\ifx\n@xt\next#1\else\xeqna#1\fi
\fi\let\next=\e@ns\fi\next}
\def\figure#1#2{\global\advance\nofigure by 1 \vglue#1%
\hyperdef\hypernoname{figure}{\the\nofigure}{}%
{\elevenpoint
\setbox1=\hbox{#2}
\ifdim\wd1=0pt\centerline{Fig.\ \the\nofigure\hskip0.5mm}%
\else\def\caption{Fig.\ \the\nofigure\quad#2\hskip0mm}
\setbox0=\hbox{\caption}
\ifdim\wd0>\hsize\noindent Fig.\ \the\nofigure\quad#2\else
                 \centerline{\caption}\fi\fi}\par}
\def\lfigure#1#2{\global\advance\nofigure by
1\vglue#1%
\hyperdef\hypernoname{figure}{\the\nofigure}{}%
\leftline{\elevenpoint\hskip10truemm  Fig.\
\the\nofigure\quad #2}} 
\def\figlbl#1{\ifdraftmode{\hfill\escapechar-1{\rlap{\hskip-1mm%
\sevenrm\string#1}}}\fi%
{\xdef#1{\noexpand\hyperref{}{figure}{\the\nofigure}%
{\the\nofigure}}}%
\edef\ewrite{\write\equa{{\string#1}\the\nofigure}%
\write\equa{}}\ewrite%
\writedef{#1\leftbracket#1}}
\def\tablbl#1{\global\advance\notable by
1\ifdraftmode{\hfill\escapechar-1{\rlap{\hskip-1mm%
\sevenrm\string#1}}}\fi%
{\xdef#1{\noexpand\hyperref{}{table}{\the\notable}%
{\the\notable}}}%
\hyperdef\hypernoname{table}{\the\notable}{}%
\edef\ewrite{\write\equa{{\string#1}\the\notable}%
\write\equa{}}\ewrite%
\writedef{#1\leftbracket#1}}

\catcode`@=12

%% file: lfont.tex
\def\sla#1{\mkern-1.5mu\raise0.4pt\hbox{$\not$}\mkern1.2mu #1\mkern 0.7mu}
\def\Dbar{\mkern-1.5mu\raise0.4pt\hbox{$\not$}\mkern-.1mu {\rm D}\mkern.1mu}
\def\Ds{\mkern-0.5mu\raise0.5pt\hbox{$\not$}\mkern-.6mu {\bf D}\mkern.1mu}
\def\Abar{\mkern1.mu\raise0.4pt\hbox{$\not$}\mkern-1.3mu A\mkern.1mu}
\def\Abbar{\mkern0.5mu\raise0.5pt\hbox{$\not$}\mkern-0.8mu{\bf A}\mkern.1mu}
\def\Bbar{\mkern-0.mu\raise0.4pt\hbox{$\not$}\mkern-.3mu B\mkern 0.6mu}
\newskip\tableskipamount \tableskipamount=8pt plus 3pt minus 3pt


\newdimen\chapskip

\font\ssbx=cmssbx10  

\font\caprm=cmr9
\font\capit=cmti9
\font\capbf=cmbx9
\font\capsl=cmsl9
\font\capmi=cmmi9
\font\capex=cmex9
\font\capsy=cmsy9
\chapskip=17.5mm
\def\makeheadline{\vbox to 0pt{\vskip-22.5pt
\line{\vbox to8.5pt{}\the\headline}\vss}\nointerlineskip}
\font\tenbi=cmmib10 
\font\ninebi=cmmib9
\font\sevenbi=cmmib7 
\font\fivebi=cmmib5
\textfont4=\tenbi
\scriptfont4=\sevenbi
\scriptscriptfont4=\fivebi
\font\headrm=cmr10

\font\sixrm=cmr6
\font\fiverm=cmr5
\font\sixmi=cmmi6
\font\fivemi=cmmi5
\font\sixsy=cmsy6
\font\fivesy=cmsy5
\font\sixbf=cmbx6
\font\fivebf=cmbx5
\skewchar\capmi='177 \skewchar\sixmi='177 \skewchar\fivemi='177
\skewchar\capsy='60 \skewchar\sixsy='60 \skewchar\fivesy='60

\def\elevenpoint{
\textfont0=\caprm \scriptfont0=\sixrm \scriptscriptfont0=\fiverm
\def\rm{\fam0\caprm}
\textfont1=\capmi \scriptfont1=\sixmi \scriptscriptfont1=\fivemi
\textfont2=\capsy \scriptfont2=\sixsy \scriptscriptfont2=\fivesy
\textfont3=\capex \scriptfont3=\capex \scriptscriptfont3=\capex
\textfont\itfam=\capit \def\it{\fam\itfam\capit} 
\textfont\slfam=\capsl  \def\sl{\fam\slfam\capsl} 
\textfont\bffam=\capbf \scriptfont\bffam=\sixbf
\scriptscriptfont\bffam=\fivebf
\def\bf{\fam\bffam\capbf} 
\textfont4=\ninebi \scriptfont4=\sevenbi \scriptscriptfont4=\fivebi
\abovedisplayskip=11pt plus 3pt minus 8pt
\belowdisplayskip=\abovedisplayskip
\smallskipamount=2.7pt plus 1pt minus 1pt
\medskipamount=5.4pt plus 2pt minus 2pt
\bigskipamount=10.8pt plus 3.6pt minus 3.6pt
\normalbaselineskip=11pt
\setbox\strutbox=\hbox{\vrule height7.8pt depth3.2pt width0pt}
\normalbaselines \rm}

%
%

\catcode`\@=11

\font\tenmsa=msam10
\font\sevenmsa=msam7
\font\fivemsa=msam5
\font\tenmsb=msbm10
\font\sevenmsb=msbm7
\font\fivemsb=msbm5
\newfam\msafam
\newfam\msbfam
\textfont\msafam=\tenmsa  \scriptfont\msafam=\sevenmsa
  \scriptscriptfont\msafam=\fivemsa
\textfont\msbfam=\tenmsb  \scriptfont\msbfam=\sevenmsb
  \scriptscriptfont\msbfam=\fivemsb

\def\hexnumber@#1{\ifcase#1 0\or1\or2\or3\or4\or5\or6\or7\or8\or9\or
	A\or B\or C\or D\or E\or F\fi }

\font\teneuf=eufm10
\font\seveneuf=eufm7
\font\fiveeuf=eufm5
\newfam\euffam
\textfont\euffam=\teneuf
\scriptfont\euffam=\seveneuf
\scriptscriptfont\euffam=\fiveeuf
\def\frak{\ifmmode\let\next\frak@\else
 \def\next{\Err@{Use \string\frak\space only in math mode}}\fi\next}
\def\goth{\ifmmode\let\next\frak@\else
 \def\next{\Err@{Use \string\goth\space only in math mode}}\fi\next}
\def\frak@#1{{\frak@@{#1}}}
\def\frak@@#1{\fam\euffam#1}

\edef\msa@{\hexnumber@\msafam}
\edef\msb@{\hexnumber@\msbfam}

\mathchardef\boxdot="2\msa@00
\mathchardef\boxplus="2\msa@01
\mathchardef\boxtimes="2\msa@02
\mathchardef\square="0\msa@03
\mathchardef\blacksquare="0\msa@04
\mathchardef\centerdot="2\msa@05
\mathchardef\lozenge="0\msa@06
\mathchardef\blacklozenge="0\msa@07
\mathchardef\circlearrowright="3\msa@08
\mathchardef\circlearrowleft="3\msa@09
\mathchardef\rightleftharpoons="3\msa@0A
\mathchardef\leftrightharpoons="3\msa@0B
\mathchardef\boxminus="2\msa@0C
\mathchardef\Vdash="3\msa@0D
\mathchardef\Vvdash="3\msa@0E
\mathchardef\vDash="3\msa@0F
\mathchardef\twoheadrightarrow="3\msa@10
\mathchardef\twoheadleftarrow="3\msa@11
\mathchardef\leftleftarrows="3\msa@12
\mathchardef\rightrightarrows="3\msa@13
\mathchardef\upuparrows="3\msa@14
\mathchardef\downdownarrows="3\msa@15
\mathchardef\upharpoonright="3\msa@16

\mathchardef\downharpoonright="3\msa@17
\mathchardef\upharpoonleft="3\msa@18
\mathchardef\downharpoonleft="3\msa@19
\mathchardef\rightarrowtail="3\msa@1A
\mathchardef\leftarrowtail="3\msa@1B
\mathchardef\leftrightarrows="3\msa@1C
\mathchardef\rightleftarrows="3\msa@1D
\mathchardef\Lsh="3\msa@1E
\mathchardef\Rsh="3\msa@1F
\mathchardef\rightsquigarrow="3\msa@20
\mathchardef\leftrightsquigarrow="3\msa@21
\mathchardef\looparrowleft="3\msa@22
\mathchardef\looparrowright="3\msa@23
\mathchardef\circeq="3\msa@24
\mathchardef\succsim="3\msa@25
\mathchardef\gtrsim="3\msa@26
\mathchardef\gtrapprox="3\msa@27
\mathchardef\multimap="3\msa@28
\mathchardef\therefore="3\msa@29
\mathchardef\because="3\msa@2A
\mathchardef\doteqdot="3\msa@2B

\mathchardef\triangleq="3\msa@2C
\mathchardef\precsim="3\msa@2D
\mathchardef\lesssim="3\msa@2E
\mathchardef\lessapprox="3\msa@2F
\mathchardef\eqslantless="3\msa@30
\mathchardef\eqslantgtr="3\msa@31
\mathchardef\curlyeqprec="3\msa@32
\mathchardef\curlyeqsucc="3\msa@33
\mathchardef\preccurlyeq="3\msa@34
\mathchardef\leqq="3\msa@35
\mathchardef\leqslant="3\msa@36
\mathchardef\lessgtr="3\msa@37
\mathchardef\backprime="0\msa@38
\mathchardef\risingdotseq="3\msa@3A
\mathchardef\fallingdotseq="3\msa@3B
\mathchardef\succcurlyeq="3\msa@3C
\mathchardef\geqq="3\msa@3D
\mathchardef\geqslant="3\msa@3E
\mathchardef\gtrless="3\msa@3F
\mathchardef\sqsubset="3\msa@40
\mathchardef\sqsupset="3\msa@41
\mathchardef\vartriangleright="3\msa@42
\mathchardef\vartriangleleft="3\msa@43
\mathchardef\trianglerighteq="3\msa@44
\mathchardef\trianglelefteq="3\msa@45
\mathchardef\bigstar="0\msa@46
\mathchardef\between="3\msa@47
\mathchardef\blacktriangledown="0\msa@48
\mathchardef\blacktriangleright="3\msa@49
\mathchardef\blacktriangleleft="3\msa@4A
\mathchardef\vartriangle="0\msa@4D
\mathchardef\blacktriangle="0\msa@4E
\mathchardef\triangledown="0\msa@4F
\mathchardef\eqcirc="3\msa@50
\mathchardef\lesseqgtr="3\msa@51
\mathchardef\gtreqless="3\msa@52
\mathchardef\lesseqqgtr="3\msa@53
\mathchardef\gtreqqless="3\msa@54
\mathchardef\Rrightarrow="3\msa@56
\mathchardef\Lleftarrow="3\msa@57
\mathchardef\veebar="2\msa@59
\mathchardef\barwedge="2\msa@5A
\mathchardef\doublebarwedge="2\msa@5B
\mathchardef\angle="0\msa@5C
\mathchardef\measuredangle="0\msa@5D
\mathchardef\sphericalangle="0\msa@5E
\mathchardef\varpropto="3\msa@5F
\mathchardef\smallsmile="3\msa@60
\mathchardef\smallfrown="3\msa@61
\mathchardef\Subset="3\msa@62
\mathchardef\Supset="3\msa@63
\mathchardef\Cup="2\msa@64

\mathchardef\Cap="2\msa@65

\mathchardef\curlywedge="2\msa@66
\mathchardef\curlyvee="2\msa@67
\mathchardef\leftthreetimes="2\msa@68
\mathchardef\rightthreetimes="2\msa@69
\mathchardef\subseteqq="3\msa@6A
\mathchardef\supseteqq="3\msa@6B
\mathchardef\bumpeq="3\msa@6C
\mathchardef\Bumpeq="3\msa@6D
\mathchardef\lll="3\msa@6E

\mathchardef\ggg="3\msa@6F

\mathchardef\circledS="0\msa@73
\mathchardef\pitchfork="3\msa@74
\mathchardef\dotplus="2\msa@75
\mathchardef\backsim="3\msa@76
\mathchardef\backsimeq="3\msa@77
\mathchardef\complement="0\msa@7B
\mathchardef\intercal="2\msa@7C
\mathchardef\circledcirc="2\msa@7D
\mathchardef\circledast="2\msa@7E
\mathchardef\circleddash="2\msa@7F
\def\ulcorner{\delimiter"4\msa@70\msa@70 }
\def\urcorner{\delimiter"5\msa@71\msa@71 }
\def\llcorner{\delimiter"4\msa@78\msa@78 }
\def\lrcorner{\delimiter"5\msa@79\msa@79 }
\def\yen{\mathhexbox\msa@55 }
\def\checkmark{\mathhexbox\msa@58 }
\def\circledR{\mathhexbox\msa@72 }
\def\maltese{\mathhexbox\msa@7A }
\mathchardef\lvertneqq="3\msb@00
\mathchardef\gvertneqq="3\msb@01
\mathchardef\nleq="3\msb@02
\mathchardef\ngeq="3\msb@03
\mathchardef\nless="3\msb@04
\mathchardef\ngtr="3\msb@05
\mathchardef\nprec="3\msb@06
\mathchardef\nsucc="3\msb@07
\mathchardef\lneqq="3\msb@08
\mathchardef\gneqq="3\msb@09
\mathchardef\nleqslant="3\msb@0A
\mathchardef\ngeqslant="3\msb@0B
\mathchardef\lneq="3\msb@0C
\mathchardef\gneq="3\msb@0D
\mathchardef\npreceq="3\msb@0E
\mathchardef\nsucceq="3\msb@0F
\mathchardef\precnsim="3\msb@10
\mathchardef\succnsim="3\msb@11
\mathchardef\lnsim="3\msb@12
\mathchardef\gnsim="3\msb@13
\mathchardef\nleqq="3\msb@14
\mathchardef\ngeqq="3\msb@15
\mathchardef\precneqq="3\msb@16
\mathchardef\succneqq="3\msb@17
\mathchardef\precnapprox="3\msb@18
\mathchardef\succnapprox="3\msb@19
\mathchardef\lnapprox="3\msb@1A
\mathchardef\gnapprox="3\msb@1B
\mathchardef\nsim="3\msb@1C
\mathchardef\ncong="3\msb@1D

\mathchardef\varsubsetneq="3\msb@20
\mathchardef\varsupsetneq="3\msb@21
\mathchardef\nsubseteqq="3\msb@22
\mathchardef\nsupseteqq="3\msb@23
\mathchardef\subsetneqq="3\msb@24
\mathchardef\supsetneqq="3\msb@25
\mathchardef\varsubsetneqq="3\msb@26
\mathchardef\varsupsetneqq="3\msb@27
\mathchardef\subsetneq="3\msb@28
\mathchardef\supsetneq="3\msb@29
\mathchardef\nsubseteq="3\msb@2A
\mathchardef\nsupseteq="3\msb@2B
\mathchardef\nparallel="3\msb@2C
\mathchardef\nmid="3\msb@2D
\mathchardef\nshortmid="3\msb@2E
\mathchardef\nshortparallel="3\msb@2F
\mathchardef\nvdash="3\msb@30
\mathchardef\nVdash="3\msb@31
\mathchardef\nvDash="3\msb@32
\mathchardef\nVDash="3\msb@33
\mathchardef\ntrianglerighteq="3\msb@34
\mathchardef\ntrianglelefteq="3\msb@35
\mathchardef\ntriangleleft="3\msb@36
\mathchardef\ntriangleright="3\msb@37
\mathchardef\nleftarrow="3\msb@38
\mathchardef\nrightarrow="3\msb@39
\mathchardef\nLeftarrow="3\msb@3A
\mathchardef\nRightarrow="3\msb@3B
\mathchardef\nLeftrightarrow="3\msb@3C
\mathchardef\nleftrightarrow="3\msb@3D
\mathchardef\divideontimes="2\msb@3E
\mathchardef\varnothing="0\msb@3F
\mathchardef\nexists="0\msb@40
\mathchardef\mho="0\msb@66
\mathchardef\eth="0\msb@67
\mathchardef\eqsim="3\msb@68
\mathchardef\beth="0\msb@69
\mathchardef\gimel="0\msb@6A
\mathchardef\daleth="0\msb@6B
\mathchardef\lessdot="3\msb@6C
\mathchardef\gtrdot="3\msb@6D
\mathchardef\ltimes="2\msb@6E
\mathchardef\rtimes="2\msb@6F
\mathchardef\shortmid="3\msb@70
\mathchardef\shortparallel="3\msb@71
\mathchardef\smallsetminus="2\msb@72
\mathchardef\thicksim="3\msb@73
\mathchardef\thickapprox="3\msb@74
\mathchardef\approxeq="3\msb@75
\mathchardef\succapprox="3\msb@76
\mathchardef\precapprox="3\msb@77
\mathchardef\curvearrowleft="3\msb@78
\mathchardef\curvearrowright="3\msb@79
\mathchardef\digamma="0\msb@7A
\mathchardef\varkappa="0\msb@7B
\mathchardef\hslash="0\msb@7D
\mathchardef\hbar="0\msb@7E
\mathchardef\backepsilon="3\msb@7F
\def\Bbb{\ifmmode\let\next\Bbb@\else
 \def\next{\errmessage{Use \string\Bbb\space only in math mode}}\fi\next}
\def\Bbb@#1{{\Bbb@@{#1}}}
\def\Bbb@@#1{\fam\msbfam#1}

\catcode`\@=12